\documentstyle[11pt,aaspp4]{article}
\begin{document}
%
%
\def\etal{{\it et al.}\ }
\def\abs#1{\left| #1 \right|}
\def\EE#1{\times 10^{#1}}
\def\gcm{\rm ~g~cm^{-3}}
\def\kms{\rm ~km~s^{-1}}
\def\ergs{\rm ~erg~s^{-1}}
\def\isotope#1#2{\hbox{${}^{#1}\rm#2$}}
\def\wl{~\lambda~}
\def\wll{~\lambda~\lambda~}
\def\HI{{\rm H\,I}}
\def\HII{{\rm H\,II}}
\def\HeI{{\rm He\,I}}
\def\HeII{{\rm He\,II}}
\def\HeIII{{\rm He\,III}}
\def\CI{{\rm C\,I}}
\def\CII{{\rm C\,II}}
\def\CIII{{\rm C\,III}}
\def\CIV{{\rm C\,IV}}
\def\NI{{\rm N\,I}}
\def\NII{{\rm N\,II}}
\def\NIII{{\rm N\,III}}
\def\NIV{{\rm N\,IV}}
\def\NV{{\rm N\,V}}
\def\NVI{{\rm N\,VI}}
\def\NVII{{\rm N\,VII}}
\def\OI{{\rm O\,I}}
\def\OII{{\rm O\,II}}
\def\OIII{{\rm O\,III}}
\def\OIV{{\rm O\,IV}}
\def\OV{{\rm O\,V}}
\def\OVI{{\rm O\,VI}}
\def\CaI{{\rm Ca\,I}}
\def\CaII{{\rm Ca\,II}}
\def\NeI{{\rm Ne\,I}}
\def\NeII{{\rm Ne\,II}}
\def\NaI{{\rm Na\,I}}
\def\NaII{{\rm Na\,II}}
\def\NiI{{\rm Ni\,I}}
\def\NiII{{\rm Ni\,II}}
\def\CaII{{\rm Ca\,II}}
\def\FeI{{\rm Fe\,I}}
\def\FeII{{\rm Fe\,II}}
\def\FeIII{{\rm Fe\,III}}
\def\FeIV{{\rm Fe\,IV}}
\def\FeV{{\rm Fe\,V}}
\def\FeVII{{\rm Fe\,VII}}
\def\CoII{{\rm Co\,II}}
\def\CoIII{{\rm Co\,III}}
\def\ArI{{\rm Ar\,I}}
\def\ArII{{\rm Ar\,II}}
\def\MgI{{\rm Mg\,I}}
\def\MgII{{\rm Mg\,II}}
\def\MgIII{{\rm Mg\,III}}
\def\SiI{{\rm Si\,I}}
\def\SiII{{\rm Si\,II}}
\def\SiIII{{\rm Si\,III}}
\def\SiIV{{\rm Si\,IV}}
\def\SI{{\rm S\,I}}
\def\SII{{\rm S\,II}}
\def\SIII{{\rm S\,III}}
\def\FeI{{\rm Fe\,I}}
\def\FeII{{\rm Fe\,II}}
\def\FeIII{{\rm Fe\,III}}
\def\FeIIV{{\rm Fe\,IV}}
\def\kI{{\rm k\,I}}
\def\kII{{\rm k\,II}}
\def\La{{\rm Ly}\alpha}
\def\Ha{{\rm H}\alpha}
\def\Hb{{\rm H}\beta}
\def\Hg{{\rm H}\gamma}
\def\He{{\rm H}\epsilon}
\def\Paa{{\rm Pa}\alpha}
\def\Pag{{\rm Pa}\gamma}
\def\Bra{{\rm Br}\alpha}
\def\Brg{{\rm Br}\gamma}
\def\Lya{{\rm Ly}\alpha}
\def\Msun{{~\rm M}_\odot}
\def\Msunyr{M_\odot~yr^{-1}}
\def\tyr{t_{\rm yr}}
%
%
\title{Late Spectral Evolution of SN 1987A: \\
II. Line Emission} 
\author{Cecilia Kozma and Claes Fransson}
\affil{Stockholm Observatory,
S-133$\,$36 Saltsj\"obaden, Sweden}
\authoremail{cecilia@astro.su.se, claes@astro.su.se}
\begin{abstract}
Using the temperature and ionization  calculated in our previous
paper, we model
the spectral evolution of SN 1987A. We find that the temperature
evolution is
directly reflected in the time evolution of the lines. In particular,
the
IR-catastrophe is seen in the metal lines as a transition from thermal
to
non-thermal excitation, most clearly in the [O I]$\wll$6300, 6364
lines. The good agreement with observations clearly confirms the
predicted
optical to IR-transition. Because the line emissivity is independent
of
temperature in the non-thermal phase, this phase has a strong
potential for
estimating the total mass of the most abundant elements. 
The
hydrogen lines arise as a result of recombinations following
ionizations in the
Balmer continuum during the first $\sim$ 500 days, and as a result of
non-thermal ionizations later.

The distribution
of the different zones, and 
therefore the gamma-ray deposition, is determined from the line profiles
of the most important lines, where possible. We find that hydrogen
extends into the core to $\lesssim 700 \kms$. The hydrogen envelope has
a density 
profile close to $\rho \propto V^{-2}$ from $2000 - 5000 ~\kms$. The
total mass of hydrogen-rich gas is $\sim 7.7 \Msun$, of which $\sim
2.2 \Msun$ is mixed within 2000 $\kms$.
The helium mass derived from the line fluxes is sensitive to
assumptions about the degree of redistribution in the line.
The mass of the helium dominated
zone is consistent with $\sim 1.9 \Msun$, with a further $\sim 3.9 \Msun$
of helium residing in the hydrogen component. Most of the oxygen-rich gas
is confined to  400 -- 2000 $\kms$, with a total mass of $\sim 1.9
\Msun$. Because of uncertainties in the modeling of the non-thermal
excitation of the 
[O I] lines, the uncertainty in the oxygen mass is considerable. In
addition, masses of nitrogen, neon, magnesium, iron and nickel
are estimated. 

The dominant contribution to the line luminosity often
originates in a 
different zone from where most of the newly synthesized material
resides. This applies to e.g. carbon, calcium and iron. The [C I]
lines, mainly arising in the helium zone,
indicate a substantially lower abundance of carbon mixed with helium
than stellar evolution models give, and a more extended zone with CNO
processed gas is indicated. The [Fe II] lines
have in most phases a strong contribution from primordial iron, and at
$t \gtrsim 600 - 800$ days this component dominates the [Fe II] lines. The
wings of the [Fe II] lines may therefore come from primordial iron,
rather than synthesized iron mixed to high velocity. 
Lines from ions with low ionization potential indicate that the UV
field below at least $1600 $ \AA~ is severely quenched by dust
absorption and resonance scattering. 

\end{abstract}



\section{INTRODUCTION}
In Kozma \& Fransson (1997; hereafter Paper I) we calculated the
temperature and
ionization in the ejecta of SN 1987A, which determine the 
 line and continuum
emission. Because one of the most important goals is to derive abundances
and masses
of the synthesized elements, these parameters are crucial for understanding the conditions in the
ejecta. As will be shown in this paper, many
lines have
strong contributions from several abundance zones, and the emission
may in fact
be dominated by zones where only a small fraction of the mass of the
element
resides. In Paper I, it was shown that the temperature of the different
abundance
regions differ considerably, which puts an analysis based on a
uniform
temperature into question. In addition, the emission from a given zone
is
influenced strongly by the composition. Even a trace amount of an
efficient
cooler, like Ca II, can  quench the emission from other lines. Finally,
there is an interaction between the different abundance zones, mediated by
the
radiation. An obvious example is the emission from the hydrogen lines,
which
before $\sim$ 400 days is powered mainly by UV emission from the other
regions
absorbed by the Balmer continuum.

In spite of these complications, there has been important progress
based on more
limited forms of analysis.
In a series of papers Li, McCray \& Xu, with coworkers, (Li
\& McCray 1992, 1993, 1995, Li, McCray \& Sunyaev 1993, Xu \etal
 1992)  analyze the most important
emission lines from SN 1987A. Additional discussions are found in
Kozma \& Fransson (1992: hereafter KF92), Wang \etal 1996,  and Chugai
\etal (1997).

 From the time evolution of the hydrogen
lines Xu \etal (1992) and KF92 find that for times earlier than $\sim$ 400 days
photoionization from 
the $n=2$ level dominates the ionization of hydrogen. In KF92 it was
shown that the UV-emission from the ejecta, emitted as a result of the
gamma-ray thermalization, could provide the necessary
source for the Balmer continuum. 
Recently, Chugai \etal (1997) have modeled the time evolution of
$\Ha$, using a time-dependent model for the hydrogen-rich gas, similar
to that in this paper and in Fransson, Houck \& Kozma (1996). They find that
they indeed get a good fit to the observations of $\Ha$ up to the last
HST-observations at 2870 days after explosion. 

An important clue to the conditions in the ejecta was obtained from an
analysis of the [O I] $\wll 6300,6364$ lines, using the fact that the
lines went from optically thick to thin.  From this
Spyromilio \& Pinto (1991) and Li \& McCray (1992)  estimate the density of the oxygen emitting gas and the filling factor and
temperature of the oxygen component.
In a further paper the [Ca II] $\wll 7291, 7324$ and Ca II $\wll 8600$ lines
are studied by Li \& 
McCray (1993). Their main conclusion is that the calcium lines do not
originate from the newly synthesized calcium, but from primordial calcium within the
hydrogen-rich regions. A similar conclusion was reached by Fransson \&
Chevalier (1989) in the context of Type Ib supernovae. 

In modeling the  infrared emission lines of iron, cobalt and nickel, 
Li, McCray \& Sunyaev (1993) conclude that the iron-rich component
must have a filling factor $\gtrsim 0.30$ in the core. However, they
do not take the contribution of other composition regions into
account. A large filling factor for the iron is also
indicated from dynamical arguments (Basko 1994, Herant \& Benz
1992). Finally, based on  modeling of the He I $\wl 1.0830 ~\mu$m and He I
$\wl 2.058 ~\mu$m lines Li \& McCray (1995) estimate the helium mass as
$\sim$ 3 $\Msun$ plus a similar amount mixed with hydrogen.

An interesting result by Chugai \etal (1997) 
is that they find that the intensity of the Fe II lines can 
be explained only if trapping of the positrons from ${}^{44}$Ti is
efficient in the iron-rich parts of the ejecta. Because  Coulomb
scattering is not efficient enough, this puts interesting constraints
on the strength of the magnetic field. At earlier epochs trapping
should be more efficient.

In this paper we exploit our results in Paper I for the calculation
of the line emission from SN 1987A, using a time-dependent formalism
with non-thermal processes included.

Our basic model, assumptions, and
physical processes included, as well as the explosion models we use,
have been discussed in Paper I. A brief summary of this is given in
section ~\ref{sec-modelsum}. 
In section ~\ref{sec-lineprofile} we discuss the
calculation and importance of the line profile for constraining the
density distribution, as well as the influence of
dust absorption.
Section ~\ref{sec-results}  contains a detailed discussion of our
results, while in section ~\ref{sec-shortcomings} we discuss the limitations
of our modeling, and the sensitivity of our results to various
assumptions and simplifications. 
In section
  ~\ref{sec-discussion} we discuss our  results, and in section
\ref{sec-conclusion} we briefly summarize  
our main conclusions.

\section{SUMMARY OF THE MODEL}
\label{sec-modelsum}
Because all details of the model are given in Paper I, we only
summarize the
main features here. All important elements are included with the
ionization
stages important in this analysis. H I, He I, O I, Ca II, and Fe I-IV
are
treated as multilevel atoms. The equations of the ionization balance
and
individual level populations are solved time-dependently to include 
freeze-out
effects. Also the energy equation is solved time-dependently, allowing
for
adiabatic cooling by the expansion. Radioactive decay of
$\isotope{56}{Co}$,
$\isotope{57}{Co}$ and $\isotope{44}{Ti}$ are responsible for the
energy input
in forms of gamma-rays and positrons. The thermalization of these is 
calculated
by solving the Spencer-Fano equation for the abundances and electron
fraction in
the zone.

As input models we use the 10H model (Woosley \& Weaver 1986, Woosley
 1988) and the 11E1
model (Nomoto, \& Hashimoto 1988, Shigeyama, Nomoto, \& Hashimoto
 1988, Shigeyama \& Nomoto  1990, Hashimoto \etal 1989)
for the abundances of the elements within the different burning
zones. The
ejecta is divided into a number of concentric zones of different
compositions,
masses and filling factors. Although simplified, this mimics the
mixing of the
different burning zones in the ejecta. Outside of the core we attach a
hydrogen
envelope with a density profile fixed to give agreement with the line
profile of
the $\Ha$ line (see section ~\ref{sec-hydrogen}).

\section{LINE PROFILES}
\label{sec-lineprofile}
The best way of constraining the emissivity variation, and therefore the
density distribution, comes from modeling of the line profiles of the
relevant lines. For a spherically symmetric distribution, it can be shown
that for an optically thin line, the emissivity, $j(V)$, is given by
\begin{equation}
j(V) \propto {1 \over V}~{dI_{\nu}(V) \over dV}  
\label{eq:lprof_1}
\end{equation}
(Fransson \& Chevalier 1989). If $j(V)$ is proportional to the gamma-ray
deposition, and if one can consider the  gamma-ray source as central this
can be transformed into 
\begin{equation}
\rho(V) \propto V~{dI_{\nu}(V) \over dV} .
\label{eq:rhocen}
\end{equation}
The first of these approximations is reasonable for many lines. In
particular, lines dominated by recombination, following non-thermal ionization,
or lines dominated by non-thermal excitation belong to this category. Examples
of the former class are the H I and He I lines, while the [O I] $\wll 6300,
6364$ lines at stages later than $\sim 1000$ days belong to the latter. If
one line dominates the cooling of a given region this will also be a good
approximation. One example of this is the [Fe II] $\wl 25.99 ~\mu$m line from the Fe
-- He zone at $t \gtrsim 800$ days.  

The assumption of a central
gamma-ray source  is doubtful at velocities outside of which there is a
substantial amount of radioactive material,  $V_{\rm Ni}$, in this case
$V_{\rm Ni} \approx 2000 \kms$. 
A general gamma-ray deposition function, $D_{\gamma}$, is discussed in KF92,
defined so that the mean intensity, $J_{\gamma}$, can be written as
\begin{equation}
J_{\gamma} = \frac{L_{\gamma} D_{\gamma}}{16 \pi^2 R_{\rm core}^2},
\end{equation}
where $R_{\rm core}$ is the core radius. For such a general deposition function
the density distribution can be obtained from
\begin{equation}
\rho(V) \propto {1 \over D_{\gamma}}~{1 \over V}~{dI_{\nu}(V) \over dV}. 
\end{equation} 
Approximations for $D_{\gamma}$ are discussed in KF92 for the case of uniform
distribution of the radioactive material, and outside of the core. Within the
core 
 one finds $J_{\gamma} \approx$ constant,
and 
\begin{equation}
\rho({V}) \propto {1 \over V}~{dI_{\nu}(V) \over dV} \quad {\rm for ~} V <
V_{\rm Ni}.
\label{eq:rhoconst}
\end{equation}
At $V \gg V_{\rm Ni}$ one has $D_{\gamma} \propto 1/V^2$ (as for a central source),
and  equation (\ref{eq:rhocen})  is recovered. The important result is here the
correspondence between the density distribution and the derivative of the line
profile with respect to velocity.

Observations indicate that dust was present in the supernova ejecta at $\sim$
530 days (Lucy \etal
1989, 1991). The onset of the dust formation is likely to have taken
place already at $\sim 350$ days and was completed at $\sim 600$ days
(Meikle \etal 1993, Whitelock \etal 1989). 
From the fact that the dust seemed to be opaque both in the optical
and IR (Spyromilio \etal 1990, Haas \etal 1990), it has been suggested
that the dust is clumpy, with very 
high optical depth in each clump. The latter is required to explain
the presence of line asymmetries even at day 1806
(Bouchet
\etal 1996). Lucy \etal (1991) estimate a covering factor $f_{\rm cov}
\approx 0.4$ of the opaque dust. A similar covering factor, $f_{\rm cov} \approx 0.55$, is
estimated by Wooden \etal (1993) from their dust temperatures and
luminosities, based on a maximum dust velocity of $V_{\rm dust}
\approx 1800 \kms$. 

The presence of dust will decrease the escaping fluxes
in lines and continua by a factor which conveniently can be parameterized by
a covering factor, $f_{\rm cov}$, applicable inside a velocity $V_{\rm dust}$. In
this paper we treat this as a pure absorption process. The
thermalization of the absorbed radiation is discussed in more detail in
connection with the photometry in in a subsequent paper. Here we only
summarize points relevant for this paper. 
To model the absorption  by the dust we assume a model
 where optically thick clumps of dust
form at 350 -- 600 days.  In the dust
formation phase it is probable
that the covering factor depends on time, while later it is likely to
be constant, with the optically thick dust clumps expanding with the
rest of the ejecta. We therefore assume a linear increase in the covering
factor from 350 to 600 days. At that time the covering factor is
$f_{\rm cov} = 0.40$. 
Dust is likely to form only in the metal-rich parts of the ejecta (Kozasa
\etal 1991). We therefore assume that only emission inside the velocity
$V_{\rm dust}$ is affected by the dust absorption. Here we take
$V_{\rm dust} = 2000 
\kms$, similar to the core velocity.

The dust may have important consequences for the UV intensity in
the ejecta. Multiple resonance scattering increases the path length of
the photons, increasing the effective probability for absorption of
the UV photons by the dust in the core. This may be very efficient in
decreasing the ionizing intensity in the ejecta, and introduces a
major uncertainty in the calculations. The effects of a lower ionizing
flux will be discussed in section \ref{sec-phot}.
Besides absorbing the radiation the dust may also cool the gas
efficiently. This was discussed in Paper I.

In our models we calculate the line profile directly from the source function,
$S(r)$, and optical depth in the line, $\tau(r)$, as function of
radius. 
The intensity at a given
velocity is in the red wing
\begin{eqnarray}
I_{\nu}(V) & = & 2 \pi [ 
\int_{V R_0 /V_0}^{R_{\rm dust}} (1 - f_{\rm cov}) S(r) (1 - e^{-\tau(r)})~
r ~dr \cr 
&& + \int_{R_{\rm dust}}^{R_0} S(r) (1 - e^{-\tau(r)})~ r ~dr]
,
\label{eq:lprof}
\end{eqnarray}
where $V_0$ is the velocity at the maximum radius of the ejecta,
$R_0$, and $R_{\rm dust} = [(V_{\rm dust} t)^2+(V R_0 /V_0)^2]^{1/2}$. We
have here neglected scattering of the background continuum, responsible for
the P-Cygni absorption. More general cases, including this effect, are discussed
in Fransson (1984). For an optically thin line without dust 
absorption equation (\ref{eq:lprof}) reduces
to 
\begin{equation}
I_{\nu}(V) = 2 \pi \int_{V R_0 /V_0}^{R_0} j(r) ~ r ~dr,
\label{eq:lprofj}
\end{equation}
where $j(r)$ is the emissivity.

\section{RESULTS}
\label{sec-results}
\subsection{Line Emission}
In order to compare our calculated line fluxes directly to
observations we redden the calculated fluxes using $E_{B-V}$=0.06 from the Galaxy
and $E_{B-V}$=0.10 from the LMC (Sonneborn \etal 1996). The extinction
curves are taken from Savage \& Mathis (1979), 
 and Fitzpatrick (1985). 
In addition, as explained in section~\ref{sec-lineprofile}, the internal dust absorption in
the ejecta is taken into account for the emission from the core region.

\subsubsection{Hydrogen}
\label{sec-hydrogen}
The hydrogen lines are discussed in detail in Xu \etal (1992) and
KF92, and we will therefore be fairly brief in this respect.
A major issue in connection with the hydrogen emission is the density
distribution in the ejecta. As we discussed in section~\ref{sec-lineprofile}, the
best way of constraining this is by the line profiles. Although a
complete investigation of this question is a major, separate issue, we
have made some experiments with different, simplified models. 

We divide the hydrogen distribution into a core contribution,
most likely caused by mixing of hydrogen into the metal core, and an
envelope component, which is 
probably relatively undisturbed by the mixing. Of these, the
core component is especially uncertain. In the simulations by Fryxell, M\"{u}ller,
\& Arnett (1991) they find a hydrogen mass of $\sim 1 \Msun$, inside of
$2000 \kms$, while Herant \& Benz (1992) find $\sim 2 \Msun$. Typical
filling factors are $20 - 40$  \%. We therefore treat the
hydrogen core mass
as a free parameter. For the envelope we use the density distribution
of the Shigeyama \&  Nomoto  (1990) 14E1 model, as well as our own,
parameterized models. 

\begin{figure}
\plotone{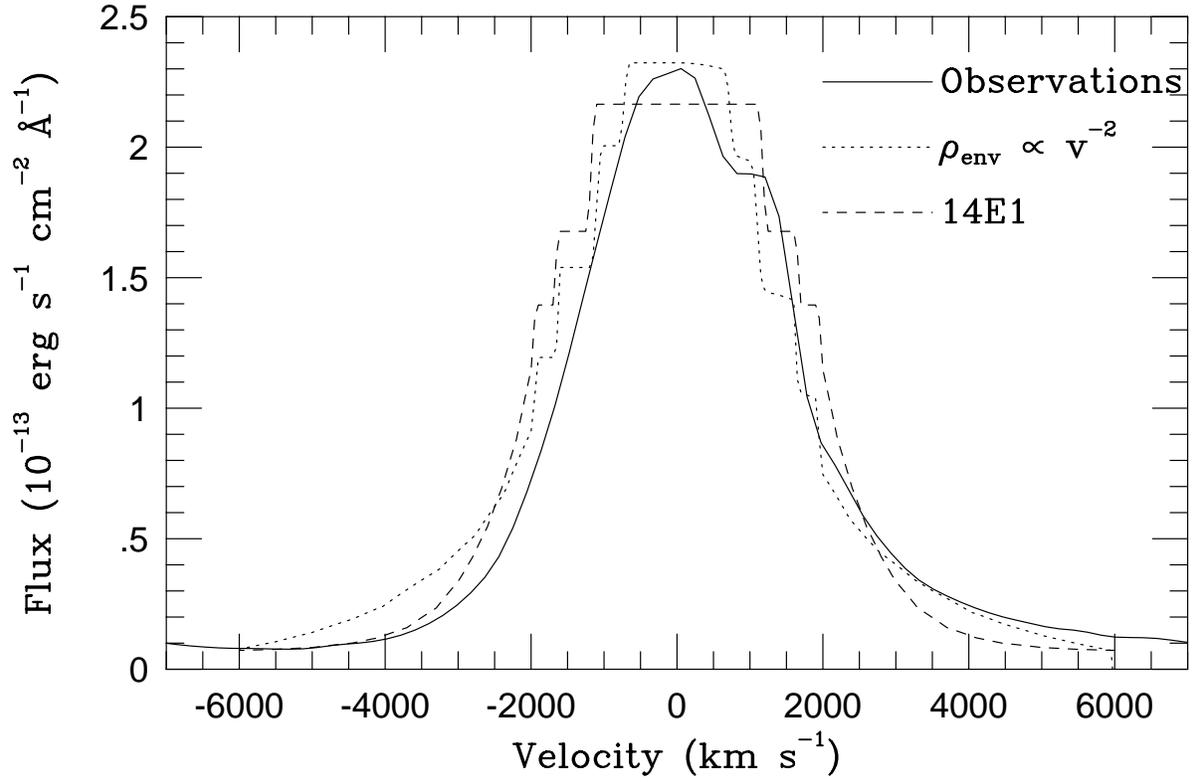}
\caption{The $\Ha$  line profile  at $\sim$ 800 days.
The observed line profile from Phillips \etal (1990) (solid line),
compared to 
two of our models for the hydrogen distribution. The bump at $\sim 933
\kms$ is due to an 
unresolved, narrow [N II] $\wl 6583$ line from the ring. Note the too
weak line wings and the too flat peak  in the 14E1 model.
\label{fig:hprof}}
\end{figure}

In
Figure \ref{fig:hprof} we show the observed $\Ha$ line profile at 804 days,
taken from Phillips \etal (1990), together with our models.
Because only the envelope contributes at velocities greater
than $V_{\rm core}$ from the line center, we first discuss the effects of
this on the line wings. The flux in the blue wing is complicated by
the fact that $\Ha$ is optically thick. Photons from the background
(i.e., continuum and other weak lines) are therefore scattered by
$\Ha$, 
and a P-Cygni absorption will result. Although we take optical depth
effects into account for the line itself, we do not include scattering
by other lines and continua. This most likely explains the asymmetry
of the blue and red wings of $\Ha$, and for this reason we concentrate
on the red wing in this discussion. Dust scattering is, however,
included in the emission from the core and behind it.
 The bump at $933  \kms$ is caused by an unresolved [N II]
$\wl 6583$ line from the circumstellar ring, as can be verified from
high resolution observations. A weaker bump at $682 \kms$ corresponds
to the [N II] $\wl 6548$ component of this doublet.

From Figure \ref{fig:hprof} it is clear that the 14E1 model gives 
too low a flux in the red wing. The reason is that the density
profile in the envelope of this model is too steep. The 11E1 model,  with a lower 
envelope mass, has an even steeper density gradient, and therefore does 
 not improve the fit. For this reason we have simply parameterized
the envelope distribution by $\rho = \rho_0~(V_0/V)^\alpha$, where $\rho_0$ is the
density at velocity $V_0$. $\rho_0$ is determined by  the total
mass of the hydrogen envelope in the  model. 
The exact
comparison is somewhat sensitive to the continuum level assumed. As
equation (\ref{eq:lprof_1}) shows, the derivative of the line profile
is, however, 
more relevant than the absolute level, and we find that a $\rho
\propto V^{-2}$ 
model, having a similar slope in the wing as the observation, is for
this reason the most satisfactory model. 
The resulting line profile at $\sim$ 800 days is shown in Figure
\ref{fig:hprof}, and 
the flatter density distribution improves the fit to the red wing
considerably. However, between $5000 - 6000 \kms$ the $\rho \propto
V^{-2}$ model has a  derivative that is too steep, and the 14E1 model
is closer to
the observed slope. The mass within this velocity region is in the
14E1 model $0.24 \Msun$, while in the  $\rho \propto
V^{-2}$ model the mass between $5000 - 6000 \kms$ is $1.8 \Msun$. The
maximum mass in this region 
should therefore be close to that of the 14E1 model, and $\lesssim 0.5
\Msun$. Our favored value of the total mass  
outside $V_{\rm core} = 2000 \kms$ is therefore $\sim 5.5 \Msun$.

Having fixed the envelope component, we now add a core contribution,
which gives the additional flux needed inside $V_{\rm core}$. The fact
that the line profile is clearly peaked ($dI_{\nu}/d \vert V
\vert > 0$) to $\lesssim 700 \kms$ shows that hydrogen is present
at least to this velocity. We find that a fairly uniform distribution
of mass $\sim 2 \Msun$ in the core between $1000 - 2000 \kms$ gives an
acceptable fit to the profile. Between $700 - 1000 \kms$ we need $\sim
0.2 \Msun$.   
The mass within the core is therefore $\sim 2.2 \Msun$, and the total 
mass of hydrogen-rich gas in the ejecta $5.5 + 2.2 = 7.7 \Msun$. Of
this $\sim 3.9 \Msun$ is hydrogen, while most of the rest is helium.  

\begin{figure}
\plotone{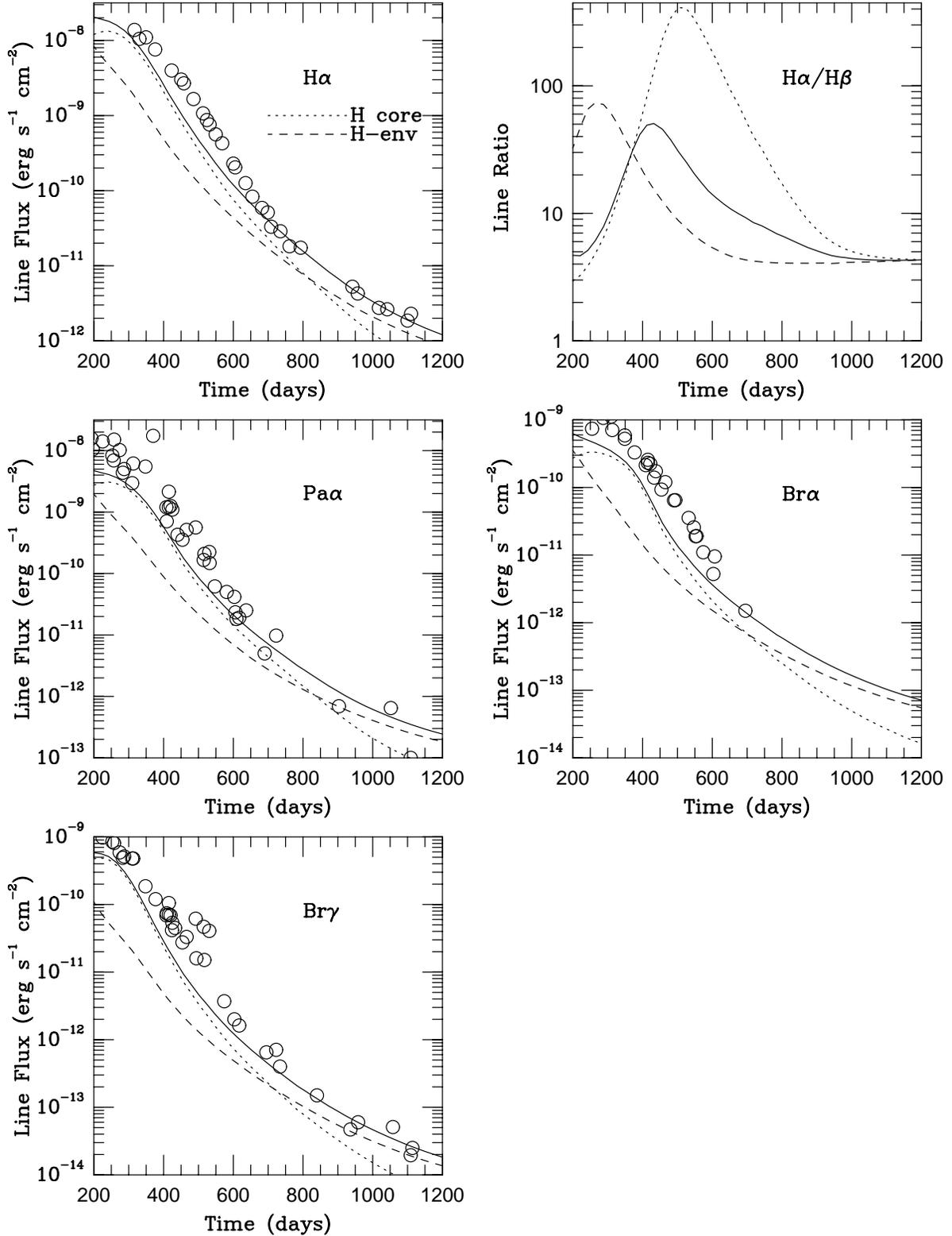}
\caption{The time evolution for different hydrogen lines (solid lines), together with observations
(open dots). Data from Danziger \etal (1991), Bouchet \&
Danziger (1993), Wooden \etal (1993), Meikle \etal (1989, 1993), and Bautista \etal (1995).
The contributions from the core
region are shown as  dotted lines and from the envelope as dashed lines.
\label{fig:phothi}}
\end{figure}

\begin{figure}
\plotone{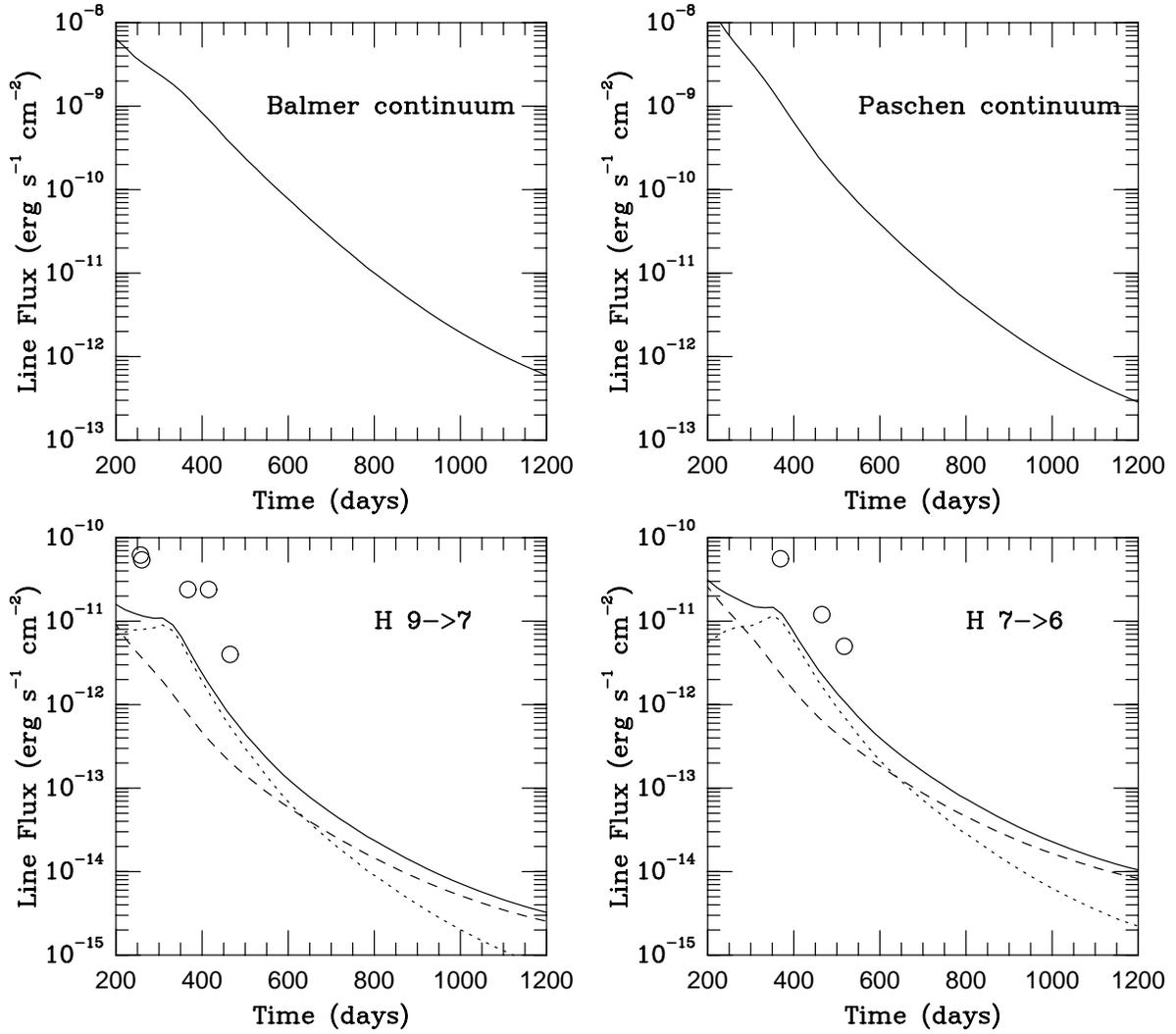}
\caption{Same as Figure 2 
but for the Balmer and
Paschen continuum, as well as for the H 9$\rightarrow$7, and 
the H 7$\rightarrow$6 lines. Observations are from Roche,
Aitken, \& Smith (1993), and Wooden \etal (1993).
\label{fig:phothi2}}
\end{figure}

With  the hydrogen density distribution determined by the line profile, we
discuss the light curves for the model with $M_{\rm H}(\rm core) \approx
2.2 \Msun$ and $\rho_{\rm env} \propto V^{-2}$.
In Figures \ref{fig:phothi} and \ref{fig:phothi2} we show the time evolution for $\Ha$, $\Paa$, $\Bra$, $\Brg$, the $\Ha$/$\Hb$ ratio, the Balmer, and
Paschen continua, and the H 9$\rightarrow$7, and H 7$\rightarrow$6 transitions.
The solid line is the total calculated line flux, using a full
hydrogen atom with all $nl$-states up to $n=20$ included (see appendix
in Paper I), which should be
compared to the observations. In the figures we also show the
contributions from the core regions (dotted line), and the hydrogen
envelope (dashed line). Of these,
the contribution from the core regions dominates up to $\sim 700-900$ days,
after which the envelope contributes most. This is a result of 
freeze-out, which is most pronounced, in the outer, low density, 
envelope regions.

In our calculations $\Ha$ is optically thick in the core up to $\sim$
1300 days. $\Ha$  becomes optically thin in the outer parts of
the envelope already at $\sim$ 550 days, and the entire envelope is thin at
$\sim$ 850 days. $\Hb$ is optically thick in the core regions up to day
$\sim 1000$, while in the outer regions of the envelope $\Hb$ becomes 
optically thin after $\sim$ 350 days. The inner envelope regions
become optically thin at  $\sim$ 700 days. The total $\Ha/\Hb$
ratio, as well as the ratios from the core and envelope regions, are
shown in Figure \ref{fig:phothi}. 
When the Balmer lines are optically thick, i.e. Case C, $\Hb$ splits
into one $\Paa$ and one $\Ha$ photon. As  $\Hb$ becomes thin
the $\Ha/\Hb$ ratio settles at a value of $\sim$ 3.7 at 1200
days, slowly increasing to $\sim$ 4.3 at 2000 days.
Taking reddening into account, the ratio
should be multiplied by a factor of 1.18, resulting in an observed $\Ha/\Hb$ $\sim$
4.3 at 1200 days, as seen in Figure \ref{fig:phothi}.

Martin (1988) and Storey \& Hummer (1995) discuss hydrogen
recombination at low temperature. Martin does not take collisions 
into account, i.e. 
 recombination in the low density limit. He compares 
calculations using an $n$-method (the different $l$-states are
populated according to statistical weights), and an $nl$-method
where he treats the $l$-substates explicitly. In Case B he finds that at 500 K
the $\Ha/\Hb$-ratio is 2.5 and 4.2 for the $n$ and $nl$-methods
respectively, showing the importance of treating the
$l$-states individually. The assumption of complete $l$-mixing (the
$n$-method), however, becomes better for higher densities. Storey \&
Hummer include both individual $l$-states, as well as collisional
processes.
For an electron density of $10^4$ cm$^{-3}$, a temperature of 500 K,
and Case B they find an $\Ha/\Hb$-ratio of 3.68. The collisional
processes makes the $n$-method a better approximation and reduces the
ratio. We find in our models that $l$-mixing is accurate 
up to $\sim 500$ days. At later time the decreasing density causes
increasing deviations from this approximation. At 1200 days this gives
 differences by $\gtrsim 50 \%$ for lines like $\Ha$ and Pa$\alpha$. At the
same epoch the $\Ha / \Hb$ ratio is a factor of about two higher in
the model with individual $l$-states. Therefore, at least at late time
an accurate calculation requires inclusion of these effects. In our calculations at 2000 days Case B is valid, but our
temperature, as well as the electron density, is lower than in the
example from Storey \& Hummer, resulting in our somewhat higher
ratio. The effect of decreasing temperature and electron density is to
increase the $\Ha/\Hb$-ratio.

Photoionization from $n=1$ is never a dominant process. 
Instead, for $t \lesssim 500$ days photoionization in the Balmer continuum
is the most important source of ionization, as
was found already in KF92 and Xu \etal (1992). Here we demonstrate this in a more
realistic context, with a more self-consistent treatment of the UV-field,
temperature and ionization.
At $t \gtrsim 500$ days non-thermal ionization from the ground state dominates
photoionization from excited levels, in agreement with KF92.

For  hydrogen we include non-thermal excitations up to $n=4$. 
For $n=2$ the contribution from non-thermal excitations
is larger than direct recombination,
 except for the outermost envelope regions at later times.
For $n=3$ and $n=4$ recombination dominates
during most of the evolution.
Later than $\sim$ 1000 days, however, for $n=3$  the two contributions are of the same
order, while for the $n=4$ level recombinations
 always dominate. 
Within the core
 two-photon emission always dominates the de-population of $n=2$ over 
Ly$\alpha$ emission. In the envelope, however, Ly$\alpha$
emission exceeds two-photon emission.

\subsubsection{Helium}
\label{sec-helium}
\begin{figure}
\plotone{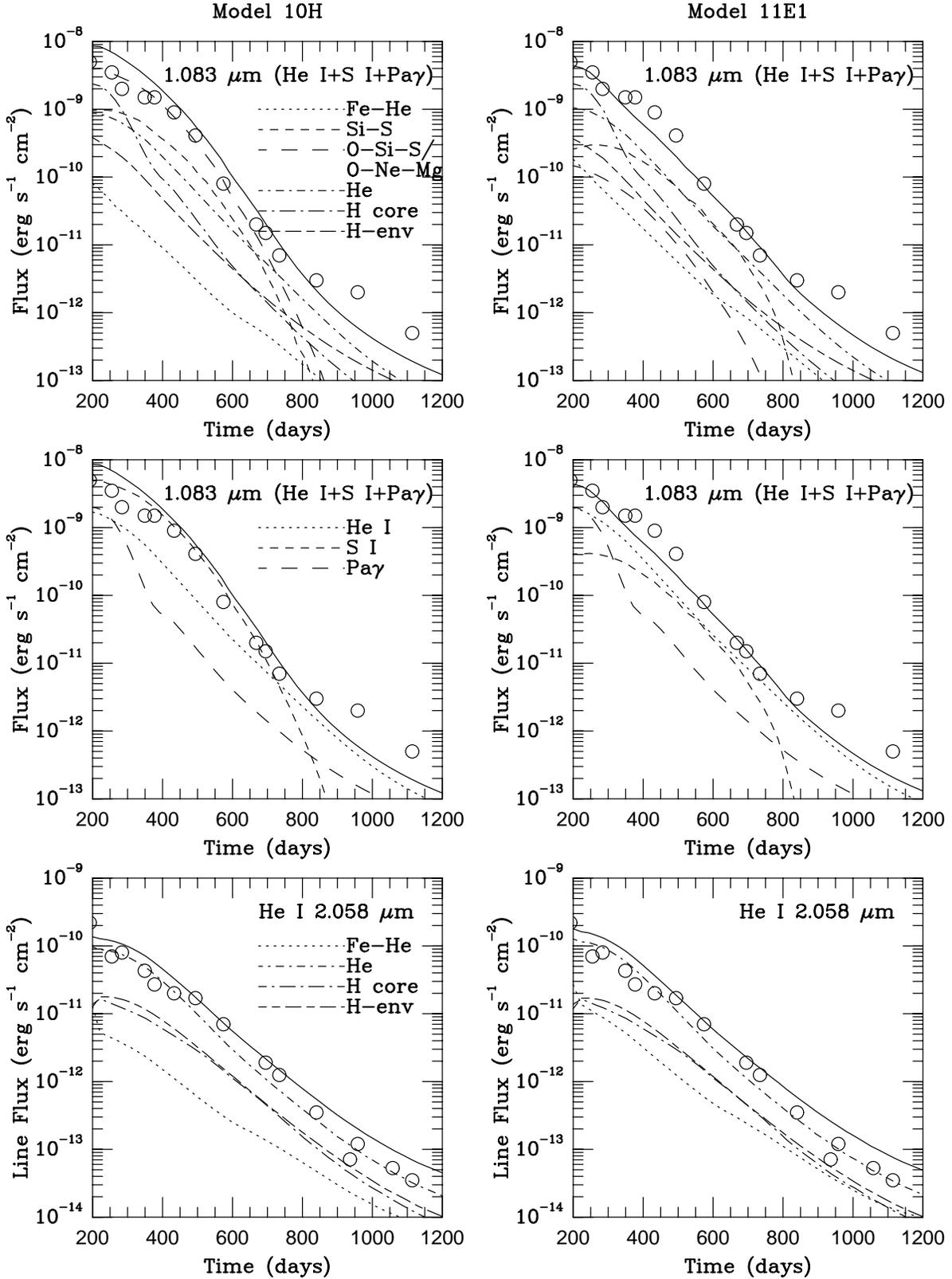}
\caption{Time evolution  of the He I $\wll 1.0830 ~\mu$m and
$2.058~ \mu$m lines. 
Calculations for both the 10H and 11E1 models are shown together with
observations. In the two upper figures the sum of the He I $\wl 1.0830 ~
\mu$m, [S I] $\wl 1.0820 ~\mu$m and Pa$\gamma$ are shown, together with the
contributions from the different composition zones.
In the middle figures the contributions from all zones due to He I
$\wl 1.0830~\mu$m, 
[S I] $\wl 1.0820~ \mu$m, and Pa$\gamma$ are shown.
The lower figures show the total emission due to He I $\wl 2.058 ~\mu$m, together
with contributions from different composition zones.
Observations are from Meikle \etal (1989), (1993), Wooden \etal
(1993), and Bautista \etal (1995).
\label{fig:phothei}}
\end{figure}

The regions contributing to the He I lines (see Fig.
\ref{fig:phothei}) are naturally the helium, but also  the iron and
hydrogen regions. 
In our calculations we find that the fraction of the deposited
non-thermal energy going into these three regions are $\sim$ 8\%, 
$\sim$ 5\%, and $\sim$ 60\%, respectively, at around 500 days, in
accordance with Table 3 in Paper I when allowance for the positron
input to the Fe -- He zone is made.
For 
determination of the helium mass the distribution in velocity is extremely
important, since this directly determines the gamma-ray deposition,
\begin{equation}
\Delta L_{\gamma} = L_{\gamma} ~\Delta \tau_{\gamma} = {L_{\gamma} ~
\kappa_{\gamma}~\Delta M(V) \over 4 \pi ~(V ~t)^2}. 
\label{eq:lgamdep}
\end{equation}
A relatively large mass, $\Delta M(V)$, at high velocity can therefore be
difficult to detect. This is further discussed in Houck \& Fransson (1996)
for SN 1993J. 

\begin{figure}
\plotone{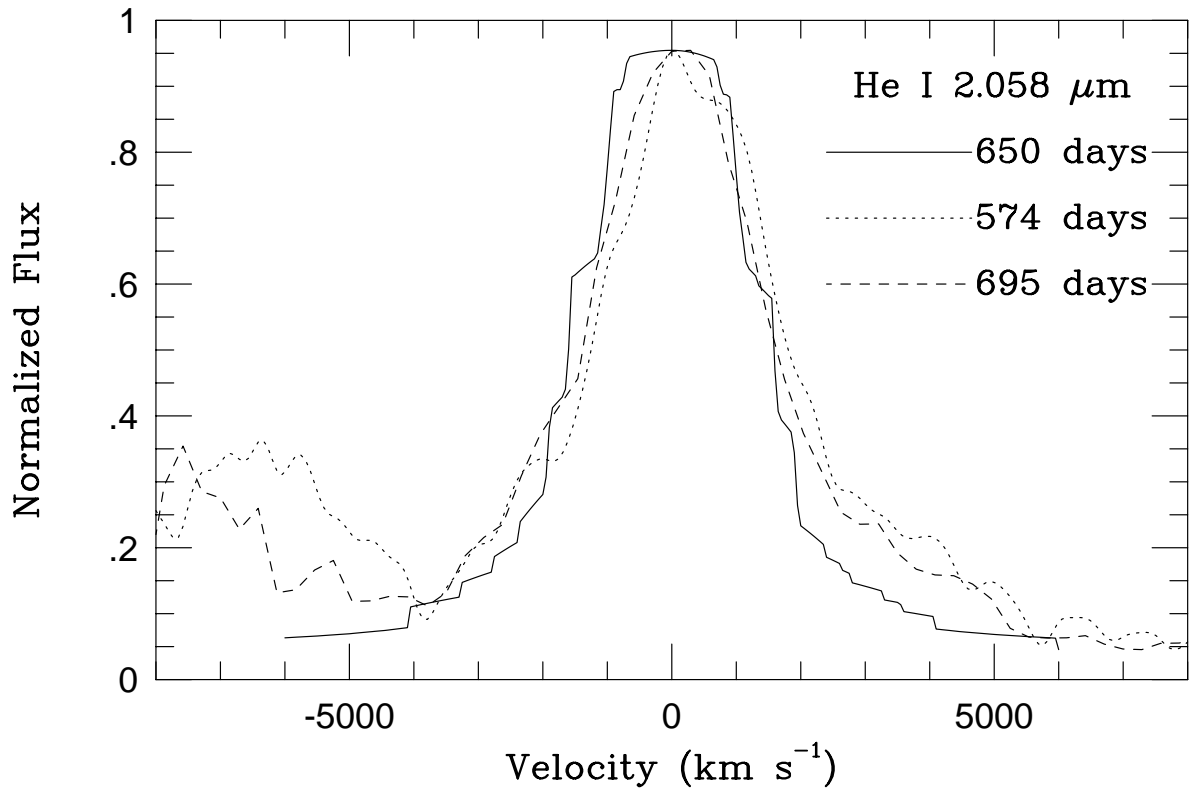}
\caption{The He I $\wl$ 2.058 $\mu$m line at 574 and 695 days from Meikle \etal
(1993), together with our calculated profile at 650 days (solid line), including dust. The
helium distribution used is discussed in the text.
\label{fig:lprofhe}}
\end{figure}

In the same way as for $\Ha$, we constrain the helium distribution from the
line profiles.  Of the He I lines the $\wl 2.058~ \mu$m line is best suited for
this type of analysis, because of the relative absence of blending with other
lines (Fig. \ref {fig:lprofhe}). For this purpose we take the observations by Meikle \etal
(1993) at 574 and 695 days, when the line
is optically thin (see below). Although the line is relatively free from blends,
there may still be weak lines superimposed. Also, the signal to noise
is limited. The helium density distribution based on this line fit is
therefore uncertain, especially for  $V \gtrsim 3000 \kms$. Because of
the change in flux during this time interval, we normalize the line
profiles to the peak flux of the line. The total line fluxes agree
well with the observations, as can be seen from Figure \ref{fig:phothei}.
We find that the helium density is relatively flat between
$1500 - 3500
\kms$, and then falling above
$\sim 4000
\kms$. Therefore, most of the mass is at velocities  $\gtrsim 3000
\kms$. The continued rise of the line profile even inside of $1000 \kms$ 
shows that there is a substantial amount of helium close to the
center. The extension to $\gtrsim 4000 \kms$ is in contrast to the model by
Li
\& McCray (1995), who use
$V_{\rm max} = 2500
\kms$. 
Because most of the helium is outside of the core,
only the fraction in the Fe -- He zone and the helium core fraction are
affected by dust absorption.
In our calculations we have 0.6 $\Msun$ of He within $2000 \kms$ and 1.4
$\Msun$ between 2000 and $4000 \kms$, i.e. a total helium zone mass of $2.0
\Msun$. This is the total mass in the helium zone, of which $\sim  1.9
\Msun$ is pure helium. In addition to this, we have $\sim 3.9 \Msun$ 
of 
helium from the hydrogen-rich regions. 

Figure  \ref{fig:phothei} shows the $ \wl 1.083 ~\mu$m and $ \wl
2.058~ \mu$m  light curves. 
The He I $ \wl 2.058~ \mu$m line is dominated by
emission from the helium regions at all times, and therefore the
differences between the models are small. 
For the $ \wl 1.083 ~\mu$m feature, on the other hand, we find that the contribution from [S I] 
$ \wl 1.0820 ~\mu$m is important, and actually dominates the emission for model 10H,
up to $\sim$ 700 days.  
At later epochs the He I line from the helium zone 
dominates  the light curve. Later than $\sim$ 1200 days the
contributions from the hydrogen 
envelope takes over. From the hydrogen envelope,  most of the
emission is due to He I, but the contribution from  Pa$\gamma$ increases
with time. 

The flux of He I $ \wl 2.058~ \mu$m  is sensitive to the
treatment of the continuum destruction probability of the $\wl 584$
line. The large abundance of carbon in the He -- C zone can cause the
$\wl 584$ line to be absorbed by C I. This process, and the competing
processes of escape and branching to the $ \wl 2.058~ \mu$m line, has
been discussed in KF92 and by Li \& McCray (1995). As was mentioned in
the discussion of the continuum destruction in Paper I, we find that with the destruction probability from
equation (32), in 
Paper I photoabsorption of the $\wl 584$ line is 
important only for $t \lesssim 300$ days.  The continuum destruction
probability was in KF92 estimated to $\sim 7\EE{-6}~(T/5000 {\rm K})^{1/2}~[X({\rm C})/0.018]$. The reason why continuum absorption is at all
important, given the branching probability of $1.1\EE{-3}$ between the
$\wl 2.058 ~\mu$m transition and the $\wl 584$ transition, is that the
optical depth of the $\wl 2.058~\mu$m line is high. At 300 days the
optical depth is in our standard model $\sim 110$, giving roughly
equal probability of branching and absorption. After this epoch the
optical depth rapidly decreases. Li \& McCray
(1995), on the other hand, find that up to $\sim 500$ days most $\wl 584$
photons are destroyed by this process in their He -- C zone. 

\begin{figure}
\plotone{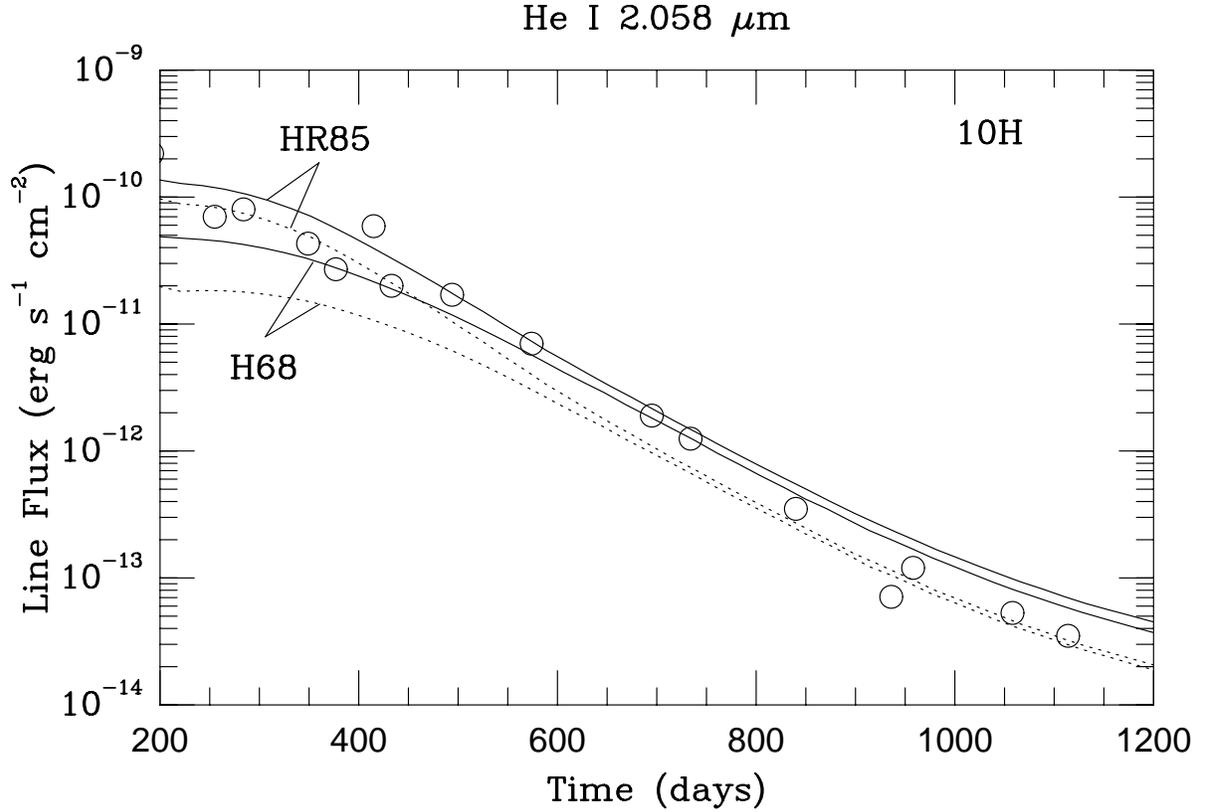}
\caption{The effect of different continuum destruction probabilities
for  He I $\wl 2.058~ \mu$m in the  10H model. The solid lines show the
total luminosity, including all zones, while the dotted lines show the
contribution from the He - C zone alone. The upper
curves, labeled HR85, give the light curves for the destruction
probability given by Hummer \& Rybicki (1985) for a Doppler profile,
assuming complete redistribution, while the lower curves, labeled
H68, gives the same for a Voigt profile and complete redistribution
from Hummer (1968).
\label{fig:photheivoigt}}
\end{figure}

To study
the effect of the form of the continuum destruction probability, and
therefore also  partial versus complete redistribution,
we show in Figure \ref{fig:photheivoigt} the light curves 
for both the case of continuum absorption calculated assuming only
absorption within the Doppler core (eq. [32], 
Paper I), and for the case of absorption
dominated by the damping wings of the line (eq. [33], 
Paper I). 
From this figure we see that the different assumptions give total fluxes
different by a factor of $\sim 2.5$ at 200 days, and a factor of $\sim
5$ for the
emission from the He -- C zone alone. The reason for the different
factors is that continuum destruction is only important in the He -- C
zone, and not in the hydrogen zones. 
The factor of $\sim 5$ can be traced directly to the continuum
destruction probability. For
the damping parameter of the $\wl 584$ line, $a \approx 2\EE{-3}$, and
a typical continuum to line opacity of $k_{\rm C}/k_{\rm L} \approx 2\EE{-6}$, the
Doppler case has a factor of $\sim 8$ lower destruction probability
compared to the damping case. In the damping case continuum
destruction is consequently important up to 500 days, in accordance
with Li \& McCray, although the optical depth in the $\wl 2.058
  ~\mu$m line is a factor 2 -- 3 smaller.
As we remarked in the discussion of the continuum destruction in Paper I, the more realistic case of
partial redistribution gives a lower importance to the line wings,
and a destruction probability closer to the Doppler case, and we
therefore believe that this case gives the best approximation to the
line flux. Chugai's (1987) partial redistribution approximation
(eq. [34], 
Paper I) gives for
$k_{\rm C}/k_{\rm L} \approx 2\EE{-6}$ only a factor $\sim 1.4$ higher
destruction probability than the Doppler case.  

The effect of the continuum absorption can also be seen in a model,
further discussed in next section, where we have replaced carbon
in the helium zone by nitrogen, with an abundance of $\sim
3.5\EE{-3}$. This model has a negligible continuum destruction of the
$\wl 584$ photons, and therefore higher $\wl 2.058~\mu$m flux at early
time. At 200 days the flux is a factor of $1.6$ higher than in Figure
\ref{fig:phothei}, decreasing to $ 1.25$ at 400 days, and $1.08$ at 600
days. This again confirms that mass estimates based on the
$\wl 2.058~\mu$m line are most reliable at $\gtrsim 600$ days. 
As expected, the $\wl 1.0830 ~\mu$m line is
not affected by 
this uncertainty. Instead, it is more sensitive to the temperature as
well as the optical depth, and in addition blending with other lines.

Observations by McGregor (1988), and Meikle \etal (1993) indicate a
large optical depth in the two helium lines for the first couple of
years, based on the asymmetry of the line profile, or
rather the blue-shifted absorption troughs. In the spectra by McGregor
the $ \wl 2.058 ~\mu$m line 
is clearly asymmetric at 437 days, indicating an optical depth
substantially larger than one. At 574 days
the observation by Meikle \etal  is consistent with the
line being either optically thick or thin, while at 695 days  the trough has
clearly disappeared.
In our calculations we find an optical depth larger than one in the helium
zones up to $\sim$ 700 days. The optical depths in the hydrogen
core, and envelope are somewhat smaller, and the $ \wl 2.058 ~\mu$m line
becomes optically thin in these regions at $\sim$ 650 and $\sim$ 400
days respectively.

For He I $ \wl 1.0830 ~\mu$m  it is difficult to extract any
information on the optical depth from the observations, due to the
blending with other lines.
In our calculations we find an optical depth greater than one in the
helium region even for $t \gtrsim 2000$ days. In the hydrogen regions the 
 $ \wl 1.0830 ~\mu$m line becomes optically thin at $\sim$ 900 days.

Li \& McCray (1995) find that in
order to fit the $ \wl 1.0830 ~\mu$m and $ \wl 2.058 ~\mu$m emission
they need $\sim 3 ~\Msun$ of nearly pure helium and
$\sim 11 \Msun$ of 
hydrogen mixed with primordial helium. They
assume a filling factor of 0.30, but point out that their calculations
are not sensitive to the choice of this parameter. Our helium mass 
is lower than Li \& McCray's. 
As we have discussed above, Li \& McCray have a substantially
larger destruction of the $\wl 584$ photons, and therefore a lower
flux in the $\wl 2.058~ \mu$m line than in our model. Consequently, at
early time the contribution from their He -- C zone is very low, and
most of their flux originates at these epochs from the He -- N zone. This
probably explains their higher helium mass. 

In most other respects, however, our calculations agree well with
those of Li \& McCray. In particular, we find the same distribution
between the various contributions to the excitation of the $2p {}^1P$
and $2p {}^3P$ levels. Non-thermal, direct excitation and recombination,
following the non-thermal ionization, give in our models roughly equal
contributions to  the $2p
{}^1P$ level, each $\sim 40 \%$. Earlier than $\sim 600$ days thermal excitations from the $2s
{}^1S$ gives an additional $\sim 10 - 20 \%$. 
The $2p {}^3P$ level has  a large contribution from
recombinations, $\sim 60 \%$ before 800 days, increasing to $\sim
97 \%$ at 1200 days. Because of the meta-stability of the $2s {}^3S$
level, thermal collisions contribute $\sim 50 \%$ earlier than 800
days. At later epochs this contribution falls rapidly because of the
adiabatic decrease of the temperature in the helium zone.

Li \& McCray  find that
in the hydrogen envelope the He I $2s ~ ^3 S$ state is depopulated by
Penning ionizations.
This is confirmed by our calculations.
Photoionization from the ground state of He I is always 
 unimportant. However, photoionization of excited levels  is important
for ionizing He I. Up to $\sim 700$ days the 
photoionization rate from excited levels in He I is somewhat higher
than, or of the same order as, the non-thermal ionization rate. After
$\sim$ 700 days the non-thermal rate slowly becomes more important.
The most important photoionization source is emission from lines in
the UV, and especially the  He I two-photon continua.

\subsubsection{Carbon and Nitrogen}
\label{sec-carbon}
\begin{figure}
\plotone{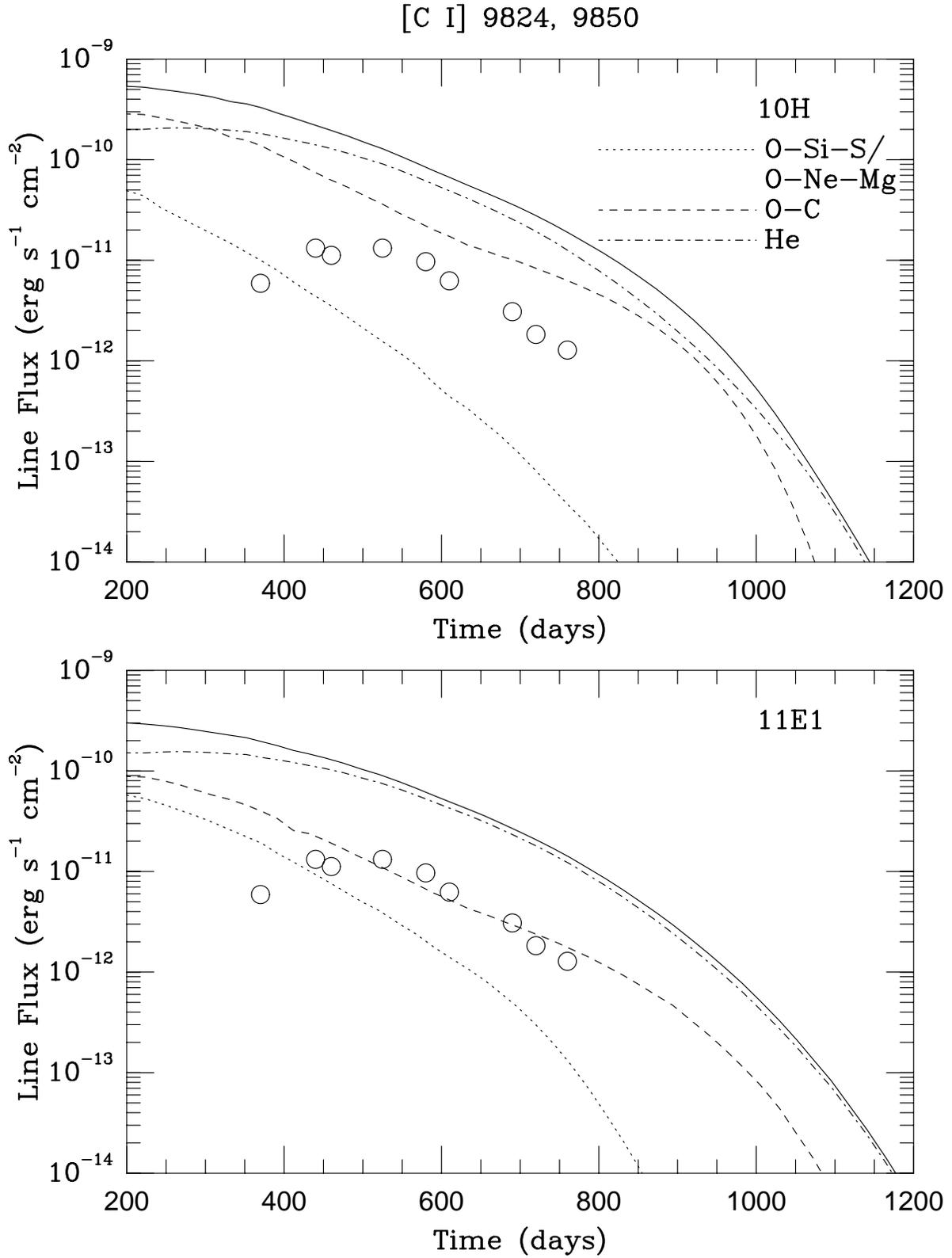}
\caption{Light curve of [C I] $\wll 9824, 9850$  for the 10H and
11E1 models. Observations are from Spyromilio \etal 1991.
\label{fig:ci9830}}
\end{figure}

Figure \ref{fig:ci9830} shows the [C I] $\wll 9824, 9850$ light curve. 
Although the shape of our light curve is in agreement with 
observations, the  model over-produces the line by a factor of $\sim
10$. 
Most of the contribution to the
$\wll 9824, 9850$ lines comes from the helium component, and 
the O -- C region. The mass of the latter varies substantially between the
11E1 and 10H models, $0.10 \Msun$ and $0.60 \Msun$,
respectively. A possible ingredient in 
reducing the line strengths 
is the influence of CO. Liu \& Dalgarno (1995)
find that while only a small fraction of the carbon goes into CO, the
cooling of the gas is increased by up to an order of magnitude. The
temperature consequently decreases at 500 days from $\sim 3000$ K  without CO, to
only $\sim 1200$ K including CO. This can easily decrease the [C I]
emission by an order of magnitude. However,  although
the [C I] emission from the O -- C region can be killed in this way,
the emission from the helium region is more difficult to quench. As
Liu \& Dalgarno point out, CO is efficiently destroyed by He II, and
little CO is expected to form in this region. One possibility is that
we have under-estimated the photoionization flux above 
11.26 eV in the model, which would explain the discrepancy. Our
neglect of UV scattering argues against this.

Because we have fixed the total gamma-ray deposition in the helium
zone from the line profile and flux of  \HeI~$\wl 2.058 ~\mu$m,
 the [C I] flux from this region should be fairly reliable. We
therefore conclude that the
most likely solution to the over-production of the [C I] line is
that the carbon mass mixed with helium is lower than in the 11E1 and
10H models ($X({\rm C}) \approx (1-2)\EE{-2}$ in the helium region in
both models). 

The enrichment of carbon in the helium shell  occurs as a result of
convection during the final helium shell burning phase
(e.g., Arnett 1996). Both the time scale and the efficiency
 are, however, uncertain due to our limited understanding of the
convection process. The amount of processed carbon mixed into the
helium shell is consequently uncertain.
To satisfy the observations, a decrease of the carbon mass in the
helium region by a factor of 5 -- 10 is required.

To check the effect of more limited mixing
of carbon into the helium shell, we have replaced the He -- C
zone by a zone with only hydrogen burning products, as given by the 
He -- N zone in the 10H model. The most important
difference is the high abundance of nitrogen, $X({\rm N}) \approx
3.5\EE{-3}$, and low carbon and oxygen abundances, $X({\rm C}) \approx
9.1\EE{-5}$ and $X({\rm O}) \approx
5.9\EE{-5}$, all by number.

\begin{figure}
\plotone{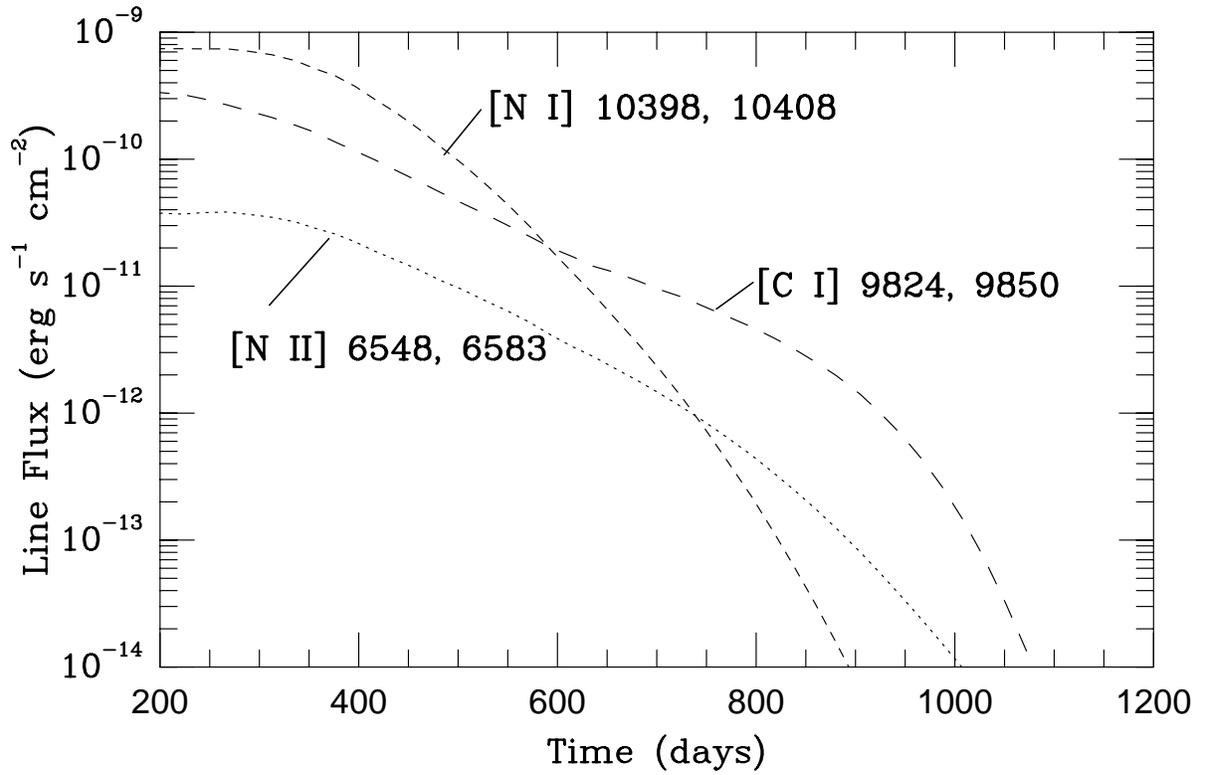}
\caption{Light curves of [C I] $\wll 9824, 9850$,  [N I] 
$\wll 10398, 10408$ and [N II] $\wll 6548, 6583$ for a model where the
He -- C zone in the 10H model has been replaced by a He -- N zone.
\label{fig:nici}}
\end{figure}

This model (Fig. \ref{fig:nici}) basically extinguishes the [C I] $\wll 9824,
9850$ emission from the helium zone, as expected. The total flux is
now close to that observed, especially taking the likely effects of
the CO-cooling in the O -- C zone into account. A test of this model
is to check if the  emission in
lines of N I and N II is now compatible with the observations.  The
strongest of the nitrogen lines is the [N I] 
$\wll 10398, 10408$ multiplet. Although not discussed previously, the
CTIO spectra by  Phillips 
\etal (1990) show a clear line at this wavelength in all spectra
covering this wavelength region. On day 786 the model gives a reddening
adjusted flux of
$\sim 6.5\EE{-13} {~\rm erg ~cm^{-2}~s^{-1}}$. Including internal dust
absorption this gives a flux of $3.9\EE{-13} {~\rm erg
 ~cm^{-2}~s^{-1}}$. On the same day the CTIO spectrum by  Phillips
\etal gives a flux of $\sim 6.9\EE{-13}  {~\rm erg ~cm^{-2}~s^{-1}}$
for the 1.04 $\mu$m line, entirely consistent within the uncertainties
of the model. The model strengths of 
[N II] $\wll 6548, 6583$ are only $\sim 1 - 2 \%$ of
the $\Ha$ line. The blending of these with $\Ha$ will therefore
effectively hide these lines. We therefore conclude that there is
 evidence from  the observations for a more
extended He -- N zone, at the expense of the He -- C zone.  

Phillips \& Williams (1991) argue that that the [C I]
$\wll 8727/9824+9850$  ratio is 
$\lesssim 0.6$ on day 589.  On the same day we find that this ratio
in the He -- C zone is $\sim 0.8$, while in the O -- C zone it is only
$\sim 0.3$. With the 'standard' model the total ratio, which is
essentially the ratio in the He -- C zone,  is in conflict with the
observations. With the He -- C zone replaced by a He -- N zone, the
total ratio is close to that in the O -- C zone, and consistent with
the observations. This provides some indirect support for our
conclusions above. 

\subsubsection{Oxygen}
\label{sec-oxygen}
\begin{figure}
\plotone{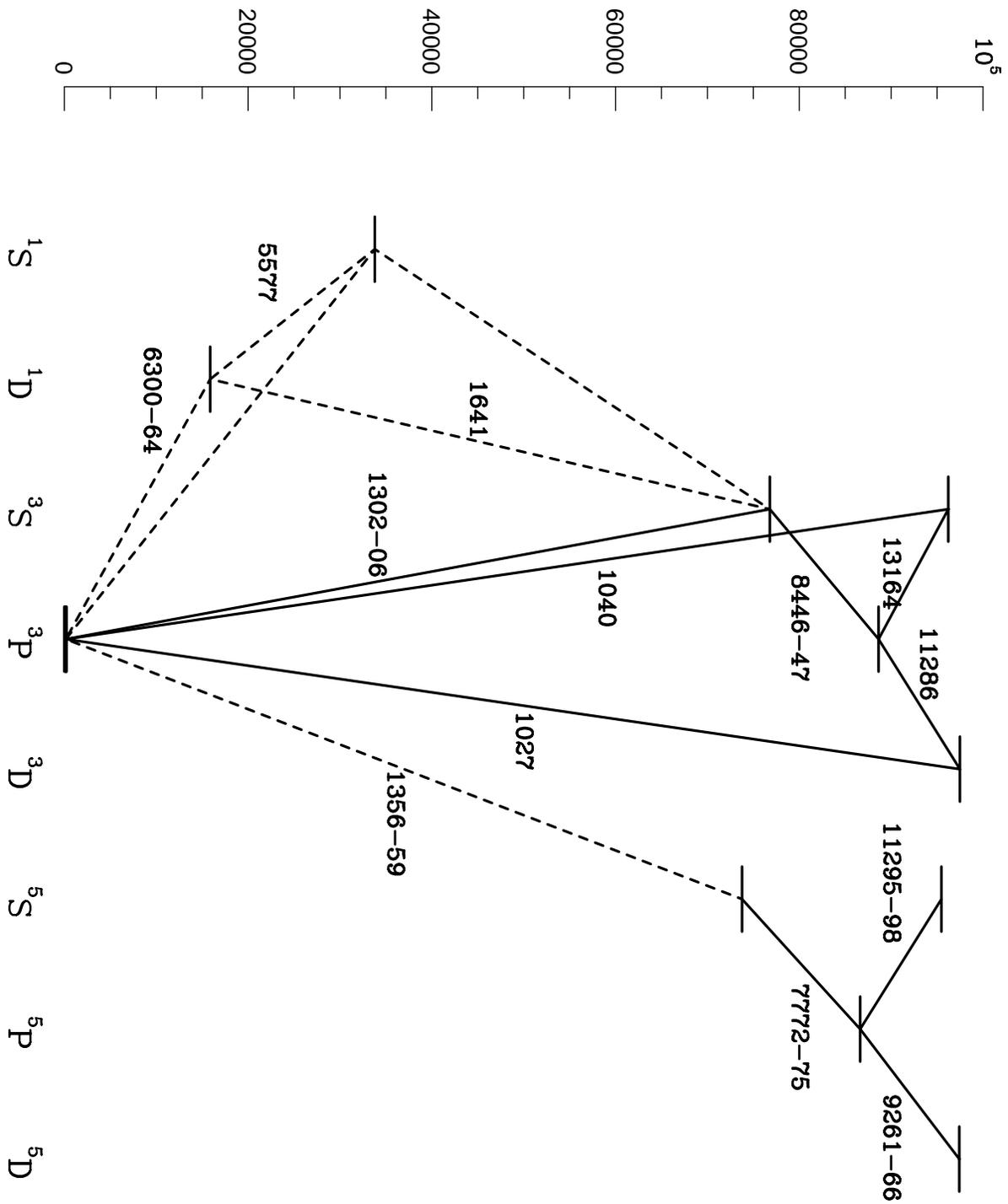}
\caption{Grotrian diagram of the levels and transitions
included in our O I model atom.
\label{fig:Ogrot}}
\end{figure}

The [\OI] $\wll$6300, 6364 lines are of special importance for the analysis,
because they are in an easily accessible, non-blended part of the
spectrum. Therefore, a  complete and accurate data set exists for
these lines. In addition, oxygen is the most abundant of the metals,
and a good probe of the progenitor evolution  and its mass (e.g.,
Thielemann, Nomoto, \& Hashimoto 1996).   
For future reference we show in Figure \ref{fig:Ogrot} the levels and transitions included
in our calculation.

\begin{figure}
\plotone{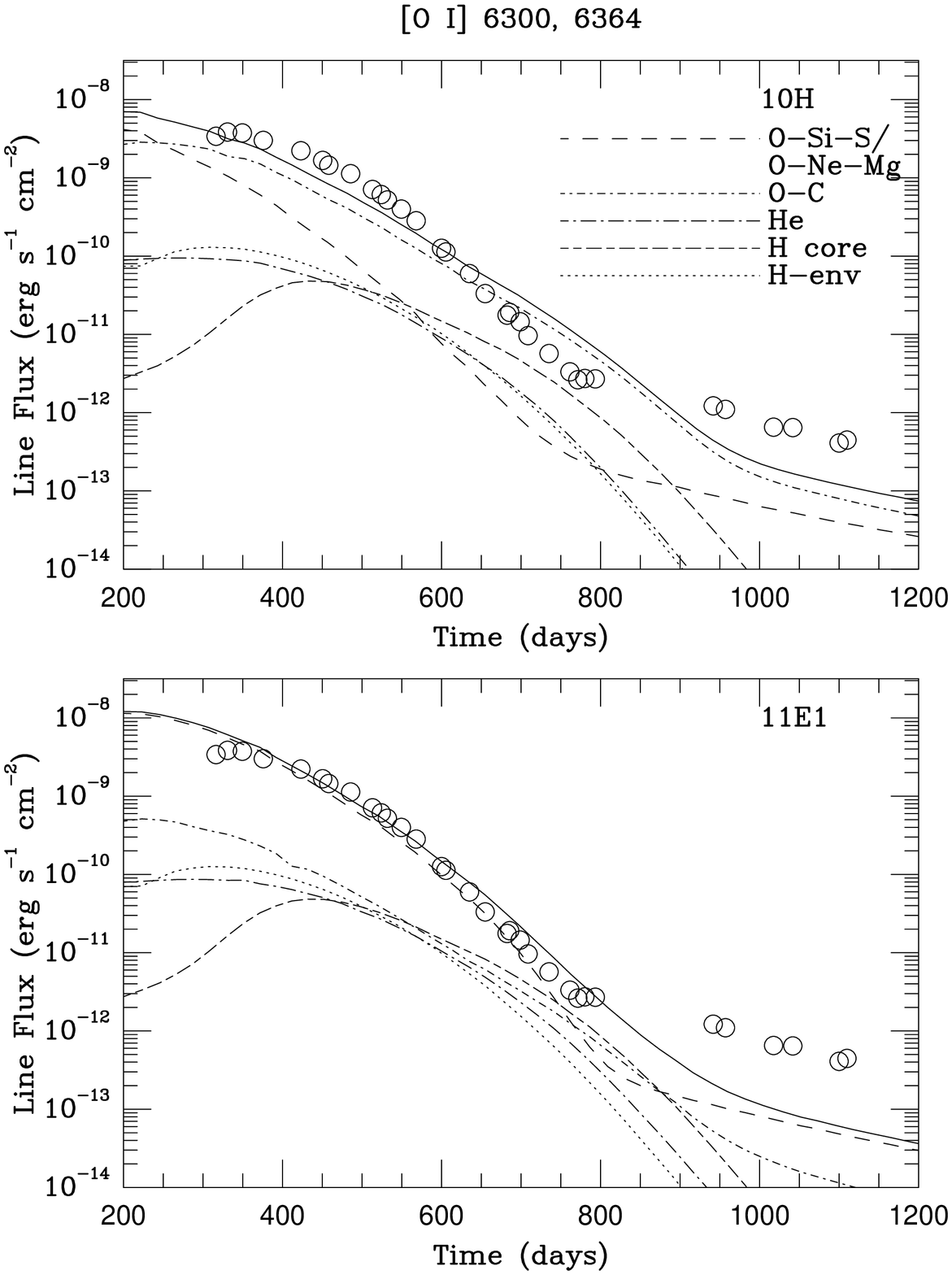}
\caption{Light curve of [O I] $\wll$ 6300, 6364. The
upper figure shows the fluxes from the 10H model and the lower from
the 11E1 model. Observations are from Danziger \etal (1991).
\label{fig:photoi1}}
\end{figure}

Starting with the 10H model, we show in Figure
\ref{fig:photoi1} 
the light curve of [\OI] $\wll$6300, 6364, together with the individual contributions
from the different zones. Observations are taken from Danziger \etal (1991). The
light curve can 
be divided into one epoch when thermal excitations of the ${}^1D$
level dominate, which lasts up to $\sim 800$ days, and one later
epoch when non-thermal excitations dominate. This transition is, as
explained in e.g. Fransson, Houck \& Kozma (1996) and as can be seen from
a comparison with Figure 2
Paper I, intimately coupled
to the temperature evolution of the core. As the oxygen-rich regions
undergo a thermal instability and cool to $\lesssim 300$ K, thermal
excitation of the  ${}^1D$ level effectively stops. The contributions
from the hydrogen and helium-rich regions are,
however, as shown in Figure \ref{fig:photoi1},
significant in the thermal phase. In fact, up to $\sim 30 \%$ of the
[\OI] $\wll$6300, 6364 emission at 600 -- 900 days comes from these
components.  Because of the adiabatic
expansion and therefore falling temperatures of the hydrogen and
helium-rich gas, the emission from both these decrease rapidly
after $\sim 800$ days. This means that
even if the non-thermal contribution from the core is small, it
dominates after $\sim 900$ days. 

In terms of the qualitative evolution we find good agreement between model calculations and
observations throughout the whole period, showing without doubt that
the IR-catastrophe really has taken place in the oxygen-rich gas. 
Quantitatively, there are, however, some disagreements. While most of the thermal phase is well reproduced, at $\sim$ 750
days the 10H model over-produces the [O
I] luminosity by a factor of $\sim$ 2. 
The 11E1 model shows  good agreement with the whole thermal part of the light curve.

The most serious flaw is in the non-thermal part.
Although agreeing qualitatively with the observations, the level of
the flat, non-thermal part of 
the curve is down by factors of 4 -- 6 in the two models. 
As a consequence of this under-production,  the
transition from the thermal to the flat non-thermal light curve occurs
$100 - 200$ days later than is observed. In this context we note that
the observed break coincides well with the calculated transition from
the O -- Si -- S zone, arguing for a larger contribution at late times
from this. 

\begin{figure}
\plotone{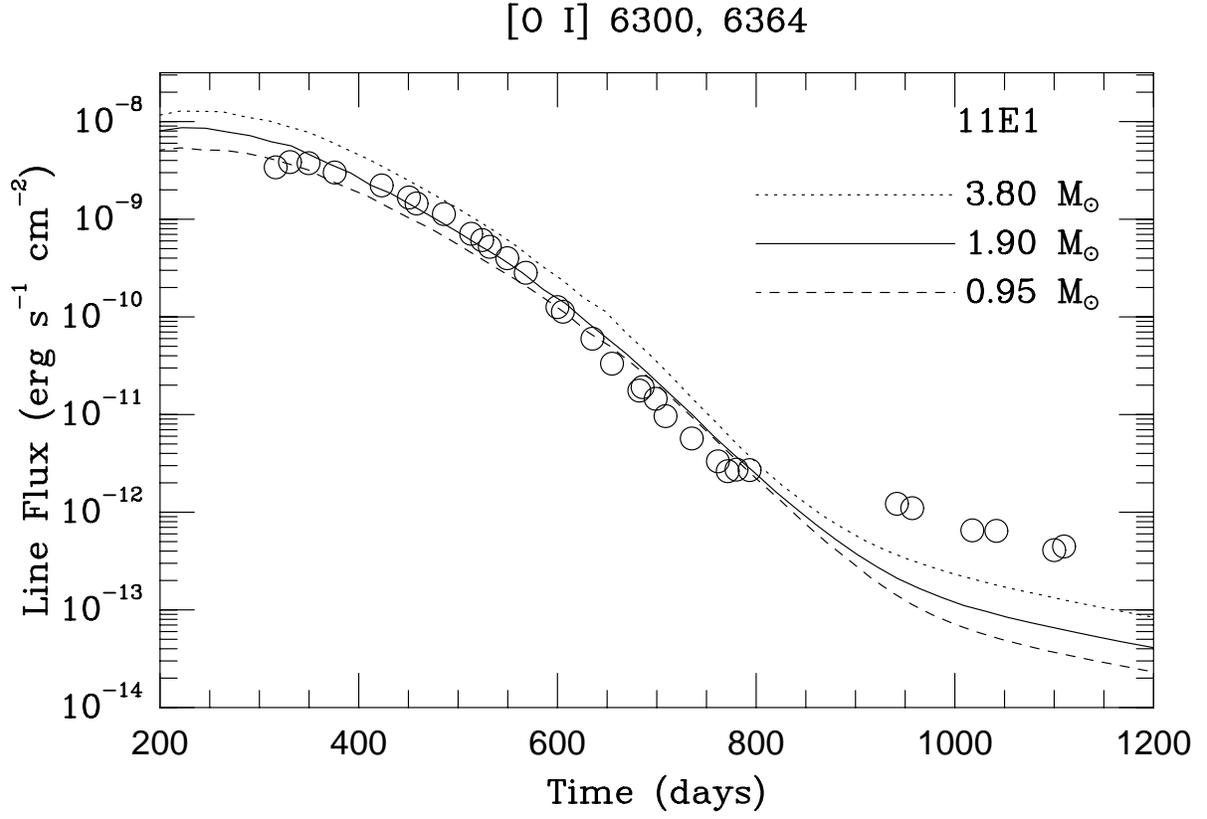}
\caption{ [O I] $\wll$ 6300, 6364 light curves for three different masses of
the oxygen zones in the 11E1 model. The
solid line shows the standard model with $M$(O) = $1.9 \Msun$. 
Note that while the thermal part of the light curve is probably bracketted by
this range, the non-thermal is under-produced even for the highest mass.
\label{fig:photoi2}}
\end{figure}

As an illustration of the sensitivity to the assumed mass of the
oxygen region, we show in Figure \ref{fig:photoi2} the effect of
varying the oxygen region mass by a factor of two from that of the
11E1 model, covering the range of $M({\rm O}) = 0.95 - 3.80 \Msun$. Other
parameters of the 11E1 model are kept constant. Here we first note
that the non-thermal part is under-produced even in the highest mass
model, 
 so increasing the oxygen mass does not solve that
problem. The thermal part of the light curve before $\sim 500$ days is
best reproduced by the 
standard 11E1 model, and the two other extremes probably bracket the
likely range of the oxygen zone mass. Between 500 -- 900 days the
difference between the models is small because of the large
contribution from the helium and hydrogen regions. 

The total oxygen mass in the 11E1 and 10H  models is roughly the
same. In spite 
of this, the flux in the 11E1 model is only $\sim 50 \%$ of that in the
10H model in the non-thermal part. In addition, although the mass of the O -- C zone is only
$\sim 0.6 \Msun$, compared to $\sim 1.2 \Msun$ in the O -- Si -- S
zone, the former contributes twice as much flux as the
latter in the 10H model.

To understand this, and the general level of the light curve we have to discuss an important technical point in
the excitation to the ${}^1D$ level. During
the non-thermal phase, in addition to direct excitation, this level receives an important contribution
from the excited 
$3s {}^3S$ level, via the $\wl 1641$ line (Fig. \ref{fig:Ogrot}). 
 Because excitation 
of the $3s {}^3S$ level is also non-thermal, mainly by recombination
following non-thermal ionizations, the basic scenario is not changed, but it
can make an important quantitative difference.   
Normally, the
de-excitation of the $3s {}^3S$ level is to the ground state, by the
$\wl
1302.2 - 1306.0$ resonance multiplet, and the
probability of a transition to the ${}^1D$ level in an individual
transition is only $A_{{}^3S_1-{}^1D_2}/\sum_{i=0}^2 A_{{}^3S_1-{}^3P_i} = 3.0\EE{-6}$. The
optical depth of the
 resonance lines are, however, very large, increasing the effective
lifetime of the ${}^3S_1$ level. The actual probability of a transition
in the $\wl 1641$ line to ${}^1D_2$, compared to ${}^3P$ is
\begin{equation}
{P_{{}^1D} \over P_{{}^3P}} =  { g_2 A_{{}^3S_1-{}^1D_2}~\lambda_{12}^3~n_1~
t \over 24 \pi~ g_{1}},
\label{eqn:brancha}
\end{equation}
 where the indices 1 and 2 refers to the
$2p^4~ {}^3P$ ground state and the $3s {}^3S$ state,
respectively. 
Inserting atomic parameters, and using the O I density derived by
Li \& McCray (1992), one finds
\begin{equation}
{P_{{}^1D} \over P_{{}^3P}} = 1.8\EE{-14}~n_1~t = 9.6\EE3~\left({t
\over 100~{\rm days}}\right)^{-2}
\label{eqn:branchb}
\end{equation}
We therefore conclude that  branching to ${}^1D$ dominates escape in
the resonance lines up to at least $\sim 3000$ days.

  A further important complication is that the resonance line photons
at 1302 -- 
1306 \AA~ may be destroyed by  photoelectric absorption by Si I,
depending on the Si I abundance in the zone. If the probability of
absorption dominates the branching probability, the
1302 - 1306 \AA~ photons are destroyed, and the branching to ${}^1D$ 
decreases drastically. This situation is similar to that earlier
discussed for helium. 
The probability for this to happen can be estimated from equation
(32) in Paper I, as
\begin{eqnarray}
P_{\rm photo} &\approx& 4.9 {\kappa_{\rm photo} \over \kappa_{\rm
line}} = 4.9~{ 8 \pi~g_1~V_{\rm
th}~\sigma_{\rm photo} \over \lambda^3~A_{21}~g_2~}~ {n_{\rm Si I}
\over n_{\rm O I}} \cr
& = & 1.2\EE{-3}~\left({T
\over 1000~{\rm K}}\right)^{1/2}~{n_{\rm Si I} \over n_{\rm O I}}  , 
\label{eqn:pphoto}
\end{eqnarray}
compared to the branching probability $P_{\rm branch} = 3.0\EE{-6}$. Therefore, if
${n_{\rm Si I} / n_{\rm O I}} \gtrsim 2.5\EE{-3}$, photoabsorption
dominates branching to the ${}^1D$ level. The triplet contribution to the 
${}^1D$ level is therefore sensitive to the chemical composition in
the oxygen-rich gas. 
 Newer models of  15 and 25 $\Msun$
stars by Woosley \& Weaver (1995) give considerably lower silicon abundances in
most of the oxygen zone. Further, the estimate in equation (\ref {eqn:pphoto})
is based on continuum absorption in the Doppler profile only. As we
have discussed for helium, absorption also in the damping
wings can, for complete distribution, increase the destruction
probability above this value. This is confirmed by the model
where we used the destruction probability from equation (33)
in Paper I. In this model the non-thermal level decreased to a
level of $\sim 70 \%$ of 
that in Figure \ref{fig:photoi1}. As we have already remarked, however, we
believe that the Doppler case is closer to the true situation of
partial redistribution. 

We can now understand the origin of the somewhat paradoxical situation
described above. The reason for the low contribution from the O --
Si -- S zone in the 10H model is the destruction of the 1302 \AA~
photons by the abundant Si I in this zone. If it had not been for this
effect the contribution from the $3s~ {}^3S$ state might have been
similar in efficiency to that of the O -- C zone, giving a flux
proportional to the oxygen mass, and increasing the total flux by a
factor $\sim 3$. 
Even in the O -- Ne -- Mg zone in the 11E1 model, the silicon abundance is
relatively high, X({\rm Si}) = $1.64 \EE{-2}$, leading to the destruction of
the 1302 \AA\ photons. In spite of the larger magnesium abundance, the
absorption by Si I dominates the Mg I absorption, because the
photoionization cross section for Si I is almost a factor of 100 higher
than for Mg I.
If for some
reason we have over-estimated the continuum optical depth at 1302 \AA,
e.g. because of too low ionization of Si I, the [O I] flux
would increase. Charge transfer between O II and Si I could  have this
effect. In  section ~\ref{sec-ct2} we discuss this quantitatively,
with negative result. 

\begin{figure}
\plotone{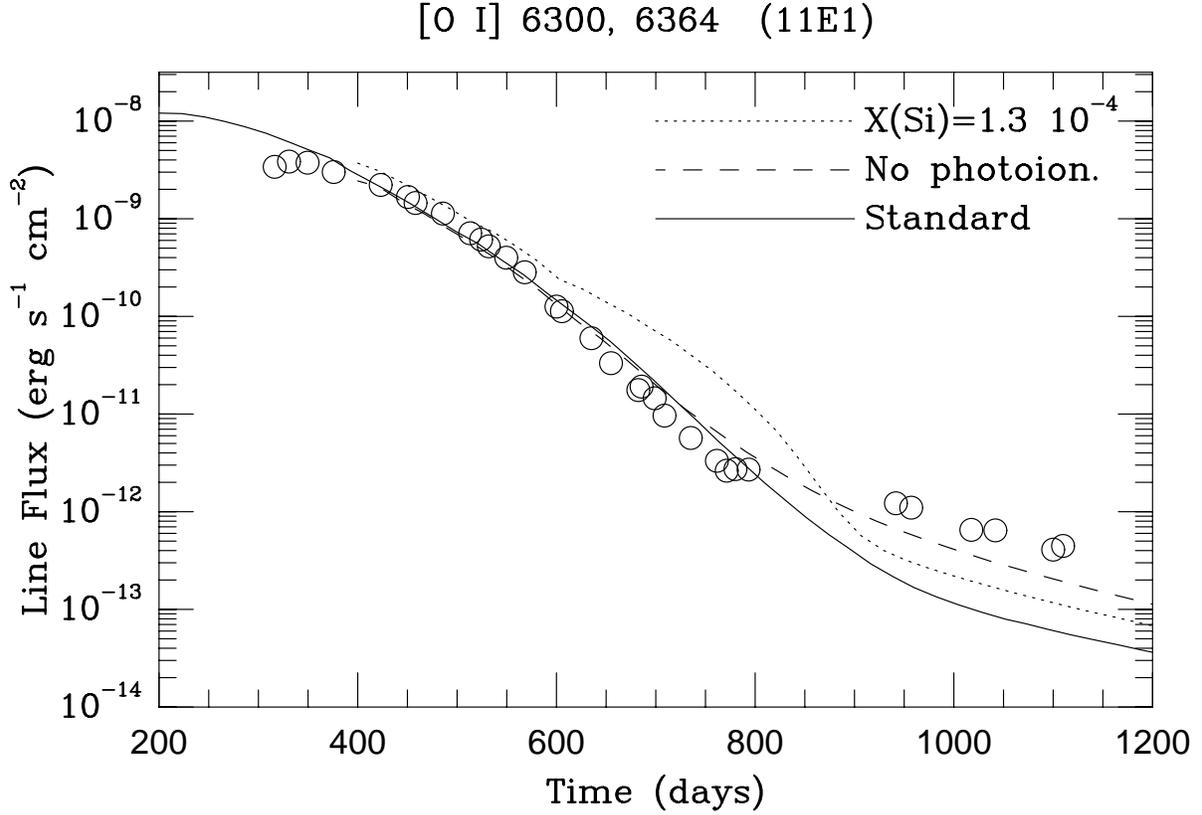}
\caption{ [O I] $\wll$ 6300, 6364 light curves for the 11E1 model, showing the
importance of the photoelectric absorption by Si I and Mg I.  The
solid line is the standard model for the O -- Ne -- Mg region  with
$X({\rm Si}) = 1.64\EE{-2}$, while the dotted line is a model with
the silicon abundance decreased to
$1.3\EE{-4}$. Note the increase in the non-thermal plateau at 
times later than $\sim$ 900 days, and the bump between 600 -- 900
days. The dashed line shows the flux for a model where the
photoelectric absorption has been switched off for both Si I and Mg
I. This increases the non-thermal flux by a factor three, compared to
the standard model.
\label{fig:photoi3}}
\end{figure}

To see the effect of a lower silicon abundance we show in Figure
\ref {fig:photoi3} the [O I] light curve for a model where we have
decreased the silicon abundance in the O -- Ne -- Mg zone in the 11E1
 model to that in the O -- C region, $X({\rm Si})=1.3\EE{-4}$, 
compared to the original
$X({\rm Si})=1.64\EE{-2}$. 
We here see that the non-thermal part indeed 
increases by a factor of $\sim 2$. However, in spite of the smaller
photoabsorption cross section of  Mg I, in this model Mg I takes 
over the role of Si I, and most of the triplet contribution
is also in this model quenched. An unwanted effect in this model is
that  there is a substantial increase in
the flux between 600 -- 850 days, which  at this epoch destroys the
agreement with the observations in the original model.  

The full extent of the photoabsorption can be seen in a model, also shown
in Figure \ref {fig:photoi3}, where we  have (artificially) 
decreased the photoabsorption of the $\wl 1302$ line to zero. Only in
this model do we get a non-thermal plateau close to the observed
level. 

To estimate the dependence on the various parameters, and therefore
more general models, it is of
interest to consider a simplified model for the non-thermal [\OI]$\wll$6300, 6364
excitation. 
If we  assume that the luminosity in these  lines is determined
only by the non-thermal excitation to the ${}^1D$ level, one can estimate the [O
I]$\wll$6300, 6364 luminosity by the following argument. If $J_{\gamma}$ is the
gamma-ray intensity, the absorbed energy in the oxygen component per unit
volume is $J_{\gamma}
\kappa_{\gamma} \rho_{\rm O}$. Of this, a
fraction $\epsilon_{\rm exc, {}^1D}$ will give rise to direct excitations of the
${}^1D$ level. In addition, a fraction $\epsilon_{\rm trip}$ will go
into excitation of the ${}^1D$ level, via the triplet levels, as
discussed above. The total energy fraction  going into the ${}^1D$
level is therefore 
$\epsilon_{{}^1D} = \epsilon_{\rm exc, {}^1D} + \epsilon_{\rm trip}$.
Because collisional de-excitations at this epoch are 
unimportant, the total luminosity in the 6300, 6364 lines is given by 
\begin{eqnarray}
L_{6300, 6364} & = & { \int \epsilon_{{}^1D}~J_\gamma~
\kappa_\gamma (r)~ \rho_{\rm O}(r) dV
\over
 \int J_\gamma~ \kappa_\gamma (r)~\rho_{\rm total}(r)~ dV }~L_{\rm bol} \cr
 & \approx &
 \epsilon_{{}^1D}~{\tau_{\gamma, \rm O} \over\tau_{\gamma, \rm total}} ~L_{\rm bol}, 
\label{eqn:lum6300}
\end{eqnarray}
where $\tau_{\gamma, \rm O}$ is the optical depth of the gamma-rays
in the oxygen and $\tau_{\gamma, \rm total}$ is the total gamma-ray
optical depth, both averaged over the ejecta according to the gamma-ray
intensity. Both $\tau_{\gamma, \rm O}$  and
$\tau_{\gamma, \rm total}$ depend on time as $t^{-2}$, and equation
(\ref{eqn:lum6300}) therefore shows that, except for the dependence of
$\epsilon_{{}^1D}$ on the electron fraction, a fixed fraction of the
bolometric luminosity is expected in the 6300, 6364 lines, in
qualitative agreement with the observations. 

From our
Spencer--Fano calculations $\epsilon_{\rm exc, {}^1D} \approx
3.0\EE{-3}~(x_e/10^{-2})^{-0.623}$ for $x_e$ in the range $\sim
3\EE{-4} - 3\EE{-2}$. At 1000 days 
we find $x_e \approx 7\EE{-3}$, so  $\epsilon_{\rm exc, {}^1D} \approx
3.7\EE{-3}$. Our 10H model has $\tau_{\gamma, \rm O} / \tau_{\gamma, \rm
total} \approx 0.18$, so a fraction $L_{6300, 6364} / L_{\rm bol} \sim 6.7\EE{-4}$ of the bolometric luminosity
should come out  as $\wll 6300, 6364$ emission. At 1000 days $L_{\rm bol}
\approx 1.0\EE{38} \ergs$ (Bouchet, Danziger \& Lucy 1991), and $L_{6300, 6364}
\approx 3.8\EE{35} \ergs$ (Danziger \etal 1991), so the observed ratio is $L_{6300, 6364} / L_{\rm bol}  \approx
3.8\EE{-3}$, while Menzies (1991) finds $L_{6300, 6364} / L_{\rm bol}  \approx
(1.2 - 2.1)\EE{-3}$ at 1000 days. This shows that unless $x_e \ll 10^{-2}$ or $\tau_{\gamma, \rm O} / \tau_{\gamma, \rm
total} \gg 0.18$, direct excitation is not sufficient.

 The maximum
contribution from the $3s {}^3S$ level can be estimated by assuming
that the 1302 -- 1306  \AA~ transitions have  very large optical
depths, and that photoabsorption is unimportant (see above). The
triplet contribution 
to the luminosity in the 6300,
6364 lines is then 
\begin{eqnarray}
L_{6300, 6364, {\rm trip}} & \approx & [f_{\rm recomb}~\epsilon_{\rm ion}~{\alpha_3
\over \alpha_{\rm tot}}~{912 \over 6300} + \epsilon_{\rm exc,
trip}~{1300 \over 6300}]~ \cr
& & {\tau_{\gamma, \rm O} \over
\tau_{\gamma, \rm total}}~L_{\rm bol}.
\label{eq:lumoitrip}
\end{eqnarray}
 Here, $ \epsilon_{\rm ion} \approx 0.4$ is the
efficiency of ionization for O I (KF92), $\alpha_3$ the effective recombination rate to the  $3s {}^3S$
level and 
$\alpha_{\rm tot}$ the total recombination rate. From calculations by
Julienne, Davies, \& Oran (1974) we estimate that $\alpha_3 / \alpha_{\rm tot}
\approx 0.22$. The factor $f_{\rm recomb}$ is the fraction of the
ionizations which recombine to O I. As we will see in section
\ref{sec-ct2}, $f_{\rm recomb}$ can be very small if charge transfer of
e.g. O II + Si I is efficient. Finally, $\epsilon_{\rm exc, trip}$ is
the fraction of the energy going into direct excitations of the
triplet levels, of which most end up in the $3s {}^3S$ state. In KF92
the total energy into excitations in the O -- Si -- S zone is $\sim
0.1$, of which $\sim 42 \%$ is to the triplets, so $\epsilon_{\rm exc,
trip} \approx 4\EE{-2}$ at $x_e \sim 10^{-2}$. The dependence on the electron
fraction is roughly  $\epsilon_{\rm exc, trip} \propto x_e^{-0.27}$, for $0.003 < x_e < 0.03$. Therefore, the maximum efficiency for
excitation from the triplets is
$\epsilon_{\rm trip} \approx 1.3\EE{-2}~f_{\rm recomb} + 8.7\EE{-3}$. The
maximum, total 
efficiency is then $\epsilon_{{}^1D} = \epsilon_{\rm exc, {}^1D} +
\epsilon_{\rm trip} \approx 1.2\EE{-2}(1 + 1.1~f_{\rm
recomb})$. With $\tau_{\gamma, 
\rm O} / \tau_{\gamma, \rm total} \approx 0.18$  and, depending on
$f_{\rm recomb}$, we find $L_{6300, 6364} \approx (2.5 - 4.9)\EE{-3}~L_{\rm bol}$, in
better agreement with the observations. This argument shows that the
triplet contribution is sufficient, and most likely necessary, to explain the observed non-thermal
level. 

The estimate above gives the luminosity for one particular
hydrodynamical structure, with $\tau_{\gamma, \rm O} /
\tau_{\gamma, \rm total} \approx 0.18$. To check the sensitivity
of this assumption we assume that all 
oxygen, with a total mass $M_{\rm O}$, is located in a shell
between velocities $V_{\rm min, O}$ and $V_{\rm max, O}$.  In this
model $\tau_{\gamma}({\rm O}) =
\kappa_{\gamma}~M_{\rm O}~G/[4~\pi~(V_{\rm max, O}~t)^2]$, where
$G = 3~[1 + (V_{\rm min, O}/V_{\rm max, O})+ (V_{\rm min, O}/V_{\rm max,
O})^2]^{-1}$,  $1 \leq G \leq 3$. With this
we get
\begin{equation}
{L_{6300, 6364} \over L_{\rm bol}} = {\epsilon_{{}^1D}~G~\kappa_{\gamma}~M_{\rm O} \over
4~\pi~V_{\rm max, O}^2~t_*^2}
\label{eq:lumoiabs}
\end{equation}
where $t_*$ is the time when the total optical depth, $
\tau_{\gamma,\rm total}$ is unity. From light curve models Woosley,
Pinto \& Hartman (1989)  find $t_* \approx 550$ days. We therefore find that in the non-thermal phase
\begin{eqnarray}
L_{6300, 6364} & = & 0.62\EE{-3}~ G~ \left({M_{\rm O} \over 1
 ~\Msun}\right)~ \cr
&& \left({V_{\rm max, O} \over 2000
\kms}\right)^{-2}  \left({t_* \over 550 ~{\rm days}}\right)^{-2}~ \cr
 \nonumber \\
 & & (1 + 1.1~f_{\rm recomb})~L_{\rm bol}.
\label{eqn:lumoiabs2}
\end{eqnarray}
The main parameters determining the direct excitation of the lines are
therefore the oxygen mass, the velocity interval of the oxygen mass, and to
a weaker degree the electron fraction in the oxygen-rich gas. Of these
the oxygen distribution is probably the main uncertainty,  introducing
 an uncertainty of up to  a factor of three. 
Equation
(\ref{eqn:lumoiabs2}) indicates that an oxygen mass of at least $1 \Msun$ is
necessary to explain the observations (for $G \approx 3$). If most of
the oxygen is in a  narrow velocity range,  $G \approx 1$, an oxygen
mass of up to $\sim 3 \Msun$ may be necessary. 
We caution, however, 
that due to uncertainties in  both the atomic physics and the hydrodynamics 
this number is not firm.  

\begin{figure}
\plotone{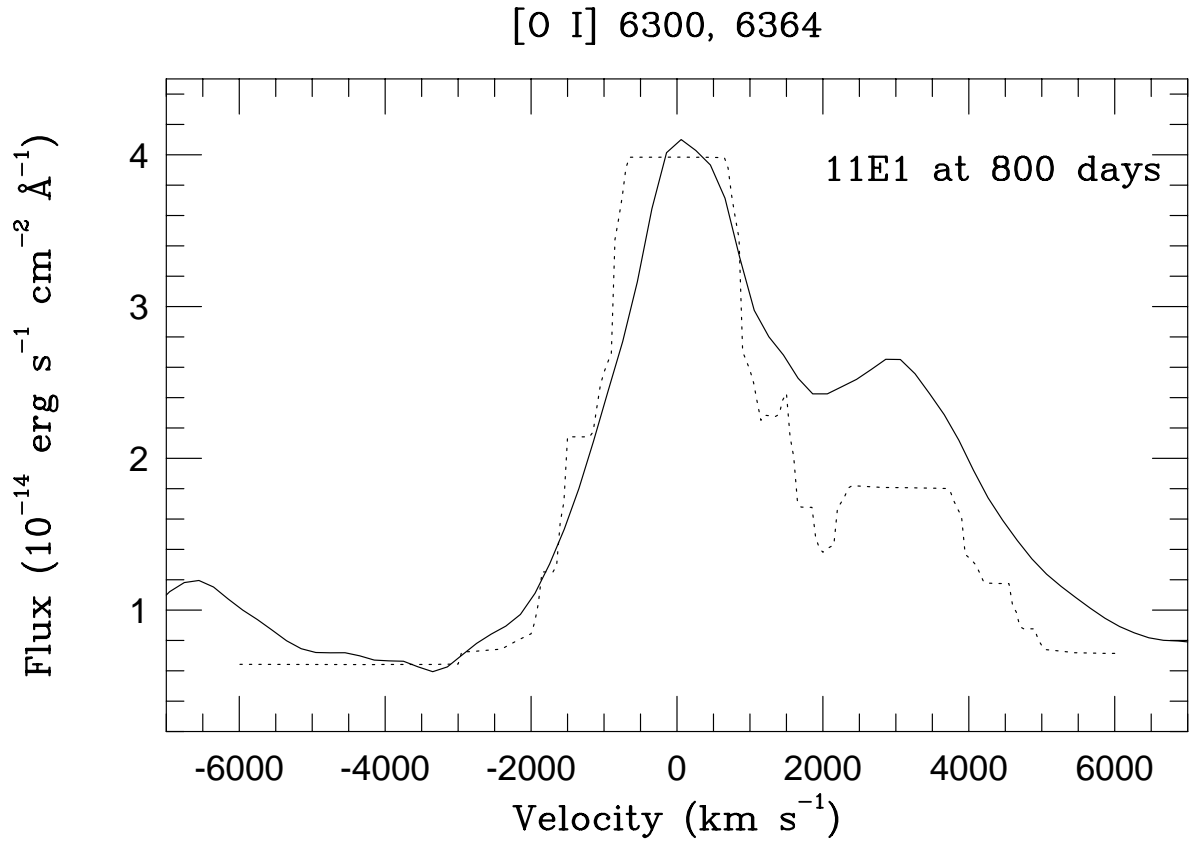}
\caption{ The [O I] $\wll$ 6300, 6364 line profile for the 11E1 model at
800 days (dotted line), together with observations from Phillips \etal
(1990) (solid line). The more peaked observed profile shows that mixing
of oxygen reaches further in, to less than $ 400 \kms$, than in the model.
\label{fig:oiprofile}}
\end{figure}

From this discussion it is obvious that there is a strong need in
constraining  the oxygen distribution from the observed line
profile of the 6300, 6364 doublet. In Figure \ref{fig:oiprofile} we
show the $\wll 6300, 6364$ lines at 800  days for the 11E1 model (with our
density distribution), including reddening and internal  dust absorption. The
observations are from Phillips \etal (1990). At this epoch the lines are
optically thin, and the doublet components can simply be added in a ratio of
3:1.  The ratio of the observed $6300 / 6364$ lines is considerably smaller
than the expected value of 3.0 in the optically thin limit. A reason
for this may be an additional contribution from the Fe I $\wl 6361$
line, which in our models is $\sim 20 \%$ of the [O I] $\wl 6364$
line. The exact level of the Fe I $\wl 6361$ line is uncertain
because of the uncertain 
UV-radiation field, and it is likely that it could explain the full
discrepancy.

In
the velocity  range $1000 - 2000 \kms$ there is a fairly good
agreement with the observed slope of the line profile, indicating a realistic
density distribution. 
At velocities $\gtrsim 2000 \kms$ the observations
show a clear wing to $\sim 3200 \kms$, considerably stronger than in
the model. There may be two reasons for this. Either the
oxygen abundance  in the hydrogen envelope is larger than assumed in
the model, or there is mixing of a small amount of processed oxygen to these
velocities. The abundances in the hydrogen envelope is set by
observations of the ring of SN 1987A (Fransson \etal 1989, Sonneborn
\etal 1996), and should be reliable. It would require very special
conditions for the oxygen abundance in the envelope to be larger than
that in the 
ring, which probably originates from a layer external to the envelope
of the progenitor. We therefore believe that the most likely solution is the
presence of some high velocity processed oxygen in the ejecta.

The observed line profile is more peaked for $V
\lesssim 900 \kms$ than the model. Our innermost oxygen zone, containing
$\sim 0.6 \Msun$, is at
$770 \kms$, and it is clear that there is oxygen present down to at
least $\sim 400 \kms$. A better fit would be obtained if the same mass
of oxygen-rich gas  was mixed uniformly between $200
- 800 \kms$. The total flux 
should be the same and the total mass of the oxygen zone therefore
similar to that in our model, $\sim 1.9 \Msun$, i.e. $\sim 1.4 \Msun$
of pure oxygen.

In the same way as for the ${}^1D$ level, one  expects non-thermal
excitations to the  ${}^1S$ level. In this case the
contribution from the triplet levels is only $\sim
2.5\EE{-3}$ of that to the ${}^1D$ level. The ratio of the
non-thermal excitations to the  ${}^1S$ and  ${}^1D$ levels is
$\epsilon_{{}^1S} \approx 0.3~ \epsilon_{\rm exc, {}^1D}$ at $x_e
\approx 7\EE{-3}$ (KF92). The ratio of the $\wl 5577$ luminosity to
the $\wll 6300, 6364$ luminosity is therefore
\begin{equation}
\frac{L_{5577}}{L_{6300, 6364}} = \epsilon_{{}^1S}~(\frac{2972}{5577})~
\frac{1}{\epsilon_{{}^1D} + \epsilon_{{}^1S}~(\frac{2972}{6300})}.
\end{equation}
If excitations via the $3s {}^3S$ level for
the ${}^1D$ level  are unimportant, $\epsilon_{{}^1D} = \epsilon_{\rm
exc, {}^1D}$, 
and the ratio
of the  [O I] $\wl 5577$ and [O I] $\wll 6300, 6364$ lines is expected
to be  $L_{5577} / L_{6300, 6364} \approx 0.14$.
In the opposite case, when the triplet contribution is important,
$\epsilon_{{}^1D} = \epsilon_{\rm exc, {}^1D} + \epsilon_{\rm trip}$, and
$L_{5577} / L_{6300, 6364} \approx 2.5\EE{-2}$.
The relative ratio can therefore give some information about
the importance of excitation via the triplet lines of [O I] $\wll 6300,
6364$. The observed flux of the $\wl$ 5577 line is unfortunately uncertain
because of  blending with other lines. An approximate analysis of
the spectrum at 804 days by Phillips \etal (1990) gives
$L_{5577}/L_{6300, 6364} \lesssim 
0.3$, as a strong upper limit. A more realistic limit, taking the
blending by other lines (the 'continuum') into account, 
reduces this upper limit by a factor of two.

The O I $\wll$ 7774, 8446, and 9265 lines arise as a result of
recombination, and in 
the case of the $\wl $ 8446 line also by Bowen fluorescence by
Ly$\beta$ (Oliva 1993). The flux of the $\wl$ 7774 line is on day 804 
$\sim 3\EE{-12} {~\rm erg ~cm^{-2}~s^{-1}}$. This is a factor of at
least four stronger than
observed in the day 804 spectrum by Phillips \etal (1990). Also the
$\wl$ 8446 line is over-produced by a similar 
factor. However, 
as suggested by Oliva, its absence can probably be explained by
scattering by the Ca II IR triplet. The velocity difference to the
closest component is only 1836 $\kms$. The $\wl 8446$
photon may therefore contribute to the excitation of the Ca II
triplet. We  discuss this further in section \ref{sec-calcium}.
The $\wl 9265$ \AA~ line could be present in the spectrum. However, the peak of the
observed feature is at 9234  \AA, which 
probably excludes it from being the O I line. The flux in this line is
otherwise close to that expected for the [O I] line. The
wavelength is consistent with Paschen 9 -- 3 $\wl 9234$. The flux of
this is, however, only expected to be $\sim 15 \%$ of that observed. There is also an
excited multiplet in S I, $4s {}^5S^o - 4p {}^5P, \wll 9212 -
9237$, coinciding with the peak of the line. This transition, which we do not include, can be expected to
be strong because of recombination from the Si -- S zone, and we think
that this may be the most likely candidate for the line. Other strong
quintet lines expected would then be the $\wl 7696$ and $\wl 8694$
lines. At the former wavelength there is a line with a flux
of $\sim 40 \%$ of the   $\wl 9234$ line, while the latter coincides
with the Ca II triplet.

Therefore, all the calculated  O I recombination lines seem too strong, although
 the $\wl 8446$ line could be scattered into 
the Ca II triplet. 
The K I resonance lines at 7664.9, 7699.0 \AA~ probably
account for the P-Cygni line at $\sim 7700$ \AA. Its width is similar
in velocity to the Na I D lines. It is, however, not likely that this doublet
could scatter the $\wl 7774$ line, as this would probably result in a much
stronger red peak of the line than observed.  Instead, its total
equivalent width is close to zero, as expected if it just scatters the
'normal' background emission. The wavelength difference
is also too large. The effective recombination rates to the O I lines from
Julienne, Davies, \& Oran (1974) are uncertain, but not by more
than a factor of two, since the total recombination rate agrees within
$\sim 40 \%$ by that found by Chung, Lin, \& Lee (1991). Instead, we
think that charge 
transfer processes, not included in our model, are responsible for the
quenching of the recombination lines. This is discussed in
section~ \ref{sec-ct2}. 

Because of the large abundance and mass of oxygen, the bound-free
emission continua of 
O I may be observable, unless quenched by charge transfer.  The
strongest of these are the recombination 
continua to the $3p{}^5P$, $3d{}^5D$, and $3d{}^3D_0$ levels, with
 edges at 4308~\AA, 8053~\AA, and 8098~\AA, respectively. On
day 800 the temperature in the oxygen zone is $\sim 1500$ K,
corresponding to a width of $\sim 660$ \AA~ at 8000 \AA, and $\sim 192$ \AA~
at 4300 \AA. For the $3d{}^5D$
and $3d{}^3D_0$ emission we estimate a flux of $4.8\EE{-15} \ergs
 ~cm^{-2}$, corrected for reddening and internal dust, at 8000~\AA. The
observed continuum level at $\sim 8000$ \AA~ is $\sim 9\EE{-15} \ergs
 ~cm^{-2}$, consistent with the O I continua. Also at 4300 \AA~
the model flux, $\sim 1.5\EE{-15} \ergs ~cm^{-2}$ is consistent with the
observations, although blending with other lines makes it difficult to
define the continuum level. Unfortunately, an unambiguous
identification of these continua from the observations is difficult,
although especially the continuum at 8000 \AA~ has no other obvious
candidate strong enough. An identification of this would make a direct
determination of the O II fraction possible, and therefore to estimate
the importance of charge transfer, as well as its temperature.

\subsubsection{Neon, Magnesium, Silicon, and Sulphur}
\begin{figure}
\plotone{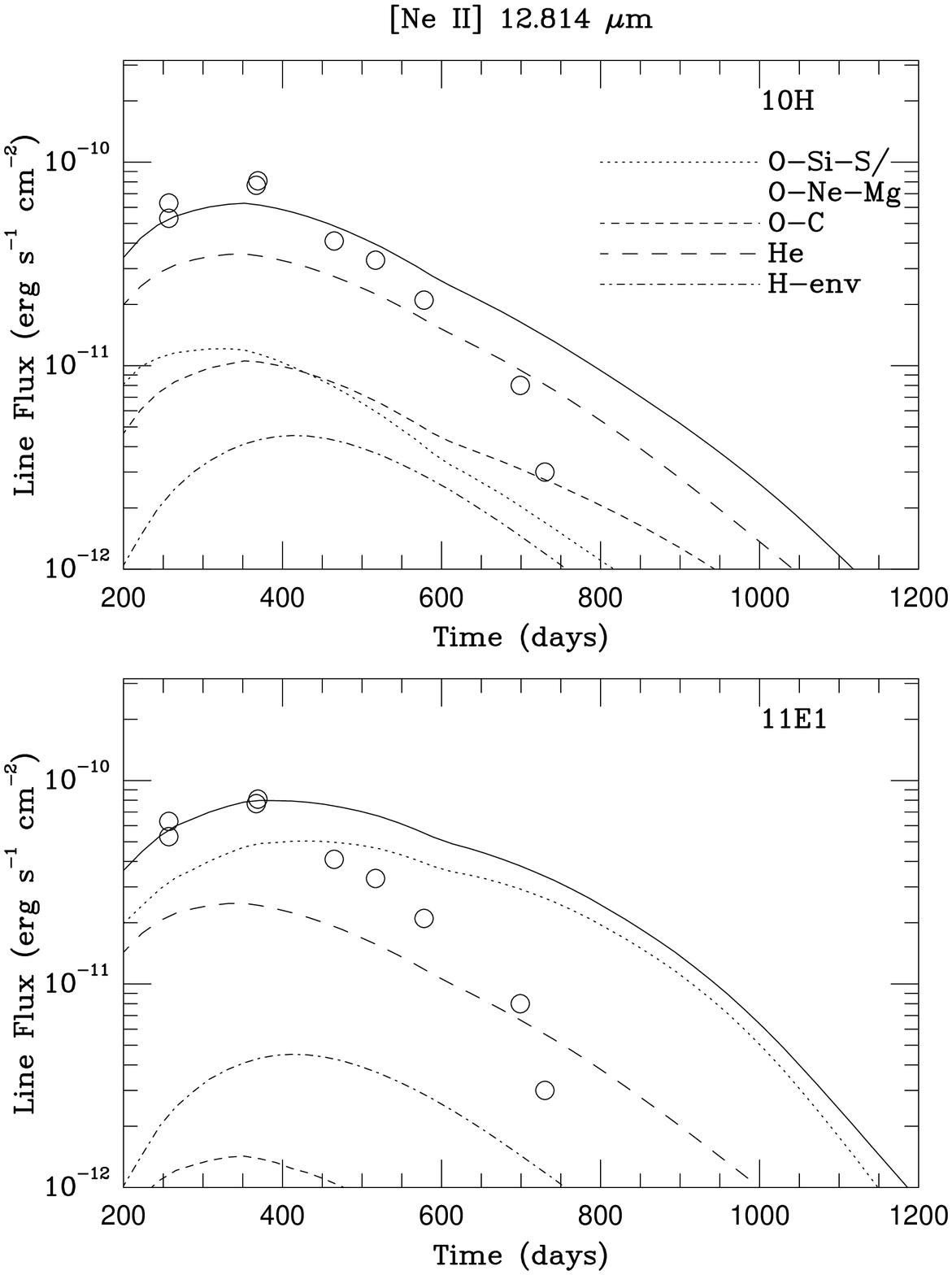}
\caption{Light curve of [Ne II] $\wl$ 12.814 $\mu$m for the 10H and the
11E1 model, respectively. Observations are from Aitken \etal (1988), 
Roche \etal (1993), and Colgan \etal (1994).
\label{fig:linne}}
\end{figure}

The neon abundance differs in the oxygen-rich zone by a factor of
$\sim 10$ between the 10H and 11E1 models. This is a result of the
different convective criteria employed in the two
models. Consequently, the emission in the $\wl 12.814~ \mu$m [Ne II] line 
differs considerably between the two models (Fig.
\ref{fig:linne}). While the agreement with 
the 10H model is satisfying up to at least $\sim 700$ days, the 11E1
model over-produces the line by a factor of $\sim 2 - 3$ at $t \gtrsim
400$ days. The last observations at $\sim 730$ days are considerably
lower  also in the 10H model. The uncertainty in the observation by
Roche \etal (1993) at this epoch, however, is considerable. 

\begin{figure}
\plotone{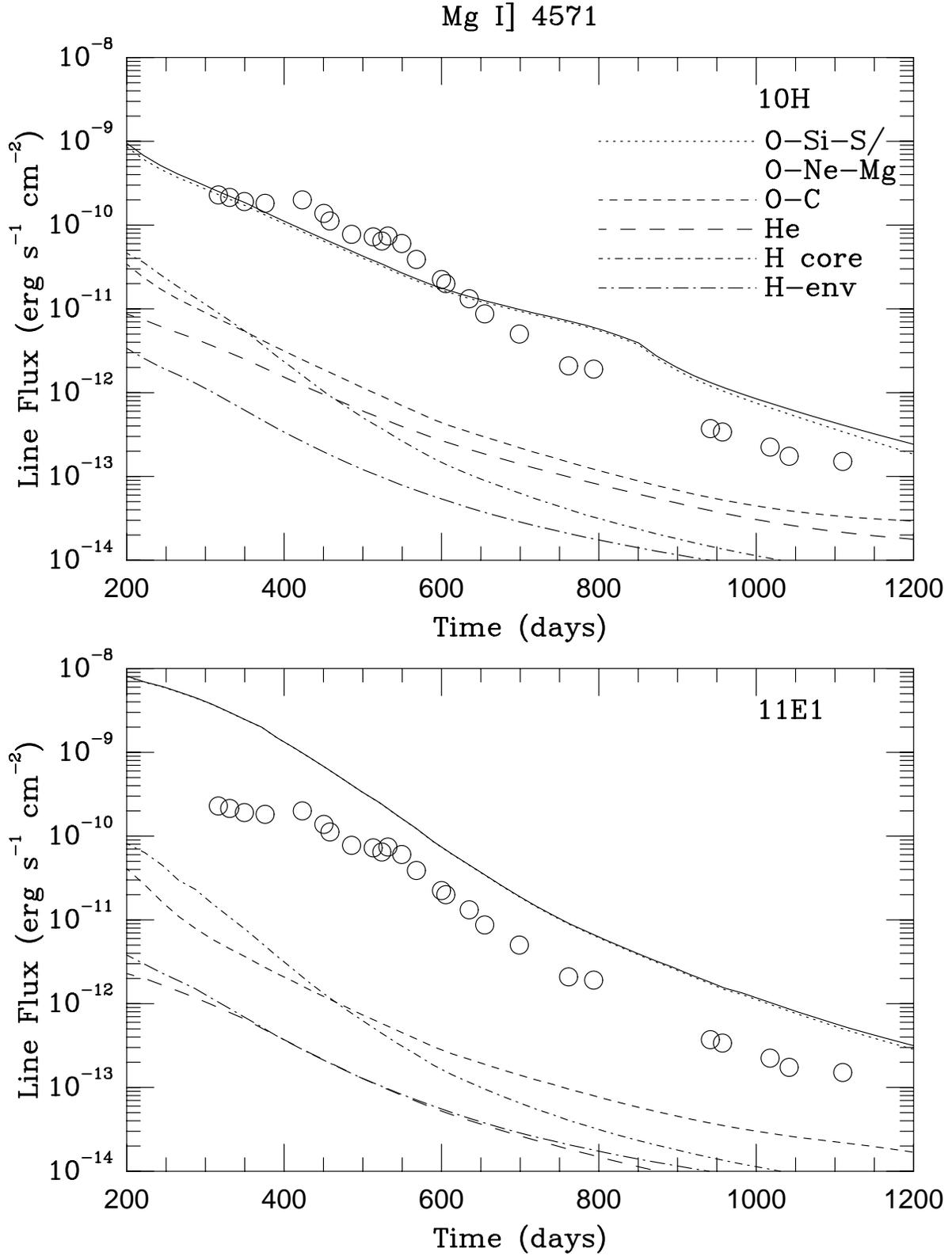}
\caption{Light curve of $\wl$ Mg I] 4571 for the 10H and the
11E1 model. Observations are from 
Danziger \etal (1991).
\label{fig:linmg}}
\end{figure}

 Mg I] $\wl 4571$  (Fig. \ref{fig:linmg}) is completely dominated by
the oxygen zone 
in both models. Because of the higher magnesium abundance in the 11E1
model, the flux is a factor of $\sim 3$ higher in this model
compared to the 10H model. The agreement with observations is 
considerably better in the 10H 
model. The Mg I] line is interesting because it is dominated by
recombination, rather than collisional excitation, as is usually the
case. It is
therefore not as sensitive to photoionization by the uncertain
UV-field as one might think (see section
\ref{sec-phot}). The effective recombination rate of the line is,
however, uncertain (see appendix in Paper I). Recombination dominates at all
epochs over both thermal and non-thermal collisional excitation. 

\begin{figure}
\plotone{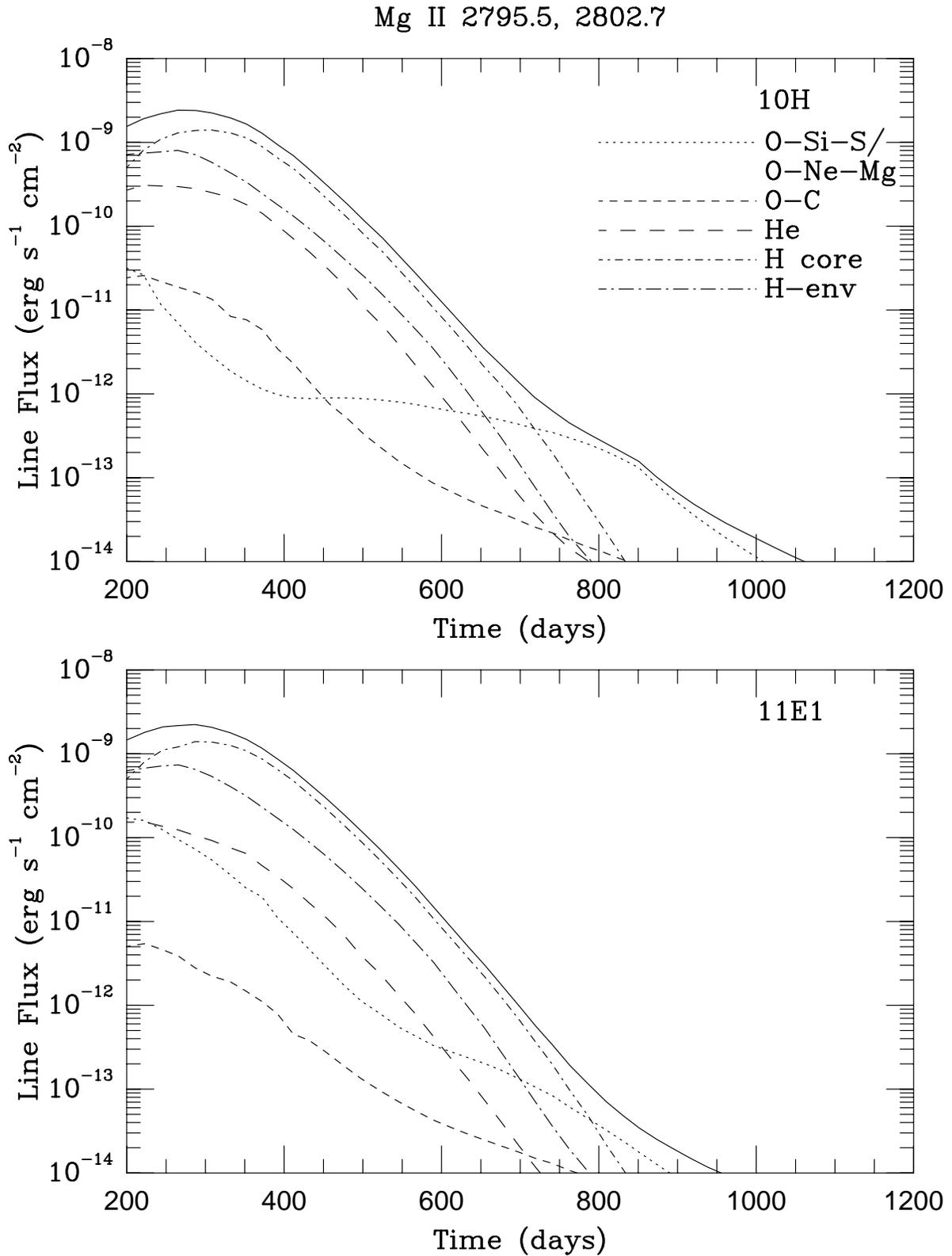}
\caption{Light curve of Mg II $\wll$ 2795, 2802 for the 10H and the
11E1 model. 
\label{fig:linmgii}}
\end{figure}

 Mg II $\wll 2795.5,
2802.7$ (Fig. \ref{fig:linmgii}) is surprisingly weak, with a flux of $\sim 10\%$ of $\Ha$, and
similar to Mg I] $\wl 4571$. Before 750 days it is excited
by thermal 
collisions, while at late time it is dominated by
non-thermal excitation. The contribution is at early time dominated by
hydrogen and helium regions, while in the non-thermal phase the 
oxygen-rich region dominates. Because of a lower Mg II fraction at
late time, the line is in this phase weaker in
the 11E1 model compared to  the 10H model, despite a higher magnesium
abundance in the former. Unfortunately,  resonance
scattering by the many  
metal lines in the UV  makes the observed luminosity at $\lesssim 1200$ days of this
line uncertain. The same is
true for the Mg I $\wl 2852$ resonance line. In section
\ref{sec-discussion} we compare it to the HST observations at 1862
days. 

\begin{figure}
\plotone{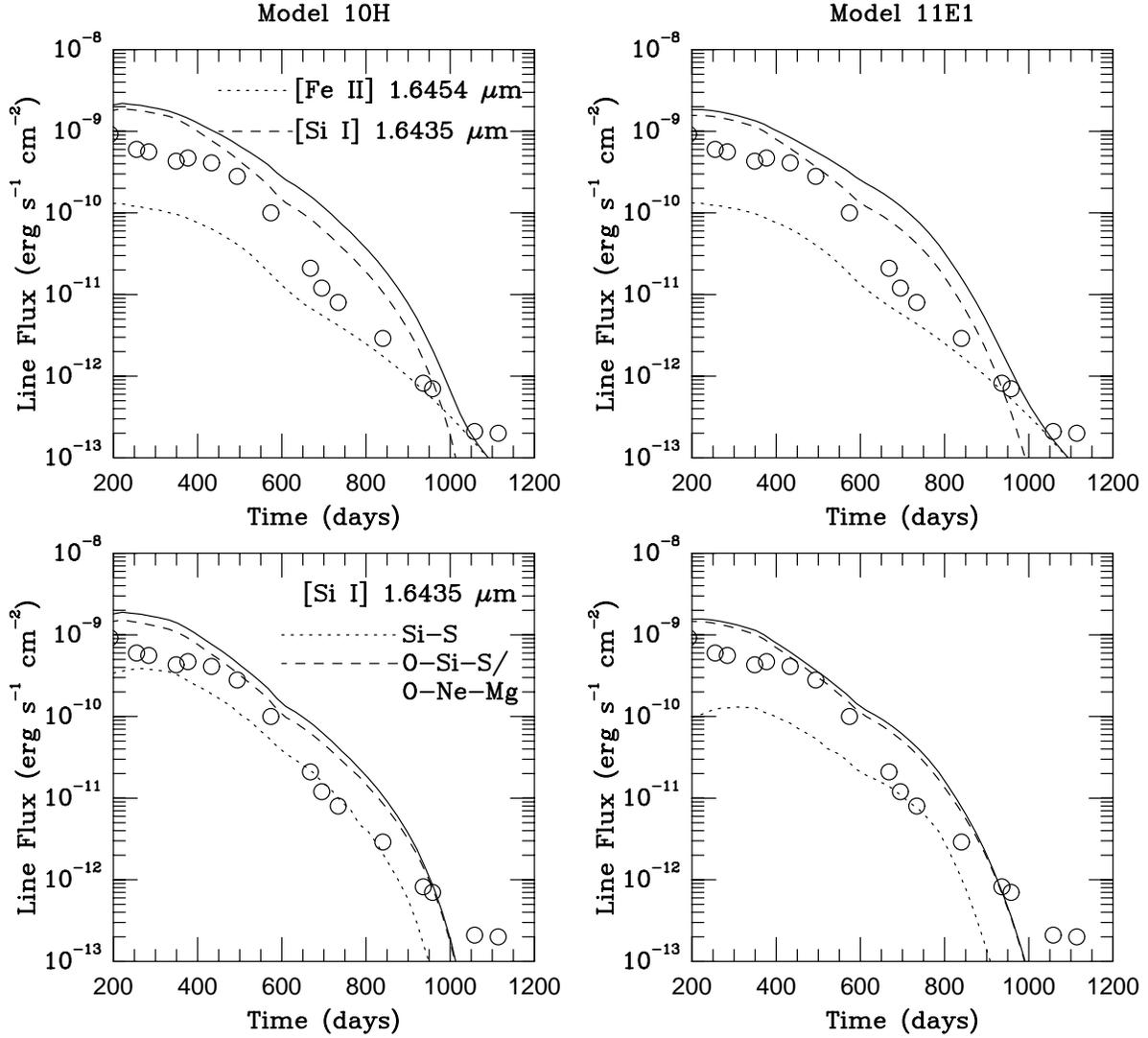}
\caption{The upper two figures show the contributions from Fe II and
Si I to the 1.64 $\mu$m feature for the 10H and 11E1 model,
respectively. The lower two figures show the contribution to the [Si I]
$\wl$ 1.6454 $\mu$m line from the different composition zones for the two models.
Observations from Meikle \etal (1989),(1993), and Bautista \etal (1995).
\label{fig:linsi}}
\end{figure}

The  1.64 $\mu$m line is most likely a blend of the [Si I] $\wl 1.6454
 ~\mu$m and [Fe II] $\wl 1.6435 ~\mu$m lines. The observed, relative
contributions are, however, uncertain. The [Si I] $\wl 1.6073
 ~\mu$m line should be a factor 2.84 weaker than the [Si I] $\wl 1.6454
 ~\mu$m line.  Also this line is  blended with [Fe II] and Br
13. Figure \ref{fig:linsi} shows the calculated fluxes in the 1.64 $\mu$m
feature, together with the individual contributions. During most of
the evolution the total flux is over-estimated by a factor 2 -- 3. 
The reason for the  early over-production may
be the large contribution from the O -- Si -- S region, which
dominates at that time. 
A lower abundance of silicon in this region, may bring the early part
into agreement with 
the observations. At $t \gtrsim 1000$ days the [Fe II] contribution from
the hydrogen region dominates  in all models.
The [Si I] $\wl 1.0991 ~\mu$m line is blended with Pa$\gamma$, and
also with the wing of the He I and [S I] lines, discussed
 below. Therefore we do
 not discuss it in any detail, and just note that the fluxes
given by Meikle \etal (1993) for this line are somewhat lower than
those from the 10H model. The Pa$\gamma$ flux should be small compared
to the [Si I] line (Fig. \ref{fig:phothei}), but the contributions from the He
I and [S I] lines large. 

As we have already discussed in section~\ref{sec-helium}, the strong
1.08 $\mu$m feature is 
most likely a blend of Pa$\gamma$, He I $\wl 1.0830 ~\mu$m and [S I] $\wl
1.0820~ \mu$m. At $t 
\lesssim 400$ days there is too high a luminosity in this feature in the
10H model 
(Fig. \ref{fig:phothei}). The good agreement with the He I $\wl
2.058~ \mu$m light curve gives us confidence that the He I $\wl 1.0830~
\mu$m is fairly accurate. The [S I] line is therefore likely to be
too strong in the 10H model. This is mainly a result of the high
sulphur abundance in the O -- Si -- S region in the 10H  model. 

 [S I] $3p^4~^3P_2 - 3p^4~^1D$ at 1.0820 $\mu$m and 
[S I] $3p^4~^3P_1 - 3p^4~^1D$ at 1.1306 $\mu$m originate from
the same upper level. The probability for  emitting a $\wl 1.0820 ~\mu$m
photon is 77 \%, and 23 \% for a $\wl 1.1306 ~\mu$m photon.
From  observations by Meikle \etal (1993) we estimate the flux in
the feature at 1.13 $\mu$m to $\sim 10^{-12}$
erg~s$^{-1}$~cm$^{-2}$ at day 695. 
Another possible candidate to this feature is O I $\wl 1.1287 ~\mu$m.
As was mentioned in section~\ref{sec-oxygen}, this line may arise as a result of the
O I -- Ly$\beta$ fluorescence mechanism (Oliva 1993).
We have not included this process in our calculations, and can
therefore not estimate the efficiency of this mechanism. 
Assuming the 1.13 $\mu$m feature to be due to [S I] $\wl 1.1306 \mu$m
results in a maximum  estimated 
flux of the [S I] $\wl 1.0820  ~\mu$m line of $\sim 3\EE{-12} $
erg~s$^{-1}$~cm$^{-2}$. In our calculations 
the flux in  [S I] $\wl 1.0820~ \mu$m is $\sim 8\EE{-12}$ and
$\sim 3\EE{-12}$ erg~s$^{-1}$~cm$^{-2}$ for the 10H and 11E1 models,
respectively. 
Also from this line the 11E1 model is therefore favored by the [S I]
emission. As mentioned in section \ref{sec-oxygen}, there may be other
allowed lines from S I in the spectrum, not included in our model.

\subsubsection{Calcium}
\label{sec-calcium} 
\begin{figure}
\plotone{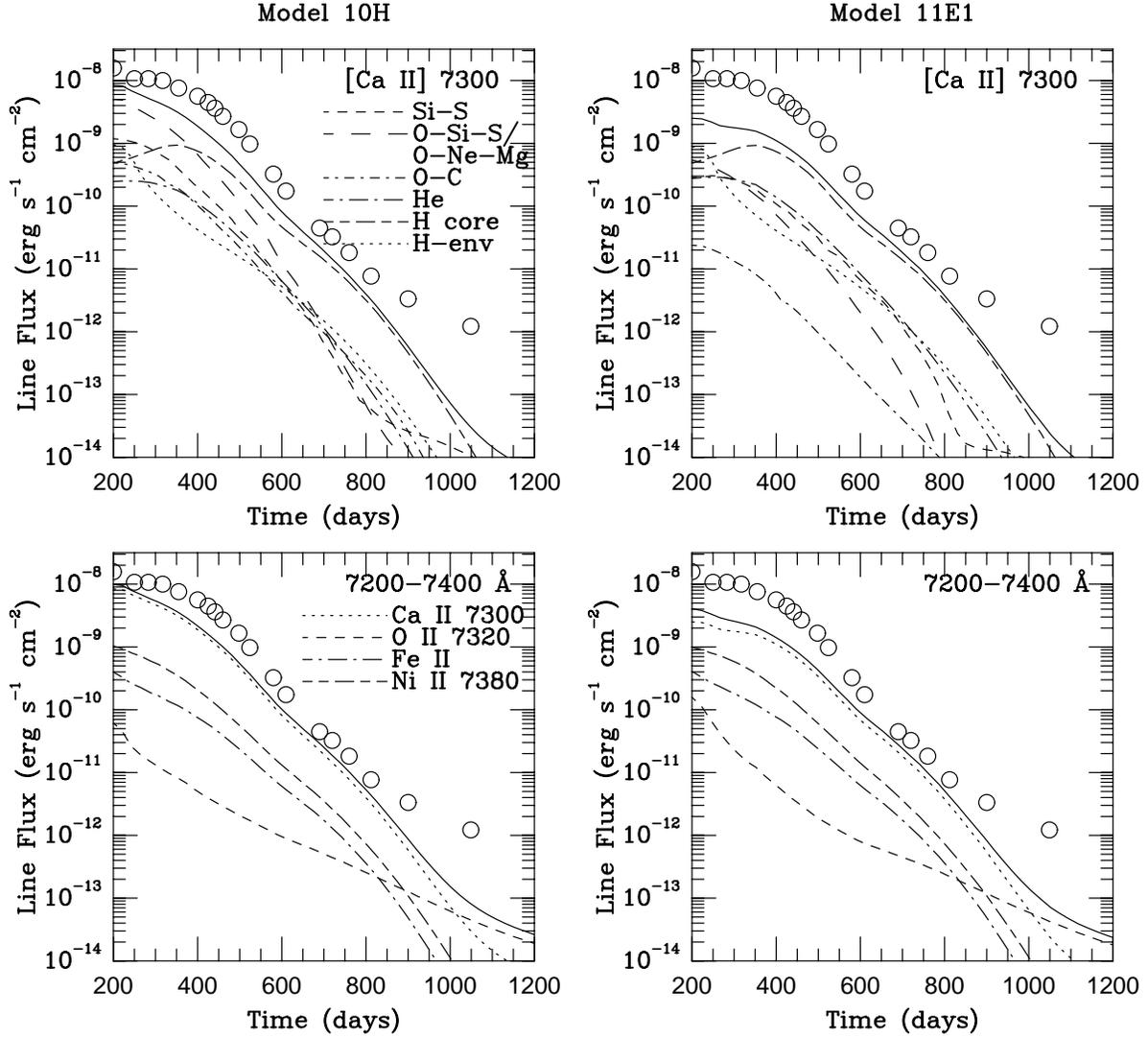}
\caption{The  upper figures show the light curve of 
[Ca II] $\wll$ 7291, 7324 for the 10H and 11E1 models,
respectively, with  contributions from the individual components. 
In the  lower figures the light curve of the 
[Ca II] $\wll$ 7291, 7324 lines, as well as for other strong lines in the
wavelength range 7200 -- 7400 \AA\ are shown. 
The curve labeled  Fe II includes several weak lines.
Observations from Spyromilio \etal (1991),  from Phillips \&
Williams (1991) and from Suntzeff \etal (1991). Note the strong
deficiency later than $\sim 800$ days, probably indicative of
photoexcitation. 
\label{fig:photcaii1}}
\end{figure}

\begin{figure}
\plotone{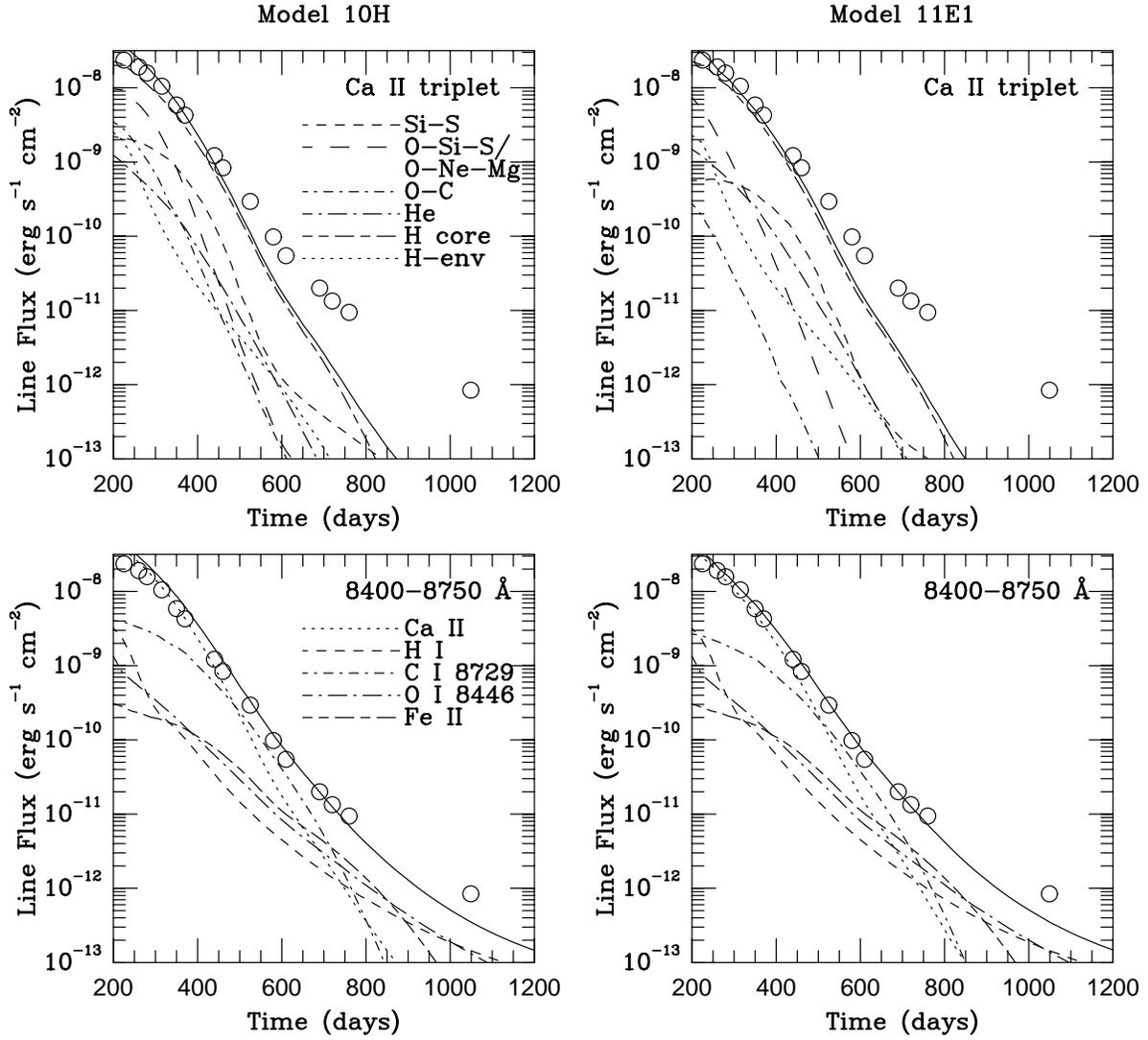}
\caption{The upper  figures show the Ca II IR-triplet with
contributions from the individual regions for 10H and 11E1,
respectively. In the lower figures the light curve of other lines
in the wavelength range 8400 -- 8750 \AA\ are shown. The curve labeled
H I contains the Paschen lines in this wavelength region.
Observations  from Spyromilio \etal (1991) and from Suntzeff \etal (1991).
\label{fig:photcaii2}}
\end{figure}

The light curves of the  [Ca II]  $\wll$7291,
7324 lines and the IR-triplet are shown in Figures \ref{fig:photcaii1}
and \ref{fig:photcaii2},
together with contributions from the different components. The overall
agreement of the $\wll$7291,
7324 light curve with observations is for the 10H model quite
satisfying, while the lines are severely under-produced in the 11E1
model. At  $t \gtrsim 900$ days there is a clear
tendency to under-produce
the luminosity in both models. This is still more pronounced with the triplet, but
here the contributions from the [C I] $\wl 8729$ and O I $\wl 8446$
lines also have to be included.  

 An
interesting point is the origin of the emission at various
phases, which is sensitive to the explosion model and differs between
the 10H and 11E1 models. At
no epoch in either model  does the Si -- S component, where the
explosively synthesized calcium
resides, dominate. Instead, either the oxygen region or the
hydrogen-rich gas contribute most of the emission in the lines.

 We agree with  Li \& McCray (1993) that little of the Ca II
emission comes from the Si -- S zone, although our conclusion  about the origin
differs somewhat from that
of Li \& McCray. These authors find that the Ca II lines all originate
from primordial calcium in 
the hydrogen and helium-rich matter, with a filling factor of $\sim
0.1$. Based on 
the likely calcium mass, they  argue that the newly synthesized calcium
captures too small a fraction of the total gamma-ray flux to explain
the observed line strengths. Their argument rests on the
assumption that calcium 
is the dominant element in the newly synthesized gas. This is not the case,
and nucleosynthesis models show that the calcium fraction is only
$\sim 0.05$,  with most of the mass in silicon and sulphur. Because of
the efficiency of Ca II as a coolant (e.g., Fransson \& Chevalier
1989), most of  the cooling of this mass may 
in fact be done by Ca II lines. The reason for the dominance of the
other regions is therefore the low total mass of the Si -- S zone,
compared to the other regions.
In the 10H model the mass of this region is $\sim 0.30
 \Msun$, and a fraction of only $\sim 5\EE{-2}$ of the
gamma-rays are absorbed in this gas. 

As found by Fransson \& Chevalier (1989), Ca II may  be
a strong coolant of the oxygen-rich gas, if mixing of calcium during the
hydrostatic pre-explosion burning is efficient. Crudely, the critical
calcium fraction for Ca II to dominate the cooling is $X({\rm Ca}) \sim 10^{-3}$. The extent of this
mixing is determined by the convection criterion used, as well as the
extent of over-shooting. In
the 10H model, where the Ledoux criterion was used and over-shooting
included,  this mixing is important, resulting in an abundance
of $\sim 6\EE{-4}$ in the O -- Si -- S zone, by
number. Consequently, at
$t \lesssim 400$ days the O -- Si -- S region dominates the $\wll$ 7291, 7324
contribution, while at later time the hydrogen component within the
core, with unprocessed calcium, 
dominates.  
Already at 700 days the temperature in the
hydrogen-rich gas is more than a factor
of two higher than in the O -- Si -- S region, explaining the higher
contribution to the Ca II emission from the former component. The
slow adiabatic 
decrease of the hydrogen temperature makes the decline of the Ca II
lines from this
component relatively slow. The temperature effect is seen even more
clearly in the triplet emission, which  is
dominated by the hydrogen core component at all times. 
    The 11E1
model, on the other hand, has very little calcium mixed  with the
oxygen. 

Li \& McCray find that in order for the Ca II lines to be
at the observed level after
$\sim 400$ days, radiative pumping of the H and K lines is necessary. 
 As is seen in Figures \ref{fig:photcaii1} and \ref{fig:photcaii2}, our light
curves agree fairly well 
with observations up to $\sim 800$ days, if all lines coinciding with
the 7300 \AA~ lines and IR-triplet are included. 
The O I $\wl 8446$ line has a velocity difference of 1836 $\kms$ from the Ca
II $\wl 8498$ component, and 3372 $\kms$ from the stronger $\wl 8542$
line. These lines are optically thick in both the hydrogen and oxygen
components, as well as in the Si -- S -- Ca region. In the H-core the
$\wl 8498$ component is optically thick up $\sim 750$ days, while the
$\wl 8542$ line is optically thick in the H-envelope at  $\sim 3400
\kms$ up to $\sim 350$ days. The O I $\wl 8446$
photons will therefore be scattered and emerge as Ca II triplet
emission up to $\sim 800$ days. This justifies the inclusion of the line in the flux from
the Ca II triplet, although in section~\ref{sec-ct2} we argue that
charge transfer probably decreases the 8446 flux dramatically. 

At times later than $\sim 800$ days there
is a clear 
deficiency. This is especially clear when observations
from CTIO at 953 -- 1149 days are included (Suntzeff \etal 1991). The
observed flux in the 7300 line on day 1046 is a factor of $\sim 10$ higher than
in our models, and the 8600 complex a factor of two, including  O I
$\wl 8446$, which may be questionable.  
Our models do include radiative excitation, but only by the
continuum. On day 1046 the two-photon continuum flux within the region
covered by 
the H and K lines, $\sim 170$ \AA, is only $\sim 6\EE{-14} \ergs ~{\rm
cm}^{-2}$, which is much lower than the total, dereddened flux in the
7300 \AA~ 
lines and IR-triplet, $\sim 2.9\EE{-12} \ergs ~{\rm cm}^{-2}$. If we,
however, for the same day extrapolate the observed average, dereddened flux on
either side of the H and K lines,
$\sim 1.3\EE{-14} \ergs ~{\rm cm}^{-2}~{\rm \AA}^{-1}$, to the wavelength of these lines, we can estimate
the total absorbed flux by the H and K lines to $\sim 2.3\EE{-12}
\ergs ~{\rm 
cm}^{-2}$, close to the total flux in the 7300 \AA~
lines and IR-triplet. Although the flux in the interior of the ejecta
may differ from the escaping flux, this strongly suggests that
radiative excitation in the H and K lines is responsible for the flux
in the 7300 \AA~
lines and IR-triplet at late epochs. The fact that the fluxes of these
lines follow the bolometric flux at a level $(L_{7300}+L_{\rm IR}) /
L_{\rm bol} \approx 7\EE{-3}$ does not mean that non-thermal excitation is
necessary, as for the [O I] lines. If the U-band flux follows the
bolometric, as is a fair approximation to the observations after $\sim 800$ days  (Suntzeff
\etal 1991), radiative excitation will also result in a nearly constant
$(L_{7300}+L_{\rm IR}) / L_{\rm bol}$. 
In fact, using the efficiency for
non-thermal 
excitation in KF92, one can estimate the non-thermal contribution to
$\lesssim 10^{-3}~(M_{\rm Si-S}/0.3 ~\Msun)$ of the bolometric, much less
than observed. 

The relative intensities of the Ca II H and K lines, the $\wll$7291,
7324 lines and
the IR-triplet are discussed under various conditions by 
Ferland \& Persson (1989),  Fransson \&
Chevalier (1989), and Li \& McCray (1993). These authors find
that above $n_e \approx 10^7 \rm~cm^{-3}$ the H and K, as well as the IR
triplet, are in LTE, and therefore mainly depend on temperature. At
temperatures below $\sim 4500$ K both the H and K and the IR
triplet decrease rapidly with temperature. At times when radiative
excitation dominates, and if thermalization of the lines
can be ignored, the ratio of the 7300 \AA~
lines and IR-triplet is expected to be $\sim 8600 / 7300 = 1.18$,
close to the observed value of $\sim 1.4$ at 1046 days.

\subsubsection{Iron, Cobalt, and Nickel}
In our model Fe I-IV are treated as multilevel
atoms with 121 levels for \FeI, 191 levels for \FeII , 110 levels for
\FeIII, and 43 levels for \FeIV. Fe V is included only with
its ground state.
The atomic data, which includes new IRON project data, are discussed in
the appendix in Paper I. 
The total recombination coefficients (Shull \& Van Steenberg, 1982), 
and also the fractions of the radiative recombination rates going to the
ground states (Woods, Shull, \& Sarazin, 1981), are probably
reasonably well known. 
However,  a major uncertainty is the individual recombination rates
to the excited levels of \FeI, and \FeII, which are largely unknown. We will
discuss the consequences of this later. 

\begin{figure}
\plotone{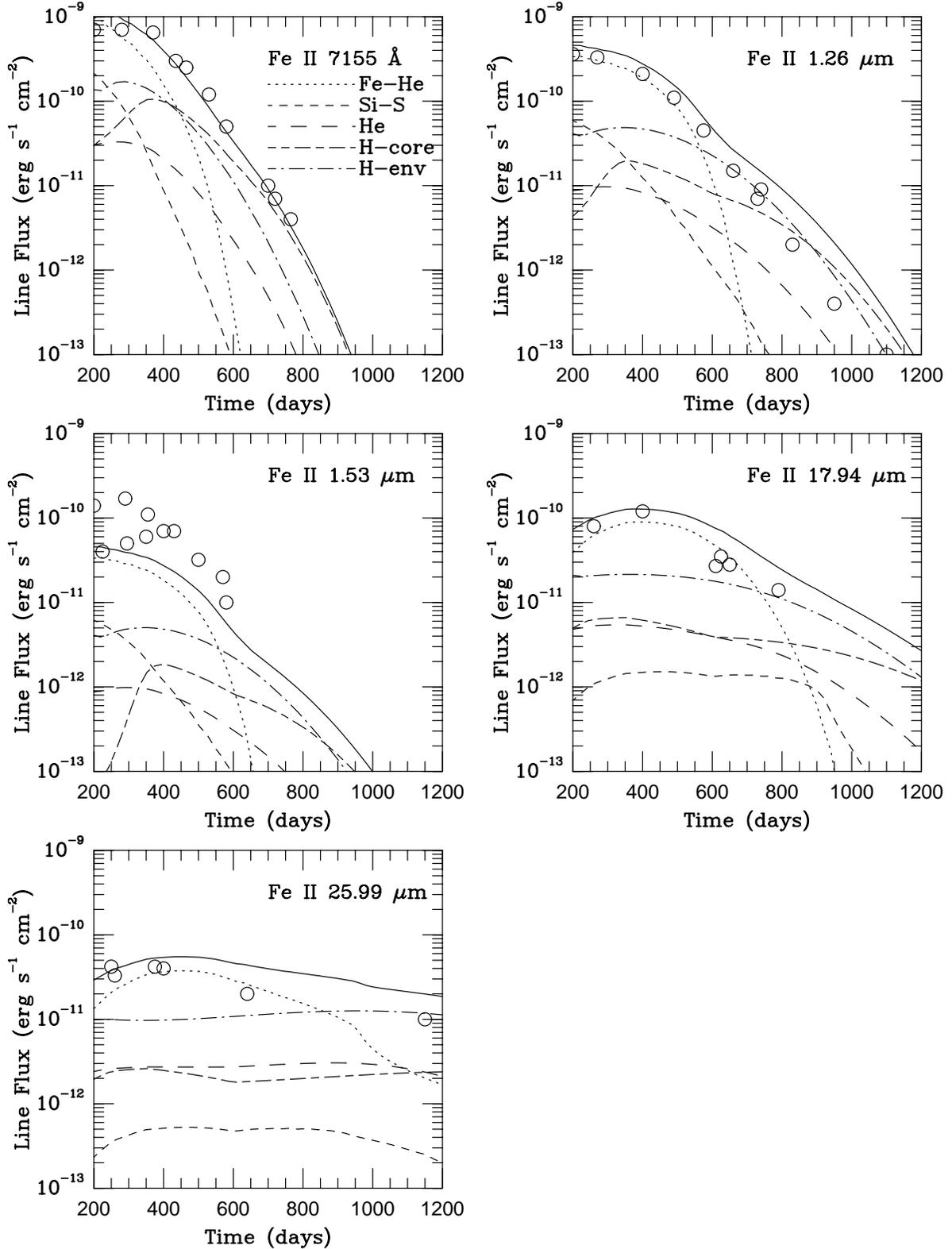}
\caption{Light curves for some of the most important Fe II lines. 
The calculations based on the 10H and 11E1 models do not differ
significantly from each other. 
The contributions from primordial and processed iron are shown
separately. Because of the IR-catastrophe the [Fe II] flux from the
iron core drops to a very low level at $t \gtrsim 600$ days, except for the 
$\wl 25.99~ \mu$m line. 
References for the observations are
given in the text.
\label{fig:photfeii}}
\end{figure}

In Figure \ref{fig:photfeii} we show the light
curves of the most important Fe II lines.
The observational data are taken from
Erickson \etal (1988), McGregor (1988), Meikle \etal (1989), 
Moseley \etal (1989), Haas \etal (1990), Varani \etal (1990), 
Spyromilio \etal (1991), Dwek \etal (1992), Jennings \etal
(1993), Colgan \etal (1994), and Bautista \etal (1995).

In general we find agreement with 
observations to within a factor of two, or better. It should be
noted that the agreement with observations at epochs earlier than
$\sim 400$ days is very good, with a possible exception of the heavily
blended $1.53~ \mu$m line. At later epochs most of the calculated
lines without correction for dust absorption
are a
factor $\sim 4$ stronger than observed. Including this correction with a factor $\sim 40 \%$, as found by
Lucy \etal (1991), gives a greatly improved agreement, although they
are still up to 
a factor of $\sim 2$ over-luminous. The importance of dust  was
earlier noted by Colgan \etal (1994) 
for the
$\wl 25.99~
\mu$m line at 640 days. 

The contributions to
especially the $\wl 7155$ line and the near-IR $\wl 1.26 ~\mu$m and $\wl
1.53 ~\mu$m lines are interesting. From being
dominated by the iron core at $t \lesssim 600$ days, the contribution from
unprocessed iron in the hydrogen  component becomes the most
important after this epoch. The same is true for the far-IR $\wl 17.94~ \mu$m
and $\wl 25.99~ \mu$m lines, although the transition for these occur later. For
these lines the hydrogen  component is important throughout the
whole period, and later than $\sim 800$ days dominates  the iron core
contribution. The rapid drop in the different lines from the core can
clearly be seen in Figure 6 of Paper I.
The dominant hydrogen contribution to the $\wl 25.99 ~\mu$m line at late
time is in line
with  the KAO-observation at 1153 days by Dwek \etal (1992), who find
an emitting iron mass of only $3.8\EE{-3} \Msun$, consistent with
primordial. 

An estimate of the luminosity of the $\wl 25.99 ~\mu$m line coming from
unprocessed gas in the hydrogen to oxygen  regions is given by 
\begin{eqnarray}
L_{26\mu} & = & 2.7\EE{36}~A_{\rm mean}^{-1}~\left({ X({\rm Fe})\over
1.7\EE{-5}}\right)~\left({M
\over 1 ~\Msun}\right)~ \cr
 & & e^{-554{\rm ~K}/T} ~\ergs,
\label{eq:lumfe26}
\end{eqnarray}
where $M$ is the total mass of hydrogen, helium or oxygen-rich gas. The optical depth of the
$\wl 25.99~ \mu$m line in these components is 
\begin{eqnarray}
\tau_{26\mu} & = & 5\EE{-2}~A_{\rm mean}^{-1}~f_{i}^{-1}~\left({ X({\rm Fe})\over
1.7\EE{-5}}\right)~\left({M \over 1 ~\Msun}\right)~ \cr
 & & \left({V \over 2000~\kms}\right)^{-3} 
 \left({T \over 4000 {~\rm K}}\right)^{-1}\left({t
\over 500 {\rm days}}\right)^{-2} \cr
 & & 
\label{eq:taufe26}
\end{eqnarray}
where $V$ and $T$ are the average velocities and temperatures, respectively, of the
unprocessed gas, and $f_{i}$ is the filling factor of the
component. We assume that $T \gg 554$ K. At
400 -- 600 days the typical temperature in the hydrogen and helium-rich
gas is
$\sim 4000$ K (Paper I). Further, our line profile fits show that most of the
hydrogen and helium-rich gas have velocities within
$\sim 4000 \kms$. Therefore, unless  clumping is high, the $\wl 25.99~
\mu$m line from the unprocessed iron is likely to be optically thin, as is also argued on observational
grounds  by Haas \etal (1990). The total luminosity of the $\wl 25.99 ~\mu$m line
on day 407 was 
$\sim 1.3\EE{37} \ergs$, decreasing to $\sim 6.3\EE{36} \ergs$ on day 640
 (Haas \etal 1990, Colgan \etal 1994). With $M \gtrsim 10 \Msun$, it is
clear that a large fraction of the $\wl 25.99 ~\mu$m line may originate from
unprocessed iron. As Figure \ref{fig:photfeii} shows, the same applies to the other [Fe
II] lines, including the $\wl 1.26 ~\mu$m and $\wl 17.94 ~\mu$m line. 

The disappearance of the iron core contribution to the near-IR [Fe II] lines
was first noted  by  Spyromilio \& Graham (1992), who correctly
attributed this to the IR-catastrophe. From their last observation on
day 734 they infer an Fe II mass of $\lesssim 0.01 \Msun$, and attribute
the rest to iron too cold to emit in the near-IR. 

While our light curves agree well with those of Li, McCray \& Sunyaev (1993)
for $t \lesssim 700$ days, a major difference between our models and those of  Li
\etal is in the behavior of the Fe II light curves at late time. The models by Li
\etal fail to explain the emission from the Fe II, Co II, and Ni II
lines for t $\gtrsim$ 2 yr, with  too rapid a drop of
especially the optical and near-IR lines at $\sim 600$ days. As a
solution they suggest 
that photoionization by the UV
continuum, especially two-photon emission, from helium surrounding the
iron may
solve the problem. We do not experience this problem, because of the
contribution to these lines from unprocessed iron in the helium and
hydrogen region. In addition, we do include photoionization in the core,
and as we discuss below, this may actually give too large an effect on the
state of ionization. If photoionization  really was needed,  there may be a 
more local source of UV-photons; according to the 10H and 
11E1 models there is a substantial amount of helium mixed
microscopically with the
iron, roughly equal abundance by number (see
Tables 1 and 2 in Paper I). This could  result in a substantial local
UV-radiation field in the iron clumps (see below).

The light curves in Figure \ref{fig:linconi_ii} of  [Co II] $\wl 10.52
  ~\mu$m, and [Ni II] $\wll 
6.634, 10.68 ~\mu$m all agree as well as can be expected, given the
quality of the atomic data. Up to $\sim 700$ days the contribution
from the Fe -- He core dominates all lines. After this epoch the Si --
S contributes most of the flux of [Ni II] $\wl 
6.634 ~\mu$m up to $\sim 1100$ days.   Primordial nickel dominates
[Ni II] $\wl 10.68 ~\mu$m  already after $\sim 600$ days,
similar to the [Fe II] lines. The [Co II] line drops rapidly after
$\sim 900$ days because of the low primordial cobalt abundance. 

While  agreement with the Fe II, Co II, and Ni II lines is
satisfying, the Fe I lines are weaker in these models by several
orders of magnitude. This is seen in 
especially the [Fe I] $\wl 1.44 ~\mu$m emission.
The under-production of the [Fe I] $\wl 1.44 ~\mu$m line is an
ionization effect, resulting from a very low predicted Fe I abundance in all
regions. The typical Fe I fraction  in the iron core at 400 -- 600 days is
only $(1-4)\EE{-5}$, with most of the ionization resulting from
photoionization by Fe II recombination lines and the He I two-photon
continuum. A 
similar effect is seen  in the [Ni I] $\wl 3.119 ~\mu$m line, which is
under-produced by a similar magnitude. In section~\ref{sec-phot} we show that
changes in the UV-field can have dramatic effects on the fluxes of these lines. 

\section{UNCERTAINTIES IN THE CALCULATIONS}
\label{sec-shortcomings}
\subsection{Filling Factors}
\label{sec-fillingfac}
The filling factors we employ for the various components are
uncertain. To test the sensitivity of our results to the
assumed filling factors, we have run a set of models where we have
varied  all filling factors by a factor of two in either direction
(always assuring that the total is one).  The range we have
investigated is therefore $f_{\rm H} = 0.075 - 0.30$,  $f_{\rm He} =
0.13 - 0.60$,  $f_{\rm O-C} = 0.03 - 0.12$,  $f_{\rm O-Si-S} = 0.06 -
0.24$,  $f_{\rm Si-S} = 0.0085 - 0.034$, and  $f_{\rm Fe} = 0.17 - 0.70$.

With regard to the temperature evolution, we find a relatively small change
up to the time of the IR-instability. The epoch when this
sets in varies in the metal-rich regions by $\sim 100$ days,
 over the whole range of  filling factors. The
IR-instability occurs  
earlier when the filling factor increases, i.e., the density
decreases. This affects the epoch of the steepest part in the light
curve of the different lines. Because of the importance of adiabatic
cooling, the temperature of the hydrogen and helium components in the
core are hardly affected at all.

Most of the hydrogen lines are insensitive to the filling factor. An
exception is $\Hb$, which in the period 200 -- 400 days increases by a
factor of $\sim 2$ when $f_{\rm H}$ decreases from 0.3 to
0.075. 
The insensitivity of the
hydrogen lines to the filling factor is in some disagreement with Xu
\etal (1992).  

The [O I] $\wll 6300, 6364$ lines show little dependence on the
filling factors in either the oxygen or hydrogen and helium-rich
regions. The non-thermal part is also unaffected by this. The
discrepancies in the light curve of this line are therefore not likely
to be connected to the assumed density, but rather to the chemical
composition or charge transfer effects. 
Also [Ca II] $\wll 7291, 7324$
show  weak dependence on the filling factors. The  exception is
the contribution from the hydrogen core, which in 
the period 400 -- 800 days  decreases by a
factor of up to three between $f_{\rm H}= 0.075$ and $f_{\rm H}=
0.3$. The lower 
filling factor would give a better representation of the light
curve. The contribution from the O -- Si -- S zone, however, differs
considerably less as $f_{\rm O}$ varies. Roughly the same  is
true for the Ca II triplet lines. 

While the optical and near-IR Fe II lines, e.g., $\wll 7155$ \AA, $1.26,
1.53, 1.64 ~ \mu$m, are insensitive to both $f_{\rm Fe}$ and  $f_{\rm
H}$, 
the far-IR lines show a higher sensitivity. Of the individual
contributions to the $\wl 17.94~\mu$m and $\wl 25.99~ \mu$m lines the hydrogen core
contribution is only weakly dependent on $f_{\rm H}$. The
Fe -- He contribution, however, decreases by a factor nearly
proportional to 
$f_{\rm Fe}$. These conclusions agree with equation (\ref{eq:lumfe26})
and equation (43) in Paper I.  
Contrary
 to  Li, McCray,
\& Sunyaev (1993) we 
find that the observations agree better with a fairly low filling
factor, $f_{\rm Fe} \sim 0.2$. 
The reason may be our inclusion of
regions other than the iron-core, and the additional Fe II emission
from these. 

Summarizing this discussion, we find that plausible variations of the
filling factors have rather small effects, at the factor of two level,
or less. Except for perhaps the far-IR [Fe II] lines, it is likely that
other uncertainties, e.g., the hydrodynamics or abundances within the
different components, are at the same level or worse. We are therefore somewhat
cautious of drawing any far-reaching conclusions about the filling
factors from this.

\subsection{Charge Transfer}
\label{sec-ct2}
Charge transfer is important for both the ionization balance and the
line emission. In particular, lines and continua arising as a result
of recombination can be severely affected. Earlier we noted a marked
over-production of the O I recombination lines. As was shown in Paper
I, the O II fraction is, however, sensitive to charge transfer with Si
I, with an uncertain rate. To see the effect of this in more detail we
have varied the rate of Si I + O II $\rightarrow$ Si II + O I in the
region $10^{-13} - 10^{-9}$ cm$^{3}$~s$^{-1}$ in our models. We find that
the O I
 $\wll 7774, 8446$ and 9265 recombination lines, as well as the O I
 continua,  all decrease by 
factors of 10 -- 100, corresponding to fluxes well below the upper
limits from the spectra. This is obvious from the O II curve in Figure
10
of Paper I. Charge transfer between O II and Si I would therefore
solve the problem with the 
too large O I recombination fluxes in the previous models. At the same time
the problem with the non-thermal, flat part of the [O I] $\wll 6300,
6364$ light curve is exaggerated, because the contribution from the
triplet levels, which to a substantial part are fed by recombination,
decreases too. 

Because of the small change in the Si I fraction, [Si I] $\wl 1.6454~
\mu$m does 
not change appreciably. 
In Paper I we noted that the increase in the Si I + O II charge transfer lead to a
decrease in the Mg I fraction. This does, however, not change 
Mg I] $\wl 4571$ 
substantially, because of the dominance of 
recombination. Because of the large contribution to
the 8600 \AA\ feature from the O I $\wl 8446$ line without O II + Si I charge
transfer (Fig. \ref{fig:photcaii2}), the flux of this decreases in
this model substantially, later than $\sim 800$ days. Photoexcitation
can probably compensate for this. 

Another interesting consequence of the O II + Si I charge transfer is that the
Fe I fraction in the O -- Si -- S zone increases by several orders of
magnitude, although still too low to give Fe I lines of sufficient
strength. The reason for this is that the UV flux in this zone to a 
large extent is determined by O I and Si I recombinations. Both 
decrease dramatically, while the Mg I recombination radiation is not
energetic enough to ionize Fe I. In principle, other charge transfer
processes in 
the other zones could possibly increase also the total Fe I
flux. 

We have also tested the influence of charge transfer from excited
states of He I, discussed by Swartz
(1994). However, here we find that they only contribute a few percent
of the total He I ionization. Charge transfer from the ground state is
also unimportant for helium.

\subsection{Photoionization}
\label{sec-phot}				
Resonance scattering in the hydrogen dominated envelope has been
discussed by Li \& McCray (1996), using a Monte-Carlo model. Although
their results are sensitive to their assumed temperature, 
 the results may be indicative to the order of
magnitude in the intensity. As input spectrum they use a pure
He I two-photon continuum, but do not include other sources like the H
I two-photon continuum or line emission, like $\La$, O I $\wll
1356, 1641$, and Mg II, as well as recombination emission from Si I
and Mg I. Nevertheless, these calculations illustrate
the importance of the UV scattering. 

In Li \& McCray's model at 200 days the $\it {emergent}$ flux is
down by a factor of $\sim 10$, while the lower temperature and density
at 800 days give a factor of $\sim 3$ suppression. The intensity
inside the ejecta should be larger than the emergent intensity. On the
other hand, scattering by the iron-rich material in the Fe -- He bubble,
not considered by Li \& McCray,
may further increase the scattering, and decrease the
intensity. Resonance scattering will decrease with time, as
is shown by the increasing UV flux in the IUE-band of SN 1987A (Pun
\etal 1995), as well as the emergence of clear lines in the
HST-spectra at $\gtrsim 1800$ days (Wang \etal 1996, Chugai \etal 1997). 

In addition to resonance scattering, which mainly redistributes the
emission in wavelength, dust absorption may have strong effects for
the UV intensity. The
resonance trapping in the UV increases the path length, compared to
the optical range, and with a covering factor of $\sim 40 \%$ the
probability for absorption may be very large. Experiments with a
simplified Monte Carlo model, similar to that in Fransson (1994),
confirms this. 

In  KF92 the different contributions to the UV emissivity were discussed in
detail, using arguments based on energy input, optical depths and
atomic physics. Most of the results in KF92 are confirmed by our
calculations, although quantitatively the emissivity differs from
KF92. The most important reason for this is the different mass
distribution in the models here, based on observed line profiles, and in KF92.

In the hydrogen zone most emission comes from H I, with $\La$ and
two-photon emission being most important. When the optical depth in
$\La$ is sufficiently large the 
photons scatter many times, and finally emerge in the two-photon
transition from the 2 $^2$S state to the ground state
(Xu \etal 1992, KF92, Fransson 1994).  
In the hydrogen-rich regions in the core we find that the two-photon continuum
dominates over $\La$ 
(i.e. A$_{2\gamma}$ $>$ A$_{\La}\beta_{\rm esc}$) at all times between 200
-- 2000 days. In the hydrogen envelope, where the density is
decreasing outward,  $\La$ dominates in the
outer zones at 500 days, and by 1200 days the $\La$
emission is dominating in the entire envelope. 
At 500 days 45 \% of the total energy absorbed by the hydrogen
envelope is emitted in the H I two-photon continuum. 

In KF92 it was found that $\sim 50 \%$ of
the energy absorbed by helium is re-emitted as He I two-photon
emission. Of the emission coming out above 11.26 eV, most is absorbed by
photoelectric absorption by the C I in the helium zone. This is caused
by the large carbon abundance in this zone. A large fraction
of this is in turn re-emitted as C I recombination lines in the UV. 
Altogether, it was estimated that $\sim 40 \%$ of the energy was emitted
in the UV between 1100 -- 3646 \AA. This high efficiency coupled to the
large relative fraction of the energy deposited in the helium zone,
means that the helium zone can be the most important source for the
UV-radiation field.

In these calculations we find that compared to KF92  a considerably
smaller fraction of 
the deposited energy goes into the helium zones, only $\sim 4 - 8 \%$,
(compared to $\sim 30 \%$ in KF92).
This is mainly due to our redistribution of the helium  to higher
velocities.
Of the total emission from the helium zones, $\sim 30 - 60 \%$ is emitted
in the UV $(\lambda < 3646$ \AA). Of this, $\sim 40 \%$ comes
from the He I two-photon continuum after $\sim 500$ days. The 1100 - 3646 \AA~
UV-efficiency is similar to the number above, $\sim 37 \% $ of the
absorbed energy.   As in KF92, we find that  the 584 \AA\ line 
is mainly recycled into the 2.058 $\mu$m line, plus two-photon
continuum, because of the higher probability of branching into the
2.058 $\mu$m line, compared to escape or continuum destruction in the
584 \AA~ line. 
The He I and H I two-photon continuum photons 
constitute the dominant sources for the UV-field in the ejecta.
 Up to $\sim$ 600 days other strong
UV-lines are Mg II $\wl$ 2800, C II] $\wl$ 2326, and C I $\wll$ 2966, 2968.

In the oxygen region recombination emission to the ground state of Si
I ($\sim 14 \%$ of the absorbed energy) and Mg I ($\sim 3 \%$) at 1521 and 1621 \AA, respectively, give most of the
emission. Also recombination to the excited states in Si I is
important ($\sim 9 \%$). Because our recombination model of Si I is
crude, we can 
not predict the detailed emissivities of these lines, although most should
emerge at $\sim 2000$ \AA. 

The UV fraction from the Fe -- He region is comparatively
low, $L(\lambda < 3646 {~\rm\AA})/L_{\rm total} \approx 0.14$. Most of this emerges
as He I recombination radiation to excited states, and He I two-photon
continuum emission ($\sim 2.6 \%$). Because of the large optical
depths most Fe II emission is degraded into optical lines. 

To investigate the uncertainty in the UV radiation field we have made
one calculation where all photoionization processes from the ground
states of the low ionization potential elements, Na I, Mg I, Si I, Ca
I, Fe I, Ni I and Co I, are switched off. 

Without photoionization we now get an Fe I fraction of $\sim 17 \%$ at
500 days, while with photoionization the Fe I fraction is $\sim
10^{-5}$.
This illustrates the sensitivity of the neutral
fraction of low ionization potential elements. The Co I and Ni I
fractions increase correspondingly.  

Our [Fe I] and [Ni I] light curves (Fig. \ref{fig:linfeni_i}) are
without photoionization only a factor of two lower
than the observations, but given the quality of the atomic data and
the sensitivity to both UV-field and filling factor, it is not too
surprising. 
Because the Fe I fraction is sensitive to the
assumed filling factor of iron, with $n({\rm Fe ~I})/n({\rm Fe}) \propto f_{\rm
Fe}^{-1}$ for  $n({\rm Fe~ I})/n({\rm Fe}) \ll 1$,  a smaller filling
factor  could  increase the fluxes. 
The [Fe II] lines decrease by $\sim 25 \%$, corresponding to the
decrease in the Fe II fraction, improving the agreement with the
observations somewhat. 

The Mg I] $\wl 4571$ line decreases by a factor of 2 -- 3 for $t \gtrsim
700$ days, in better agreement with the observations. The reason is
that radiative recombination 
is responsible for $\sim 98 \%$ of the flux of this line at 800
days. Because the fraction of Mg II at 800 days has decreased from $X({\rm
Mg~ II}) \approx 1.00$ to  $X({\rm Mg~ II}) \approx 0.28$,
the Mg I] flux  decreases.  At $t \lesssim 500$ days the Mg I] flux is
instead higher in the model without photoionization. This is 
consistent with a large recombination contribution to the line, which
does not vary much in the two models, because $X({\rm Mg~ II}) \approx
1.0$. In the non-photoionization model  $X({\rm Mg~ I}) \approx
1.1\EE{-2}$, while it is $\sim 7.3\EE{-5}$ with
photoionization. Without photoionization there is therefore a
substantial collisional contribution to the line, explaining the
increase. 

Summarizing this, there are  strong indications that the
UV-field, below at least $\sim 1600$ \AA, in the ejecta is substantially weaker than in our standard
models, because of resonance scattering, in combination with 
branching into the optical, or, especially efficient, dust
absorption. We emphasize that this uncertainty mainly affects ions with
low ionization potentials, and abundances. Dominating ions, like H I,
He I, O I,
Mg II, Ca II and Fe II, are only affected marginally.

\section{DISCUSSION}
\label{sec-discussion}
Summarizing these calculations, we find both successes and failures.
Starting with the former, we find a general good qualitative agreement  with the
observed evolution of most lines. 

The general behavior of the hydrogen and helium lines is well reproduced,
although this has already been shown by previous calculations
(Xu \etal 1992, KF92). Here, we have added the time dependence, which is
necessary for $t \gtrsim 800$ days. In KF92 we found a deficit in the flux at
late time. This has now disappeared, partly because of the freeze-out
(see also Chugai \etal 1997).
In addition, 
we  include a more realistic temperature dependence, as well as a more
consistent calculation of the intensity of the radiation in the Balmer
continuum, which in turn determines the Balmer line fluxes. 

The effects of including the time dependence are most pronounced  for
lines excited by thermal collisions and  recombination lines, coming
from regions where 
adiabatic cooling and freeze-out of the ionization are important. 
These effects cause the electron density, as well the temperature, to be larger than in steady
state (Paper I).
Lines dominated by recombination are  affected mainly by the higher
 electron 
density compared to steady state, increasing their luminosities.
Lines formed
mainly by non-thermal excitation are almost
unaffected. Consequently, the time dependence is seen clearest for the
[Fe II] lines at late stage, when the hydrogen and helium regions
dominate. In particular, the optical and near IR-lines, e.g. $\wll
7155$~\AA~ and $1.26 ~\mu$m, increase
at 1000 days by factors 3 -- 4, compared to steady state. Also the
$\wll 17.94, ~25.99 ~\mu$m lines increase by $50 - 100 \%$.  The H I and
He I lines, however, increase by only $\sim 50 \%$, at the same epoch, while
[O I] $\wll 6300, 6364$ is almost unaffected. 

A new feature of this paper is the use of the line profiles for constraining
the distribution of the various components. This was discussed already in
Fransson \& Chevalier (1989) and in FK92, but mainly from a qualitative point
of view. Here, we have shown how both the hydrogen, helium and oxygen
distributions can be determined from the observations. In particular, the
penetration of a substantial hydrogen mass to $\lesssim 700 \kms$ has been
established. This has earlier been proposed in connection to light curve
calculations (e.g., Shigeyama \& Nomoto 1990). Also oxygen and helium
are mixed to low velocities, in 
the case of oxygen to $\lesssim 400 \kms$. In the other direction,
synthesized iron is mixed outwards to at least 2000
$\kms$. Conclusions about mixing of newly synthesized iron to even higher
velocities should, however, be taken with caution, because of the strong
contribution of primordial iron to the lines. Because of the low
resolution of the spectra used, it would be desirable to repeat  this
analysis with data of higher quality.

When discussing the masses of individual elements, the luminosity of the
different lines are of obvious interest. However, the deposition of the
gamma-rays depends on the distance from the radioactive source.
While the deposition per unit mass is fairly uniform inside the radioactive
source, i.e. inside
$\sim 2000 \kms$, it decreases like $V^{-2}$ outside. The mass outside the
radioactive source is therefore not well constrained, unless the  wings of the
lines are well reproduced. Neglect of this leads to an under-estimate of the
mass. 
 
Of the masses determined, the hydrogen zone mass is probably the most
accurate, 
both because of good observations, and because of well-understood physics. The
total mass we find, $\sim 7.7 \Msun$, in the hydrogen component is probably accurate
to $\pm 2 \Msun$. Of this $\sim 3.9 \Msun$ is pure hydrogen. 

 The quality of the line profile of 
 the least blended of the He I lines, $\wl 2.058 ~\mu$m, is unfortunately not
ideal. Especially the wings are not very well defined. Therefore, although
the mass of the helium-rich component inside $\sim 4000 \kms$ is
consistent with $\sim 2
\Msun$, there could be a substantial additional mass at velocities
$\gtrsim 3000 \kms$. There is also an uncertainty in the helium mass
 connected to the form of the continuum destruction probability
 used. As illustrated in Figure \ref{fig:photheivoigt}, there may, depending on this, be a factor of up
to five uncertainty in the flux at early time. At $ \gtrsim 500$ days this
 uncertainty is considerably smaller, and this part of the light curve
 is therefore better suited for helium mass determinations.   We
again note that $\sim 3.9 \Msun$ of helium resides in the 
hydrogen-rich gas. 

In both the 11E1 and 10H models most of the [C I] emission originates from
regions outside of the O -- C shell. Although this is not the case in
the model with the He -- C shell replaced by a He -- N shell, the
influence of CO as a 
coolant in the O -- C shell makes mass estimates based on the [C I]
lines questionable.
In the model where we replaced carbon by nitrogen in the helium
shell, as we argue is
indicated, the [N I] $\wl$ 1.04
$\mu$m line is dominated by the He -- N region. The nitrogen mass in
this model is
$\sim 3.4\EE{-2} \Msun$, of which 
$\sim 2.4\EE{-2} \Msun$ is in the He -- N zone.

The strength of [Ne II] $\wl 12.814 ~\mu$m is in the 10H model dominated
by the
helium zone, while in the 11E1 model the O -- Ne
-- Mg zone takes over. The flux in the [Ne II] line was in the latter
over-produced by a factor of 2 -- 3  later than 500 days . Therefore,
the neon mass in the 11E1 model, $\sim 0.25 \Msun$, is too large, and
can be seen as a solid upper
limit. The neon mass in the oxygen zone of the 10H model, $\sim 2\EE{-2}
\Msun$, together with $\sim 4\EE{-2} \Msun$ in the helium zone, gives
an acceptable fit to the light curve. Allowing for a factor of two
uncertainty in the model, the neon mass in either of these regions, but not
both,   can 
possibly be increased by a factor of two. A best estimate is therefore
$M({\rm Ne}) \approx 6\EE{-2} \Msun$, and most likely $\lesssim 0.1 \Msun$.

 Masses based on Mg I] $\wl 4571$ have
earlier been questioned, based on the sensitivity on the far-UV radiation
(c.f., Fransson \& Chevalier 1989). The fact that the line is dominated by
recombinations from Mg II makes this determination considerably more reliable.
The magnesium mass is likely to be similar to that in the 10H model, $\sim
2.2\EE{-2} \Msun$. The main uncertainty in this comes from the adopted effective
recombination rate of the $\wl 4571$ line.

As we have discussed, and agreeing with Li \& McCray (1993), calcium is not well
constrained, because of the dominance of the zones outside of the Si -- S zone for the
emission in the Ca II lines.
 
The iron mass should be close to the value used here, certainly within a
factor of two. This mass, as well as the
${}^{56}$Co mass, is however, more accurately determined from the
bolometric light curve.  
The amount of stable nickel, ${}^{58}$Ni and ${}^{60}$Ni,  is in our model 
$\sim 0.006 \Msun$ and given the uncertainty in the atomic data for Ni
II, this is consistent with the observations. 

We now turn to a number of more specific points. Of
special interest is to examine the sensitivity of the line emission to the
nucleosynthesis. For this purpose we have for many of the lines compared the
11E1 model by Nomoto
\etal and the 10H model by Woosley. 
First we note that the most apparent discrepancy is in [C I] $\wll
9825, 9850$, which is similar in both models. This over-production, together
with the observed presence of [N I] $\wll 10398, 10408$ lines,  
indicates a less 
extensive  zone of He -- C, replaced by a He -- N zone of CNO-burning
products. This illustrates that although different models agree, there
may be systematic effects due to uncertainties in the input  physics
in the models.  

The main differences between the two models are
found in the Si -- S, and oxygen-rich zones. In the 11E1 model, neon and magnesium
are, next to oxygen, the most abundant elements, while silicon and sulphur
replaces these in the 10H model. The effects of this are most
prominent in the light curves of the [O I] $\wll 6300, 6364$, [Ne II]
$\wl 12.814~ \mu$m, Mg I] $\wl 4571$, [Si I] $\wl 1.6454~ \mu$m, and [S I]
$\wl 1.0820~\mu$m lines. 

Starting with oxygen, the thermal part of the light curve is well reproduced in both
models, although the 11E1 model gives a somewhat better fit.  When it
comes to the non-thermal part,  
both models under-produce the light curve by a large factor, although
10H somewhat 
less than 11E1. The problem with the [O I] line is  discussed further
below.

[Ne II] $\wl 12.814~ \mu$m clearly favors the 10H model,
because of the smaller neon mass in the oxygen zone, while 
Mg I] $\wl 4571$ agrees on the average better in the 10H model. This
applies especially to the 10H model where we 
decreased  the photoionization rate to zero (section \ref{sec-phot}).  
In the 11E1 model the light curve  has the right form, but is at all
epochs a factor 
of 2 -- 3 to strong. 

Because of the dominance of the zones outside of the Si -- S zone for
the [Ca II] emission, these lines are not as sensitive to the differences
in the Si -- S zone between the two models, as one could otherwise
have hoped. Because of the larger Ca II contribution from the O -- Si
-- S zone in the 10H model earlier than 400 days, this model gives a
higher flux and better agreement with the observations compared to the
11E1 model. 

While the 
contribution from the Si - S zone to the [Si I] - [Fe II] $\wl 1.64 ~\mu$m
blend varies considerably, the [Si I] emission is dominated by the
oxygen zone and does not differ much.  The [S I] $\wl 1.0820~\mu$m line
is in 10H 
over-produced by a factor of $\sim 2$ at 200 -- 300 days, and favors
therefore the 11E1 model.

Summarizing this evaluation, there is no clear winner. While
the [S I] lines favor 11E1, [Ne II], Mg I] and Ca II
lines agree better
with 10H. The early evolution of the important [O I] lines may favor
the 11E1 model, but 
because of the serious discrepancy at late time, this is not obvious.
Most discrepancies are at the factor of two level, and in
view of other uncertainties it is not clear how significant the
differences are. One may also take the opposite view and conclude that
within this factor there is good agreement with most abundances in the
models. 

The low level of the non-thermal plateau of the [O I] $\wll 6300,
6364$ lines poses a  problem for our 'standard' models. We have
already discussed various ingredients in a solution to this
problem. From our analytical discussion it is obvious that an
appreciable contribution to the ${}^1D$ excitation from the triplets
is needed. Although the energy to the triplets is sufficient, and
branching to the singlets is probable, photoabsorption of the  $\wl
1302$ line by Mg I and Si I decreases this mode by a large factor.  As
Figure \ref{fig:photoi3} shows, only a decrease 
in the photoabsorption  to near
zero has an appreciable effect on the non-thermal level. Without a
strong decrease of both the magnesium and silicon abundance this is
difficult to achieve, unless we have under-estimated the ionization of
Mg I and Si I by a large factor. In the explosion models either magnesium or
silicon, or both, tend to
have a high abundance in the inner oxygen region. 
It is somewhat ironic that the most simplified model, with only oxygen,
is the most successful model for the non-thermal phase.
We have, however, also
seen that the maximum triplet contribution from recombination  in most
cases is accompanied by too strong O I recombination 
lines, and we have argued that charge transfer with Si I may be a
solution to this. 

Another possible solution to the [O I] problem would be that the electron
fraction is much lower than we have calculated, because the excitation
efficiency increases with decreasing electron fraction. We have already 
experimented with a higher density (i.e. lower filling factor), and
this did not have much effect, because $x_e \propto n^{-1/2}$. Charge transfer can probably only change
$x_e$ by factors of a few, because the recombination rates of Mg II, Si
II, and Ne II are similar to that of  O II.  The UV field does
not influence the O I emission appreciably. 
 
A lower abundance of silicon in the O -- Si -- S region would help
improving both the [Si I] $\wl 1.6454 ~\mu$m line, and also increase the
non-thermal plateau of the [O I] $\wll 6300, 6364$ lines.  
In fact, a general decrease of the convective mixing of elements, like
silicon in the O -- Si -- S region and carbon in the He -- C region, in
connection with the shell burning stages, would probably improve the
agreement with the observations considerably. 
It is interesting that more recent explosion models by Woosley \& Weaver
(1995) give lower silicon abundances. Their 15 $\Msun$ model has
$X({\rm Si}) \sim 1.2\EE{-3}$ and the 25 $\Msun$ model 
$X({\rm Si}) \sim 3.8\EE{-3}$, both lower than in the 11E1
model. These models also show
that the extent of the He -- C zone can vary substantially. In the 15
$\Msun$ model the He -- C zone accounts for less than half of the
helium zone, with the rest is He -- N. In the 25 $\Msun$ model nearly
all of the helium zone is in the form of He -- C. The neon abundance
is in these models higher than even in the 11E1 model, possibly
creating problems.

We also remark that the exact form of the continuum destruction
probability is important, 
and unfortunately uncertain. This is
illustrated by the difference  between
equations (32), (33) and (34)  
in Paper I, where $\beta_{\rm c}$ varies by factors of ten or more, depending
on the value of $k_{\rm C} / k_{\rm L}$. A further study of this for the conditions
relevant for the ejecta would be of great interest.

Because most published observations were obtained during the first four
years, we have in this paper concentrated on the evolution up to $\sim
1200 $ days. Our models are, however, calculated up to 2000 days, and we 
therefore briefly 
compare these with the HST observations by Wang \etal (1996) at 1862
days. At this epoch blending in the UV is less severe than at earlier
epochs. Meaningful estimates of the fluxes of e.g. Mg I $\wl 2852$ and
Mg II $\wl 2800$, as well as the Balmer continuum,  can therefore be
obtained. From a fit to the Balmer continuum Wang \etal find a
temperature at 1862 days of $500 \pm 100$ K. The temperature we find
in the hydrogen-rich gas
decreases from $\sim 425$ K in the core to $\sim 390$ K in the inner
	envelope and $\sim 220$ K at the highest velocities (Figures
2 and 7 in Paper  I). Given the
observational uncertainties, especially in the level of the
'continuum', we find that this
is in reasonable agreement with the observations.

In Table \ref{tab:hst} we give the observed fluxes at 1862 days from
Wang \etal 
compared to the reddening adjusted fluxes from our 10H model at the
same epoch. 
We find the Balmer continuum flux to be a factor of two larger than 
observed. The latter may, however, be a lower limit because of the
high level of the background flux assumed. A factor of two higher flux
could well
be consistent with the observations.
Our $\Ha$ flux is  very close to the observed value, implying that our
 electron density is close to that derived from the observations. Wang
 \etal estimate from $\Ha$ an electron
density  of $\sim 6\EE3 \rm ~cm^{-3}$,
while we find  $4.4\EE3 - 1.6\EE4 \rm ~cm^{-3}$ in the different
regions. The [O I] $\wll 6300, 6364$ flux is in the model a factor of two
 lower than the observed. Although the
discrepancy is smaller than at earlier epochs, there is 
still a substantial uncertainty in the model flux because of the
effects discussed earlier in this section. The Mg I $\wl 2852$ and Mg II $\wll
 2796, 2803$ lines are not resolved in the observations. The total
 flux of the blend is a factor of $\sim 2.6$ higher than the
 calculated value. In the future we hope to make a more detailed
 comparison with these observations, as well as those even later in
 Chugai \etal (1997). 

\section{CONCLUSIONS}
\label{sec-conclusion}
For convenience we here summarize our main results.
\begin{enumerate}

\item From the evolution of especially the [O I] and [Fe II] lines
there is overwhelming evidence that the IR-catastrophe has occurred in the
metal-rich  regions at $\sim 700$ days.

\item The nearly constant ratio  of the [O I] $\wl
6300, 6364$ luminosity and  the bolometric luminosity at $t \gtrsim 800$
days shows that
non-thermal excitation dominates this part of the light curve, which
implies a very low temperature. The same applies to e.g. $\Ha$,
He I $\wl 2.058 ~\mu$m,
and  Mg
I] $\wl 4571$, although here recombination following
non-thermal ionization dominates.
								
\item For reliable mass estimates, constraints from the line profiles
have to be taken into account. Even faint line wings at high velocity may
correspond to comparatively large masses.
The most reliable masses are therefore those within $\sim 3000 \kms$. 

\item We find that the density in the hydrogen envelope is
considerably flatter than in published  models, decreasing like $\rho
\propto V^{-2}$, out to $\sim 5000 \kms$. The 
total hydrogen zone mass is 
$\sim 7.7 \Msun$, of which $\sim 3.9 \Msun$ is  hydrogen. The 
 mass of the hydrogen-rich gas within 2000 $\kms$ is $\sim 2.2 \Msun$. 

\item The helium mass is $\sim 5.8 \Msun$, of which $\sim 1.9 \Msun$
is in the 
helium zone and $\sim 3.9 \Msun$ in the hydrogen zone. Although the He
I $\wl 2.058 ~\mu$m line profile is uncertain, models and observations
indicate that the helium
emitting mass  may extend to $\sim 4000 \kms$. In the center it
penetrates to $\lesssim 500 \kms$. Before $\sim 500$ days the light
curve, and therefore the mass estimate, is sensitive to assumptions
about complete versus partial redistribution.
	
\item Our best determinations of the other masses are $M$(N) $\approx 3.4\EE{-2}
\Msun$, $M$(Ne) $\approx 6\EE{-2} \Msun$, $M$(Mg) $\approx 2.2\EE{-2} \Msun$,
$M$(${}^{58}$Ni + ${}^{60}$Ni) $\approx 6\EE{-3}
\Msun$. The total masses of carbon, sulphur and calcium are not well determined,
because of the low contributions from zones where these
elements are most abundant.

\item The oxygen distribution extends to $\lesssim 400 \kms$, and is
consistent with $\sim 1.9 \Msun$ of oxygen enriched gas. Because of 
uncertainties connected with the non-thermal part of the light curve
there could be a considerable uncertainty in this number, although
Figure \ref{fig:photoi2} gives the likely range.  
The [O I] line therefore remains an unsolved problem
and a better
understanding of  this line is clearly needed.

\item The extent of convective mixing of carbon into the helium-rich
region is likely to be  more limited than stellar
evolution calculations show. Instead, a more massive zone of
CNO-burning products is indicated by the observations. 

\item Most lines, including the O I, Ca II, and Fe II lines, have
contributions from more than one component. Even trace amounts in zones with a large
gamma-ray deposition may give substantial contributions to the total line
fluxes. 

\item The [Fe II] lines have a large contribution from primordial
iron, and at late epochs this component dominates, because of the
IR-catastrophe in the iron core. The extended wings of the [Fe II]
lines are probably produced by primordial, rather than synthesized, iron.

\item Both charge transfer and photoionization effects are important
for the strengths of lines from trace ions. The observations indicate
strong suppression of the UV flux by dust absorption and
resonance scattering. 

\end{enumerate}

Finally, after this paper was submitted we received a preprint by de Kool, 
Li, \& McCray (1997) where a similar analysis and calculation is
performed. In spite of detailed differences, it is comforting that 
the  results of these two independent calculations with regard to both 
the physical conditions and the observed emission are very similar. 

\section*{ACKNOWLEDGMENTS}
We are grateful to many people for discussions at various phases of
this project. In particular, John Houck and Peter Lundqvist have
provided much
advice on atomic data, as well as discussions.
Dima Verner has provided  recombination coefficients to excited levels
in He I,  Peter Meikle and Mark Phillips have made
observational data available in readable format. We
have also had many helpful discussions and comments on these papers by Eddie
Baron,  Roger Chevalier, Robert Cumming, Leon Lucy, Ken
Nomoto, Takashi
Kozasa, Jesper Sollerman, and Jason Spyromilio, and especially by
Dick McCray and  Hongwei Li. This research was
supported by   the Swedish 
Natural Sciences Research Council 
and the G\"oran Gustafsson Foundation  for Research in Natural Sciences and
Medicine.


\begin{deluxetable}{lll}
\tablewidth{20pc}
\tablecaption{Modeled and observed HST-fluxes at 1862 days. \label{tab:hst} }
\tablehead{
\colhead{Line}   & \colhead{Model\tablenotemark{a}} & \colhead{Observation\tablenotemark{a}}
}
\startdata
$\Ha$ & 18.8 & 16.6 \nl
Balmer cont.  & \phn 7.2 & \phn 3.1 \nl
[O I] 6300, 6364 & \phn 1.2 & \phn 2.9 \nl
Mg I] 4571 &  \phn4.6 & \phn 2.2 \nl
Mg I 2852 &  \phn0.58 & \phn 1.7\tablenotemark{b} \nl
Mg II 2796, 2803 & \phn 0.07 & \nl
\tablenotetext{a}{Observed fluxes are from Wang \etal (1996), while the
modeled are from the 10H model, adjusted for reddening. All fluxes are
in units of $10^{-14} \ergs ~\rm cm^{-2}$.}
\tablenotetext{b}{Total flux for Mg I $\wl 2852$ and Mg II $\wll 2795, 2803$.
}
\enddata
\end{deluxetable}

\newpage

\clearpage

\clearpage

\begin{figure}
\plotone{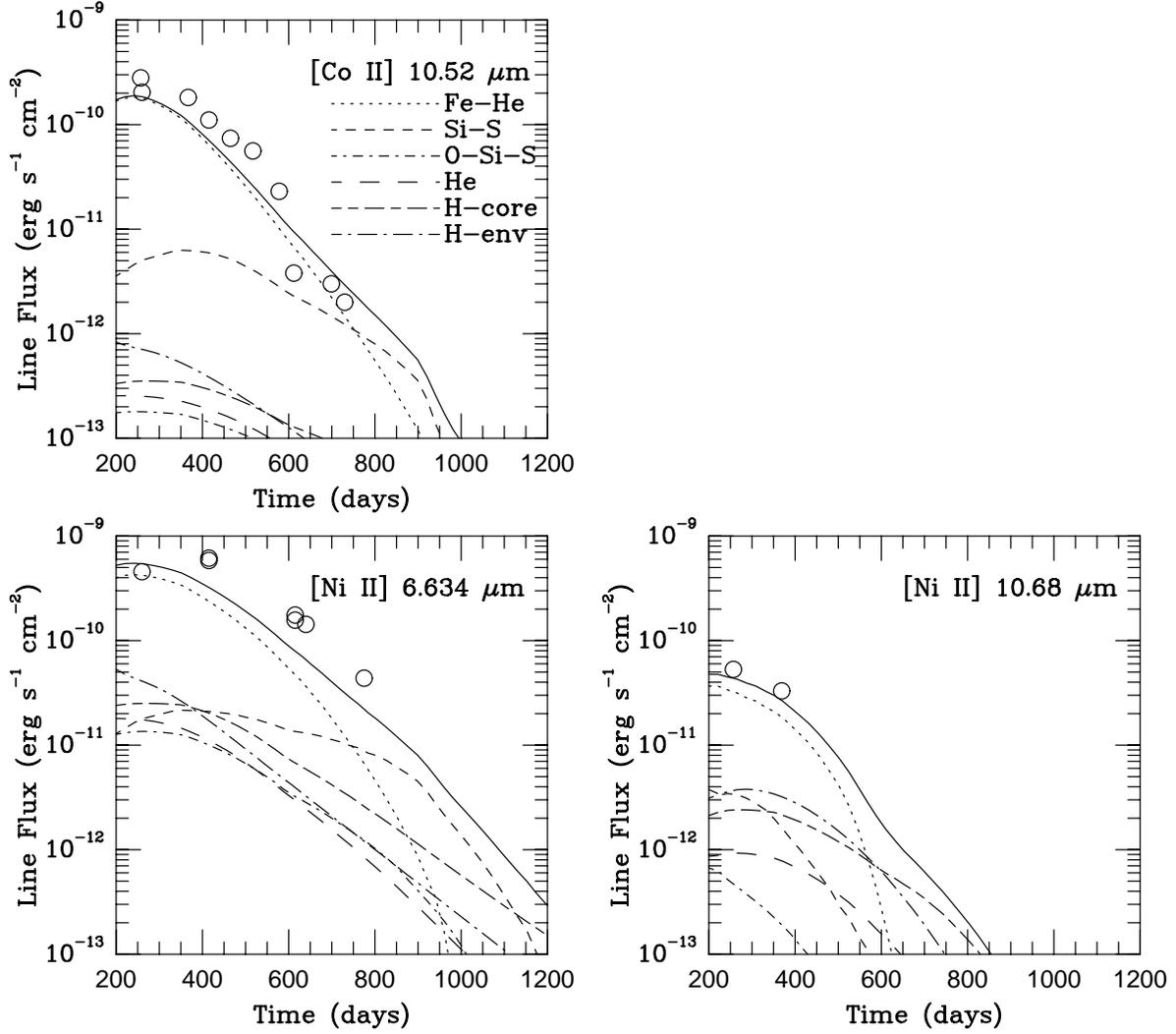}
\caption{Light curves of [Co II] $\wl$ 10.52 $\mu$m,
[Ni II] $\wl$ 6.634
$\mu$m, and [Ni II] $\wl$  10.68 $\mu$m.
The calculations based on the 10H and 11E1 models do not differ
significantly from each other. Observations are from 
Aitken \etal (1988), Jennings \etal (1993), Roche \etal (1993), 
Wooden \etal (1993), Colgan \etal (1994). 
\label{fig:linconi_ii}}
\end{figure}

\begin{figure}
\plotone{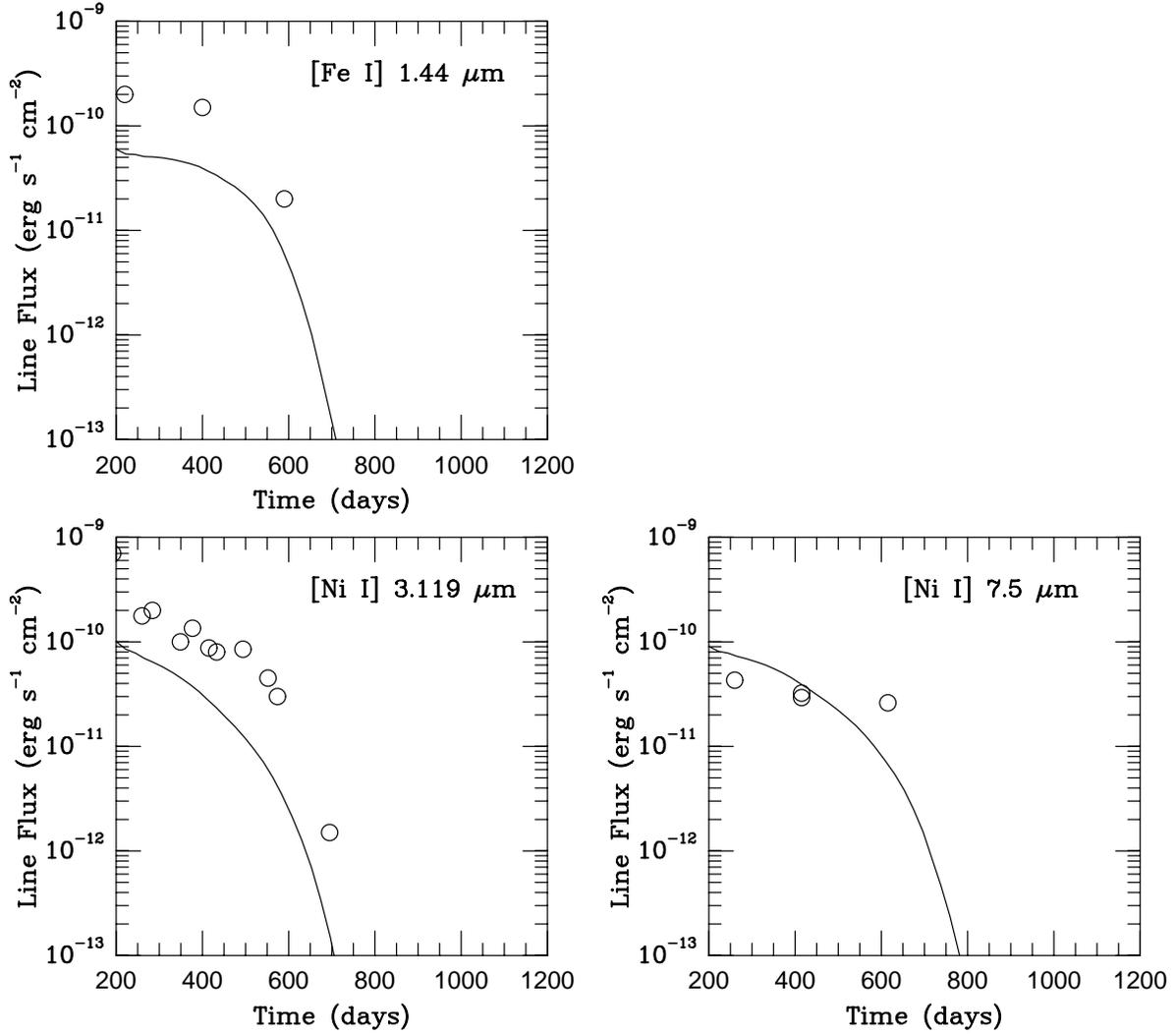}
\caption{Light curves of [Fe I] $\wl$ 1.44 $\mu$m, [Ni
I] $\wl$ 3.119 $\mu$m, and
[Ni I] $\wl$ 7.505 $\mu$m.
Note that in modeling these lines {\em no photoionization} of these
elements is included.
The calculations based on the 10H and 11E1 models do not differ
significantly from each other. Observations for [Fe I] $\wl$ 1.44 $\mu$m are
taken from Oliva (1992) and Oliva,  Moorwood, \&  Danziger  (1989).
Observations for the [Ni I] lines are from Meikle \etal (1989, 1993),
and Wooden \etal (1993).
\label{fig:linfeni_i}}
\end{figure}

\end{document}